\documentclass[PAPER,cernpreprint,coverpage=false,atlasdraft=false,UKenglish, texmf, orcidlogo]{atlasdoc}
\usepackage[backend=biber,biblatex=true,minimal]{atlaspackage}
\usepackage[labelformat=simple]{subcaption}

\usepackage{caption}
\usepackage[style=plain]{floatrow}
\usepackage{graphicx}
\usepackage{booktabs}
\usepackage{csvsimple}
\usepackage{cleveref}
\usepackage{siunitx}
\usepackage{atlasbiblatex}

\usepackage{floatrow}
\usepackage{atlasphysics}
\graphicspath{{logos/}{figures/}}

\usepackage{ANA-STDM-2020-31-PAPER-defs}

\hypersetup{pdftitle={ATLAS document},pdfauthor={The ATLAS Collaboration}}

\AtlasTitle{Measurement of the Lund jet plane in hadronic decays of top quarks and $W$ bosons with the ATLAS detector}

\AtlasVersion{3.3}
\AtlasAbstract{The Lund jet plane (LJP) is measured for the first time in $t\bar{t}$ events, using 140~$\mathrm{fb}^{-1}$ of $\sqrt{s} = 13$~TeV $pp$ collision data collected with the ATLAS detector at the LHC. The LJP is a two-dimensional observable of the sub-structure of hadronic jets that acts as a proxy for the kinematics of parton showers and hadron formation. The observable is constructed from charged particles and is measured for $R=1.0$ anti-$k_t$ jets with transverse momentum above 350 GeV containing the full decay products of either a top quark or a daughter $W$ boson. The other top quark in the event is identified from its decay into a $b$-quark, an electron or a muon and a neutrino. The measurement is corrected for detector effects and compared with a range of Monte Carlo predictions sensitive to different aspects of the hadronic decays of the heavy particles. In the $W$-boson-initiated jets, all the predictions are incompatible with the measurement. In the top quark initiated jets, disagreement with all predictions is observed in smaller subregions of the plane, and with a subset of the predictions across the fiducial plane. The measurement could be used to improve the tuning of Monte Carlo generators, for better modelling of hadronic decays of heavy quarks and bosons, or to improve the performance of jet taggers.
}

\AtlasRefCode{STDM-2020-31}

\PreprintIdNumber{CERN-EP-2024-169}

\arXivId{2407.10879}

\HepDataRecord{157586}

\AtlasJournalRef{\EPJC 85, 416 (2025)}
\AtlasDOI{10.1140/epjc/s10052-025-13924-5}

\begin{document}

\maketitle

%
%

%
%

%
%

%

%
%
%
%
%
%
%
%
%
%

%
%
%
%

%
\section{Introduction}
Jets, the sprays of hadrons resulting from the fragmentation of highly energetic quarks and gluons, are among the most commonly observed radiation patterns in the detector experiments at the Large Hadron Collider (LHC)~\cite{Evans:2008zzb}. They are frequently produced in the decays of heavy particles such as the top quark or the Higgs, $W$ or $Z$ bosons, whose properties are studied in detail at the LHC experiments to test the predictions of the Standard Model. The quantum chromodynamic (QCD) radiation that gives rise to jets, the showers of quarks and gluons and their subsequent recombination into hadrons, cannot be described by perturbative QCD and must be modelled using phenomenological models in Monte Carlo (MC) simulations. The Lund jet plane (LJP)~\cite{lundplane2018dreyer} is an observable that reconstructs the hierarchy of QCD emissions within jets, thus providing insights into the process of jet formation. The LJP belongs to a category of jet substructure (JSS) observables that make use of the jet algorithms that are used to reconstruct the four-momentum of the particle that initiated the parton shower. This paper presents the first measurement of the LJP for jets initiated by top quarks and $W$ bosons, using a dataset comprising $140~\mathrm{fb^{-1}}$ of $pp$ collision data collected by the ATLAS detector at the LHC at a centre-of-mass energy of $13$~TeV. The kinematics of parton showers and hadronisation in these jet topologies are of particular interest due to the large masses of the top and bottom quarks as well as the colour-singlet nature of the $W$ boson.

At the very large centre-of-mass energies achieved by the LHC, unprecedented quantities of heavy particles, including top quarks, have been produced. While the hadronic production of top quarks has been measured with great precision and is well described by perturbative QCD~\cite{TOPQ-2019-23,CMS-TOP-20-001}, hadronic top quark decays are more challenging to measure and model due to the presence of highly boosted decay products. Insights into this process can be gained through measurements of JSS observables, including the LJP, that are sensitive to the different kinematic regimes of parton showers and hadronisation. Recent analyses by ATLAS~\cite{STDM-2017-34,mod:TOPQ-2019-22} and CMS~\cite{CMS-TOP-17-013} have measured jet production cross-sections differentially in a wide range of JSS observables, using jets initiated by top quarks or by $W$ bosons from top decays. This measurement complements those published results with a detailed study of the LJP, a JSS observable that has found many new applications in recent years, as described below.

In LHC experiments, jets are typically reconstructed using sequential recombination algorithms. These algorithms perform a recursive sequence of pairwise recombinations of the four-momenta of final-state particles to construct the jet momentum. The most common jet algorithms of this type belong to the $k_t$-family~\cite{ktalgo1993ellis}, including the anti-$k_t$~\cite{Cacciari:2008gp} and the Cambridge-Aachen (CA)~\cite{caalgo1997dokshitzer} algorithms. Jet algorithms of this kind are particularly suited to hadron collider physics due to their insensitivity to soft and collinear QCD emissions (infrared and collinear safety) \cite{qcdhandbook1993sterman}. To define the size of the reconstructed jets, the jet algorithms require a choice of the radius parameter $R$.

The theoretical basis for the LJP observable is the Lund plane \cite{lundplane1989andersson}. In the development of parton-shower models, where quarks evolve down to the hadronisation energy scale through a succession of soft-gluon emissions, it is useful to plot the range of transverse momenta and angles available to each emission. This two-dimensional (2D) plane spanned by the range of momenta (or alternatively momentum fractions) and angles is called the Lund plane. The tree-like structure of $k_t$-type jet algorithms is motivated by the branching nature of QCD and can loosely be thought of as mimicking the evolution of the parton shower in reverse, with the finished jet representing the highly energetic particle that initiated the shower and the jet inputs representing the outgoing hadrons. Following this analogy, the LJP is constructed from the transverse momenta and angles of proto-jets, the intermediate steps of the algorithm. By following the higher-\pT proto-jet at each step in the clustering tree, one obtains the entries of the primary LJP, which for brevity is simply called the LJP in this paper. %

The first measurement of the LJP was performed by ATLAS using fairly narrow jets with a radius parameter $R=0.4$ in $139~\mathrm{fb^{-1}}$ of dijet events at $13$~TeV~\cite{STDM-2018-57}. This result has been used to test analytic predictions \cite{ljpdensitytheory2020lifson} using all-order resummations of logarithmic terms in perturbative QCD. A similar measurement for $R=0.4$ and $R=0.8$ jets has also been published by CMS \cite{CMS:2023lpp}. The Lund subjet multiplicity, an observable derived from the LJP but with greater sensitivity to higher-order parton-shower effects, was also measured recently by ATLAS~\cite{STDM-2023-07}. At the ALICE experiment, which is able to probe jets at lower $p_\text{T}$ than ATLAS and CMS, LJPs have been used to measure the properties of parton showers without contamination from hadronisation effects~\cite{deadcone1991dokshitzer,deadcone2022alice}. JSS observables have long played an important part in the development of jet-tagging algorithms, particularly for the identification of top-quark- and $W$-boson-initiated jets because these usually exhibit a distinctive multi-pronged internal structure. The taggers often combine innovative machine-learning methods with jet substructure concepts~\cite{CMS-JME-18-002,JETM-2018-03,mltoptaggers2019kasieczka} including the LJP~\cite{lundplaneqgtagging2022dreyer,lundNet2020dreyer}. The LJP was also exploited recently in CMS to improve the calibration of multi-pronged jets produced by highly boosted objects~\cite{CMS-DP-2023-046}.

The jets studied in this analysis are $R=1.0$ (\largeR) jets with transverse momentum (\pT ) larger than $350$~GeV. A single jet is studied in each event. The targeted events contain top quarks from $t\bar{t}$ pairs that decay into a $W$ boson and a $b$ quark, with one of the $W$ bosons decaying hadronically into jets, and the other into an electron or muon and a neutrino. The selected events are separated in two categories, targeting different decay topologies in the \largeR\ jet used to measure the LJP. Events are assigned to either the `top jet' or `$W$ jet' category depending on the mass of the \largeR\ jet and its distance from the nearest $R=0.4$ (\smallR) jet that is identified as coming from a $b$-quark decay ($b$-tagged jet). %

Following the introduction to the measurement in this section, the definition of the LJP is given in Section~\ref{sec:LJPdescription}. A description of the ATLAS detector is presented in Section~\ref{sec:detector}. In Section~\ref{sec:datamcsamples} the dataset and MC simulated samples are described in detail, as are the reconstructed objects and applied selections in Section~\ref{sec:selections}. The procedure for estimating contributions from background processes is described in Section~\ref{sec:backgroundestimation}, while the details of the LJP measurement are given in Section~\ref{ljp:measurement}. Uncertainties associated with the measurement are discussed and presented in Section~\ref{sec:uncertanities}. The results are presented in Section~\ref{sec:result}, followed by the conclusion in Section~\ref{sec:conclusion}.


%
\section{The Lund jet plane}
\label{sec:LJPdescription}
The LJP is a rich representation of the kinematic structure of the radiation that constitutes a jet. The original definition of the Lund plane was based on the theoretical kinematics of quark and gluon emissions inside a parton shower. Colour confinement prevents us from observing these emissions directly; however, a proxy for their kinematics can be constructed from jets, allowing the radiation patterns inside the jet to be interpreted similarly to a parton shower. The jet algorithms used at the LHC are informed by the physics of parton showers, in order to be insensitive to soft and collinear emissions, that do not modify the kinematics of the outgoing hard partons. One jet algorithm that is of particular interest for studying jet substructure is the CA algorithm, which imposes an angle-ordered hierarchy on the sequence of recombinations. This structure is analogous to the angular ordering that emerges when considering a series of gluon emissions in a parton shower, a well-known effect and fundamental property of all quantum field theories. The CA algorithm begins by combining pairs of four-momenta of the final-state particles, or their equivalent objects in the detector, into \emph{proto-jets}, based on the following metric:
\begin{equation*}
d_{ij} = \frac{\Delta R^{2}_{ij}}{R^{2}},
\end{equation*}
where $\Delta R_{ij}$ is the distance in the $(y,\phi)$ plane, defined below, between particles $i$ and $j$, and $R$ is a parameter of the algorithm. For each particle $i$ the algorithm evaluates the minimum of all $d_{ij}$. If this is smaller than one, the four-momenta of particles $i$ and $j$ are combined into a \emph{proto-jet}, otherwise particle $i$ is removed from the list. The procedure is applied recursively to each pair of proto-jets $i,j$ until there are none left with $\Delta R_{ij} < R$.

\begin{figure}[h]
\centering
\includegraphics[width=0.9\textwidth]{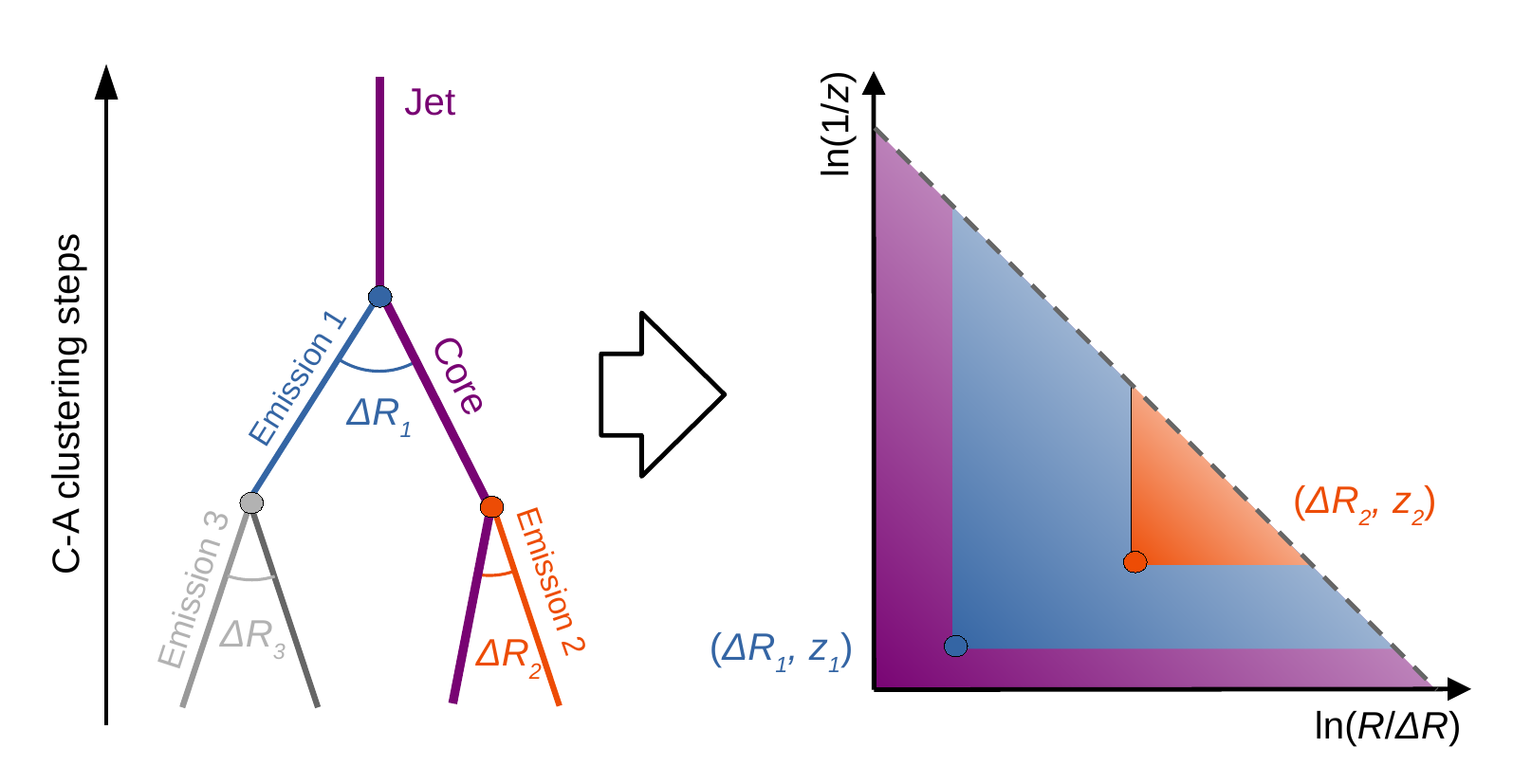}
\caption{Illustration showing the process by which the LJP is constructed. The CA algorithm clusters pairs of proto-jets according to angular separation $\Delta R$. At each step, the higher-$p_\text{T}$ proto-jet is considered to be the core and the lower-$p_\text{T}$ one the emission. The variables $\ln( R / \Delta R)$ and $\ln (1/z)$ are plotted on the LJP for each emission originating from the core branch at each step. Of the three emissions shown, only emissions 1 and 2 are part of the (primary) LJP. The size of the phase space in which subsequent emissions may appear is indicated by the coloured areas. The dashed line indicates the boundary of the low-$k_t$ regime beyond which the number of emissions is strongly suppressed.}
\label{fig:lpconstruction}
\end{figure}

The procedure for reconstructing the Lund jet plane is illustrated in Figure~\ref{fig:lpconstruction}. The steps of the clustering sequence are followed in reverse, starting with the finished jet, and then considering the pairs of proto-jets in the previous steps. Each pair of proto-jets is used to fill a point in the LJP with the following coordinates:
\begin{align} \nonumber
\Delta R^2 &= (y_i - y_j)^2 + (\phi_i - \phi_j)^2, & z &= \frac{p_\text{T}^j}{p_\text{T}^i + p_\text{T}^j},
\end{align}
where $y$ is the rapidity, $\phi$ is the azimuthal angle in the reference frame of the jet, and $p_\text{T}$ is the transverse momentum.\footnotemark{} The indices $i,j$ denote the two `proto-jets', such that $p_\text{T}^{j}<p_\text{T}^{i}$. In analogy with a parton shower, the lower-$p_\text{T}$ proto-jet is commonly referred to as the \emph{emission}, while the higher-$p_\text{T}$ one is called the \emph{core}. The CA-clustering sequence is traversed by repeatedly unclustering the core branch until it is represented by a single jet constituent, adding a point to the LJP at each step. This definition of the plane is often referred to as the primary LJP, since child branches of the emissions are not considered. These emissions would be included in what is called the secondary LJP or even higher-order planes. These higher-order LJPs are not considered in this measurement because they are made up of very soft emissions and are much less densely populated than the primary LJP.

The LJP coordinates $\Delta R$ and $z$ are dimensionless quantities invariant under Lorentz boosts along the $z$-axis and represent the emission angle and transverse momentum fraction. Since the emissions evolve through a wide range of angles and energy fractions, it is convenient to parameterise the LJP using the logarithms $\ln(R/\Delta R)$ and $\ln(1/z)$. In the final measurement, the histograms of the LJP coordinates for all emissions are normalised using the total number of jets $N_\text{jets}$ and the bin widths to give the density of emissions
\begin{equation}
\label{eq:ljpdensity}
\frac{1}{N_\text{jets}}\frac{\dif^2 N_\text{emissions}}{\dif \ln(R/\Delta R)~\dif \ln(1/z)}.
\end{equation}

The jets studied in this analysis originate from the decays of top quarks and $W$ bosons, and consequently contain the particle spray produced by the showering of the heavy particles and the subsequent hadronisation of the decay products. This implies that one of the proto-jet pairs encountered in the LJP declustering will approximately coincide with the outgoing quarks from the heavy-particle decay. That proto-jet pair will have a different momentum-sharing fraction and angular separation than those where the core and the emission correspond approximately to a gluon splitting or the radiation of a gluon from a quark. This produces a structure in the LJP for hadronic decays of top quarks and $W$ bosons that is distinctive from that encountered in studies of jets originating from gluons or light quarks~\cite{STDM-2018-57,CMS:2023lpp}.
This measurement, similarly to all previous LJP measurements, employs only charged particles to reconstruct the plane. The measured LJP differs only slightly from the observable for all final-state particles, as QCD is approximately symmetric under isospin transformations.

\footnotetext{The coordinate system used at ATLAS is right-handed, with the origin located at the expected \emph{interaction point} (IP) in the centre of the detector. The $x$-axis points from the IP to the centre of the LHC ring, the $y$-axis points upward and the $z$-axis points along the direction of the beam pipe. Cylindrical coordinates ($r$, $\phi$) are used in the $x{-}y$ plane, where $\phi$ is the azimuthal angle around the beam pipe. The \emph{pseudorapidity} is defined in terms of the polar angle $\theta$ from the $z$-axis as $\eta = -\ln[\tan (\theta/2)]$. The pseudorapidity is equivalent to the rapidity $y = (1/2) \ln[(E + p_z)/(E - p_z)]$ in the limit of massless particles. Differences in rapidity are invariant under a Lorentz boost along the $z$-axis.}


%
\section{ATLAS detector}
\label{sec:detector}
The ATLAS experiment~\cite{PERF-2007-01} at the LHC is a multipurpose particle detector
with a forward--backward symmetric cylindrical geometry and nearly \(4\pi\) coverage in
solid angle. %
It consists of an inner tracking detector surrounded by a thin superconducting solenoid
providing a \SI{2}{\tesla} axial magnetic field, electromagnetic and hadron calorimeters, and a muon spectrometer.
The inner tracking detector covers the pseudorapidity range \(|\eta| < 2.5\).
It consists of silicon pixel, silicon microstrip, and transition radiation tracking detectors.
Lead/liquid-argon (LAr) sampling calorimeters provide electromagnetic (EM) energy measurements
with high granularity.
A steel/scintillator-tile hadron calorimeter covers the central pseudorapidity range (\(|\eta| < 1.7\)).
The endcap and forward regions are instrumented with LAr calorimeters
for both the EM and hadronic energy measurements up to \(|\eta| = 4.9\).
The muon spectrometer surrounds the calorimeters and is based on
three large superconducting air-core toroidal magnets with eight coils each.
The field integral of the toroids ranges between \num{2.0} and \SI{6.0}{\tesla\metre}
across most of the detector.
The muon spectrometer includes a system of precision chambers for tracking and fast detectors for triggering.
A two-level trigger system is used to select events.
The first-level trigger is implemented in hardware and uses a subset of the detector information
to accept events at a rate below \SI{100}{\kHz}.
This is followed by a software-based trigger that
reduces the accepted event rate to \SI{1}{\kHz} on average
depending on the data-taking conditions.
An extensive software suite~\cite{ATLAS:2024vdo} is used in data simulation, in the reconstruction
and analysis of real and simulated data, in detector operations, and in the trigger and data acquisition
systems of the experiment.

%


%
\section{Data and simulated event samples}
\label{sec:datamcsamples}
\subsection{Data sample}
The measurement is performed on data collected by the ATLAS detector at the LHC from 2015 to 2018 in $pp$ collisions at $\sqrt{s}=13$~TeV. The data must satisfy quality requirements, ensuring that all detector subsystems were fully operational during data taking ~\cite{DAPR-2018-01}. The data sample corresponds to an integrated luminosity of $140.1 \pm 1.2~\mathrm{fb^{-1}}$~\cite{DAPR-2021-01}. The combined integrated luminosity and associated uncertainty is determined primarily by the LUCID-2 detector~\cite{LUCID2}, complemented by measurements using the inner detector and calorimeters.

The events were preselected using a combination of single-lepton triggers~\cite{TRIG-2016-01,TRIG-2018-05,TRIG-2018-01} that require an event to have at least one electron or muon candidate with $\pt > 26$~GeV (or $\pt > 24$~GeV and $\pt > 20$~GeV, respectively, for electrons and muons in 2015 data). Leptons passing the lowest-$p_\text{T}$ electron trigger or muon trigger are required to be isolated from other detector signals. These triggers are complemented by higher-$p_\text{T}$ triggers without such isolation requirements.

\subsection{Monte Carlo simulations}
Monte Carlo simulations are employed in different stages of the analysis: to quantify the acceptance, efficiency and resolution of the detector and the reconstruction and selection applied, and to estimate the uncertainties due to the object calibrations and the modelling of signal and background processes.
Detector effects were incorporated using a simulation of the ATLAS detector~\cite{SOFT-2010-01} based on \GEANT~\cite{Agostinelli:2002hh}.

In all generator configurations excluding \SHERPA, the decays of bottom and charm hadrons from top quarks were simulated using the \EVTGEN\ program~\cite{Lange:2001uf}.

In all simulated event samples, the effect of other interactions in the same, previous and next bunch crossings (pile-up) was incorporated by overlaying the hard-scatter event with a number of simulated inelastic $pp$ events from the \PYTHIA[8.186]~\cite{Sjostrand:2007gs} MC generator using a set of tuned parameters called the A3 tune~\cite{ATL-PHYS-PUB-2016-017}.

\subsubsection{Monte Carlo signal samples}
These samples are used at different stages of the analysis, as illustrated in Sections~\ref{ljp:measurement} and~\ref{sec:uncertanities}, and also for comparisons with the measured LJP in Section~\ref{sec:result}.

The signal is defined as events containing prompt \ttbar pairs. Both single top and \ttbar\ production in association with a vector boson or Higgs boson are treated as irreducible backgrounds (see Section~\ref{sec:backgroundsamples}). Only samples containing events with at least one lepton from a top or anti-top quark decay are used to model the \ttbar signal.

\paragraph{\POWPY[8]} The nominal sample, which is used to model the signal and derive corrections for detector effects, was simulated at next-to-leading-order (NLO) precision in QCD using the \POWHEGBOX[v2]~\cite{Frixione:2007nw,Nason:2004rx,Frixione:2007vw,Alioli:2010xd} matrix-element generator. The \NNPDF[3.0nlo]~\cite{mod:Ball:2014uwa} set of parton distribution functions (PDFs) was employed, the \hdamp\ parameter was set to 1.5\,\mt~\cite{ATL-PHYS-PUB-2016-020}, where a top quark mass \mt\ of 172.5~GeV is assumed and the \pThard parameter is set to = 0~\cite{ATL-PHYS-PUB-2023-029}.
The functional form of the renormalisation and factorisation scales (\muR and \muF) was set to that of the default scale $\sqrt{m_t^2 + p_{\textrm T}^2}$.
Parton showering (PS) and hadronisation were modelled with the \PYTHIA[8.230] generator~\cite{Sjostrand:2014zea} using the A14 tune~\cite{ATL-PHYS-PUB-2014-021} and the \NNPDF[2.3lo]\ set of PDFs~\cite{mod:Ball:2012cx}. The same \PYTHIA[8] tune and PDF set was used for all processes simulated using this generator, unless otherwise specified.

Several alternative \POWPY[8] samples are generated modifying specific settings compared to the nominal.

\paragraph{\POWPY[8] \hdamp var.} The \hdamp\ parameter in \POWHEG\ regulates the amount of radiation against which the \ttbar\ system recoils. To assess the associated uncertainty, an additional \ttbar\ sample with $\hdamp = 3.0\,\mt$ is used~\cite{ATL-PHYS-PUB-2018-009}.

\paragraph{\POWPY[8] (ISR Up/Down, FSR Up/Down)} To estimate the uncertainty associated with the modelling of initial-state radiation (ISR) in \ttbar\ production, the scale parameters \muR and \muF in the signal \POWPY[8] sample are simultaneously varied by a factor of two or a half, while also changing the Var3c eigentune of the A14 \PYTHIA[8]\ tune~\cite{ATL-PHYS-PUB-2014-021}. Similarly, to estimate the effect of the final-state radiation (FSR) uncertainty, variations of the scale parameter $\muR^\textrm{FSR}$ by a factor of two or a half are included. The A14 tune includes a nominal $\alphas^\textrm{FSR}(m_{Z})$ value of 0.127. Changing the $\muR^\textrm{FSR}$ scale results in $\alphas^\textrm{FSR}(m_{Z})$ values of 0.114 and 0.142 for the \emph{FSR Up} and \emph{FSR Down} samples, respectively. All components of this uncertainty are obtained by applying the \PYTHIA[8] internal reweighting feature.

\paragraph{\POWPY[8] (MEC Off)} The nominal \POWPY[8] sample includes matrix-element corrections (MEC) to the parton showers in both production and decay. These corrections exploit tree-level matrix elements that describe the emission(s) of interest to improve the accuracy of the prediction and approximate the emission at NLO precision~\cite{Frixione:2023hwz}. An additional sample with the MECs turned off was produced to quantify the effect of these corrections in comparison with the measured LJP.

\paragraph{\POWPY[8] (RTT)} Another \POWPY[8] sample, called recoil-to-top (RTT), was generated to estimate an associated uncertainty. This sample employs a different strategy for the treatment of recoil against gluon emissions from $b$-quarks originating in top quark decays. In the nominal setting, only the $b$-quark makes up the recoil system, while the RTT configuration uses the on-shell $W$ boson and the top quark itself~\cite{PhysRevD.100.076006}.

\paragraph{\POWPY[8] (\pThard)} To evaluate the uncertainty from the procedure for matching the hard-scatter process with the parton shower, an alternative sample was produced with the \pThard parameter~\cite{10.21468/SciPostPhys.12.1.010} in \PYTHIA[8.230] set to 1. This parameter changes the scale used to veto emissions in the PS to avoid overlaps between the phase spaces of the outgoing partons and the initial emissions of the \PYTHIA shower, and as a consequence it influences both the phase space available for emissions and the related kinematic observables, such as the number of jets or the momentum of the \ttbar\ system.

\paragraph{\POWHER (7.0, 7.1, 7.2)} A sample produced with the \POWHEG\ matrix-element generator and the alternative \HERWIG[7.2]~\cite{herwig72releasenotes2020bellm} parton-shower and hadronisation model is used to estimate the associated modelling uncertainty. In this alternative sample, the \HERWIG[7.2] default set of tuned parameters and the \MMHT[lo] PDF set~\cite{Harland-Lang:2014zoa} are used. Additional MC samples are used to perform comparisons with the measured LJP; these employ \HERWIG[7.0.4] and \HERWIG[7.1.3] with the \MMHT[lo] PDF set and the default tunes for \HERWIG[7.0]~\cite{Bellm:2015jjp} and  \HERWIG[7.1]~\cite{Bellm:2017jjp} respectively. The various \HERWIG[7] generator versions differ in terms of tunes of the hadronisation parameters, simulation of multiple-parton interactions and choice of evolution variable in the parton shower.

\paragraph{\MGNLOPY[8]} Samples from additional \ttbar\ matrix-element generator set-ups are compared with the measurement. One such sample was obtained using the \MGNLO[2.6.0] generator~\cite{Alwall:2014hca} (later referred to as \AMCatNLO) with the \NNPDF[3.0nlo] PDF set for the hard-scatter simulation, the same functional form of \muR and \muF as used in \POWHEG, and the shower starting scale set to $\mu_{q} = H_\mathrm{T}/2$, where $H_\mathrm{T}$ is defined as the scalar sum of the $p_\text{T}$ of all outgoing partons. This hard-scatter generator was interfaced to \PYTHIA[8]. The MEC corrections are disabled in this sample.

\paragraph{\SHERPA[2.2.10]} A sample from a generator chain completely different from those above was obtained using \SHERPA[2.2.10]~\cite{Bothmann:2019yzt}. The sample uses NLO-accurate matrix elements for processes with up to one additional parton, and LO-accurate matrix elements for up to four additional partons. These were calculated using the Comix~\cite{Gleisberg:2008fv} and \OPENLOOPS~\cite{Buccioni:2019sur,Cascioli:2011va,Denner:2016kdg} libraries. The central energy scale in this sample has the functional form $\mu^{2} = m_t^{2} + 0.5\times(p_{\mathrm{T},t}^{2} + p_{\mathrm{T},\bar{t}}^{2})$. These were matched with the \SHERPA\ parton-shower simulations~\cite{Schumann:2007mg}. The CKKW matching scale for additional emissions was set to 30~\GeV.

In the \POWHEG\ and \MADGRAPH generators, top quarks are decayed at leading order with the narrow-width approximation and dedicated corrections to preserve spin correlations and include off-shell effects~\cite{Frixione:2007zp}. All \ttbar samples are normalised to the cross-section $\sigma(t\bar{t})_{\mathrm{NNLO+NNLL}} = 832 \pm 51$~pb calculated at next-to-next-to-leading order (NNLO) in QCD including the resummation of next-to-next-to-leading logarithmic (NNLL) soft-gluon terms, calculated using the \TOPpp[2.0] program~\cite{Beneke:2011mq,Cacciari:2011hy,Baernreuther:2012ws,Czakon:2012zr,Czakon:2012pz,mod:Czakon:2013goa,Czakon:2011xx}.

\subsubsection{Monte Carlo background samples}
\label{sec:backgroundsamples}
The production of a top quark in association with a $W$ boson ($tW$) was modelled using the \POWHEGBOX[v2] generator at NLO in QCD, using the five-flavour scheme and the \NNPDF[3.0nlo] set of PDFs. The diagram-removal (DR) scheme~\cite{Frixione:2008yi} was used to remove interference and overlap with \ttbar production, while the diagram-subtraction (DS) scheme~\cite{Frixione:2008yi,ATL-PHYS-PUB-2016-020} was employed to estimate the uncertainty in the interference treatment. Single top quark $t$-channel and $s$-channel production was modelled using the \POWHEGBOX[v2] generator at NLO in QCD, with the \NNPDF[3.0nlo] set of PDFs. The four-flavour scheme was used for the $t$-channel, while the five-flavour scheme was employed in the $s$-channel modelling. Parton showering and hadronisation in the single top-quark processes was modelled with \PYTHIA[8.230] using the A14 tune.

Production of a $W$ or $Z$ boson in association with jets was simulated using the \SHERPA[2.2.1]~\cite{Bothmann:2019yzt} generator, using NLO matrix elements for up to two partons, and LO matrix elements for up to four partons, calculated with the Comix~\cite{Gleisberg:2008fv} and \OPENLOOPS~\cite{Buccioni:2019sur,Cascioli:2011va,Denner:2016kdg} libraries.
These were matched with the \SHERPA\ parton-shower simulations. The samples are normalised to the NNLO predicted total cross-section~\cite{Anastasiou:2003ds}.

The production of \ttZ\ and \ttW\ events was modelled at NLO in QCD using the \AMCatNLO[2.3.3] generator, while the \ttH\ process was modelled using \POWHEGBOX[v2]. Parton showering and hadronisation in these processes was performed using \PYTHIA[8.210] (\ttZ\ and \ttW) or \PYTHIA[8.230] (\ttH).
The diboson final states ($VV$,$V=Z/W$) were modelled using the \SHERPA[2.2.2] generator, including off-shell effects and virtual Higgs boson contributions where appropriate.


%
\section{Object definitions and event selection}
\label{sec:selections}
Tracks are reconstructed by applying an iterative algorithm to hits in the inner detector~\cite{ATL-PHYS-PUB-2015-051}. They are required to have $\pt > 500$~MeV and $|\eta|<2.5$, and must pass quality requirements based on the number of layers crossed in the silicon tracking detectors. Additionally, to reduce pile-up contamination, the tracks are required to be matched to the primary vertex by having a transverse impact parameter $|d_0|$, relative to the beam line, of less than $2$~mm and a longitudinal impact parameter $z_0$, relative to the primary vertex, satisfying $|z_0 \sin(\theta)| < 3$~mm, where $\theta$ represents the track's polar angle. The primary vertex is defined as the vertex with the largest $\sum p_{\mathrm{T}}^2$ of the associated tracks among all the vertices with at least two associated tracks with $p_{\mathrm{T}} > 500$~MeV.

\LargeR\ jets are reconstructed using the \antikt algorithm~\cite{Cacciari:2008gp} with radius parameter $R=1.0$, as implemented in FastJet~\cite{Fastjet}. The jets are built from topological clusters of energy deposits in the calorimeter (\emph{topoclusters}) which are calibrated using the local hadronic cell weighting scheme~\cite{PERF-2014-07}. The \emph{trimming} algorithm~\cite{Krohn2010}, with parameters $R_{\mathrm{trim}}=0.2$ and $f_{\mathrm{trim}} < 0.05$, is applied to the jet inputs to mitigate the effects of pile-up and underlying event on the reconstructed jet. All \largeR\ jets are required to have $|\eta| < 1.1$ and $\pT > 200$~GeV. The \LargeR\ jets are defined in the manner above for this measurement because they have been widely studied in ATLAS analyses and the calibration of their energy scale and resolution, with associated systematic uncertainties, is well understood~\cite{JETM-2018-02}.

\SmallR\ jets are reconstructed by applying the \antikt algorithm with radius parameter $R=0.4$ to tracking and calorimetry information, combined using the particle-flow algorithm~\cite{PERF-2015-09}. In order to suppress jets originating from pile-up, \smallR\ jets with $p_\mathrm{T} < 60$~GeV are required to be associated with the primary vertex according to the jet-vertex-tagger algorithm (JVT)~\cite{PERF-2014-03}. All \smallR\ jets used in the analysis are required to have $p_{\mathrm{T}} > 30$~GeV and $|\eta|<2.4$. The \smallR\ jets are identified as originating from a $b$-hadron by using a multivariate algorithm~\cite{FTAG-2019-07} that combines information about large-impact-parameter tracks, the topological decay chain, and displaced decay vertices. The working point used has an efficiency of 77$\%$ in simulated \ttbar events and a corresponding rejection factor of 5 or 200 for jets initiated by $c$-quarks or light quarks and gluons, respectively. The energies of \smallR\ and \largeR\ jets are corrected by using energy- and $\eta$-dependent calibration factors derived from simulation and in situ measurements~\cite{JETM-2018-05,JETM-2018-02}. %
Electrons are reconstructed from a combination of inner detector and electromagnetic calorimeter (ECAL) information~\cite{EGAM-2018-01}. The reconstruction quality of the electron candidates is classified using a statistical likelihood-function. The electrons must satisfy a stringent reconstruction quality requirement (defined as the `TightLH' working point in Ref.~\cite{PERF-2017-01}) to be considered in the event. Electron candidates must also be isolated from surrounding tracks and topoclusters (defined as the `FCTight' working point in Ref.~\cite{PERF-2017-01}). Electron tracks are matched to the primary vertex by requiring a transverse impact parameter significance $|d_0/\sigma(d_0)| < 5$ and $|z_0 \sin(\theta)| < 0.5$~mm. Topoclusters associated with electron candidates are required to lie within the region $|\eta| < 2.5$, excluding the barrel--endcap transition region $1.37 < |\eta| < 1.52$ in the ECAL. To match the \pt requirement of the trigger and ensure a high trigger efficiency, reconstructed electrons are required to have $\pt > 27$~GeV.

Muons are reconstructed by combining inner-detector tracks with tracks reconstructed in the muon spectrometer~\cite{MUON-2018-03}. Their inner-detector tracks are matched to the primary vertex by requiring $|d_0/\sigma(d_0)| < 3$ and $|z_0 \sin(\theta)| < 0.5$~mm. The muons must be well reconstructed (at the `Medium' working point, as defined in Ref.~\cite{MUON-2018-03}) and must satisfy the same stringent isolation requirements as the electrons. Like the electrons, the muons are required to have $|\eta| < 2.5$ and $\pt > 27$~GeV.

The missing transverse momentum $\vec{p}_\mathrm{T}^\mathrm{miss}$, with magnitude \met, is defined as the negative sum of the transverse momenta of the reconstructed and calibrated physical objects, as well as a `soft term' built from all other tracks that are associated with the primary vertex~\cite{JETM-2020-03}.

To avoid double-counting of energy, an overlap removal procedure is applied to the reconstructed objects. Overlapping objects are removed from the event in the following order: electrons sharing a track with a muon; the closest \smallR\ jet within $\Delta R = 0.2$ of an electron; electrons within $\Delta R = 0.4$ of a \smallR\ jet; \smallR\ jets within $\Delta R = 0.4$ of a muon if they have at most two associated tracks; and muons within $\Delta R = 0.4$ of a \smallR\ jet.

All events employed in the analysis must satisfy a preselection requiring a primary vertex, exactly one muon or one electron, $\met > 20$~GeV and $\met + m_\mathrm{T}^W > 60$~GeV.\footnote{$m_\mathrm{T}^W = \sqrt{ 2 p_{\mathrm{T}}^\ell \met \left( 1 - \cos \Delta \phi ( \vec{p}_\mathrm{T}^{\ell}, \vec{p}_\mathrm{T}^\mathrm{miss} ) \right)}$} Events must also contain at least one $b$-tagged \smallR\ jet $j_{b1}$ close to the lepton $\ell$, satisfying $\Delta  R(\ell, j_{b1}) < 1.5$. All events are required to have one \largeR\ jet $J$ with $p_{\mathrm{T}} > 350$ GeV separated from the lepton by $\Delta R(\ell,J) > 2.3$. In each event, only the leading (i.e.\ highest-\pt) \largeR\ jet is used to reconstruct the LJP.

Lastly, the \largeR\ jet is classified as belonging to one of two exclusive categories based on its compatibility with the kinematics and topology of a hadronically decaying top quark or $W$ boson. The `top jet' selection requires the leading \largeR\ jet to have a mass $m_J > 140$~GeV and requires a second $b$-tagged \smallR\ jet $j_{b2}$ to be contained within the \largeR\ jet, with a distance requirement $\Delta R(J, j_{b2}) < 1.0$. The `$W$ jet' selection only requires the \largeR\ jet to have $60 <m_J< 100$~GeV, with no requirement placed on the presence of additional $b$-tagged \smallR\ jets in the event.


%
\section{Background estimation}
\label{sec:backgroundestimation}
Most of the background processes are estimated using MC simulations, as has been described in Section~\ref{sec:datamcsamples}. The only exception is the contribution from misreconstructed events in which no real leptons are produced promptly in the hard-scatter process. This category contains `fake-lepton' events where a lepton is radiated in a hadron decay chain or where a jet is misreconstructed as a lepton. The estimation is performed using a data-driven approach called the matrix method~\cite{mod:EGAM-2019-01}. This technique has been applied in other analyses sharing similar final states~\cite{TOPQ-2019-23,mod:TOPQ-2019-22}. The method is based on the definition of \emph{loose} and \emph{tight} criteria for lepton reconstruction quality and isolation. The tight requirements are reported in Section~\ref{sec:selections}. Loose leptons do not have any isolation requirement and use the \emph{MediumLH} working point for electron identification. The fake-lepton background is estimated by weighting the data events containing loose leptons by factors based on the known efficiencies for a real or fake lepton to meet the tight criteria. These efficiencies are evaluated using event topologies enriched in real or fake leptons respectively. The fake-lepton background, which is mainly due to multijet events, comprises less than 1\%\ of the predicted sample for both the $W$ and top jets.

The number of events passing the selections in data, and the predictions from the MC samples, is reported in Table~\ref{tab:yield}. From these numbers it follows that the data sample has a purity\footnote{Defined as $(\mathrm{Data}-\mathrm{Background})/\mathrm{Data}$.} of $96\%$ in the top jet selection and $82\%$ in the $W$ jet selection. The main source of background is the single-top process, which contributes $48\%$ and $60\%$ of the background in the top jet and $W$ jet selections, respectively.

\begin{figure}[h]
\def\sfwidth{ 0.495\textwidth }

\centering
\begin{subfigure}[b]{\sfwidth}
\centering
\includegraphics[width=\textwidth,trim={0.3cm 0 1.5cm 0},clip]{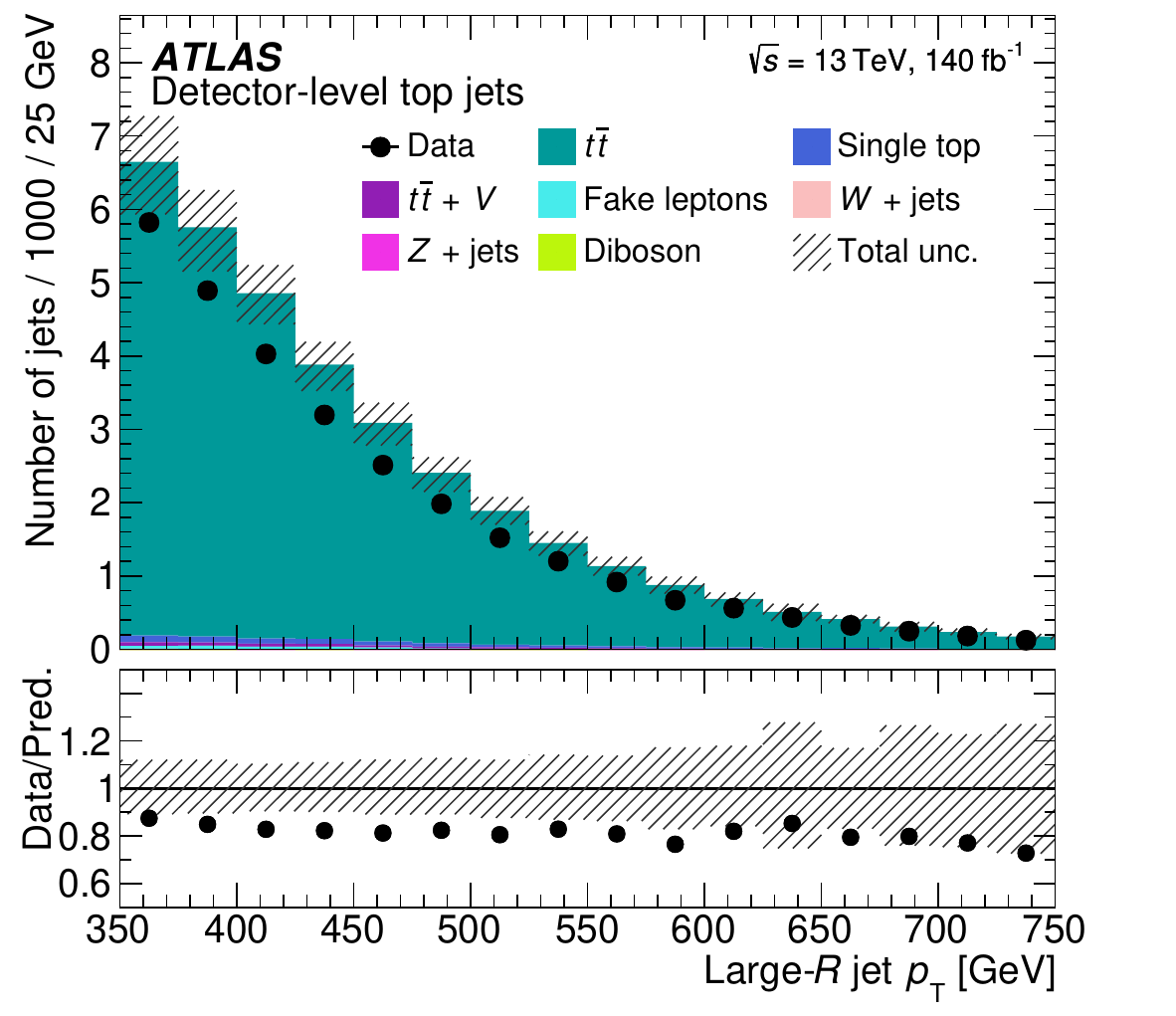}
\caption{}
\label{fig:recoplot:a}
\end{subfigure}
\begin{subfigure}[b]{\sfwidth}
\centering
\includegraphics[width=\textwidth,trim={0.3cm 0 1.5cm 0},clip]{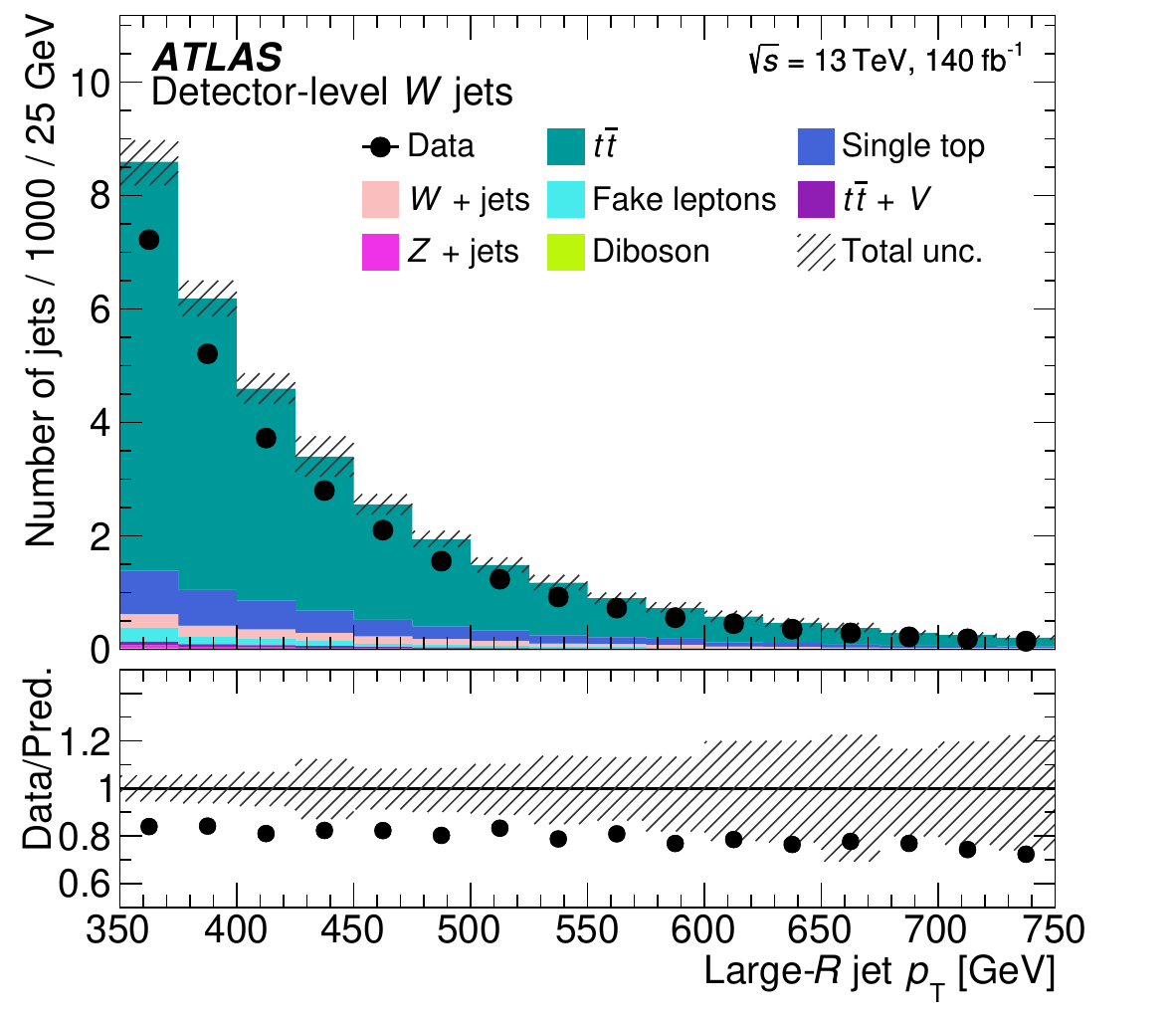}
\caption{}
\label{fig:recoplot:b}
\end{subfigure}
\begin{subfigure}[b]{\sfwidth}
\centering
\includegraphics[width=\textwidth,trim={0.3cm 0 1.5cm 0},clip]{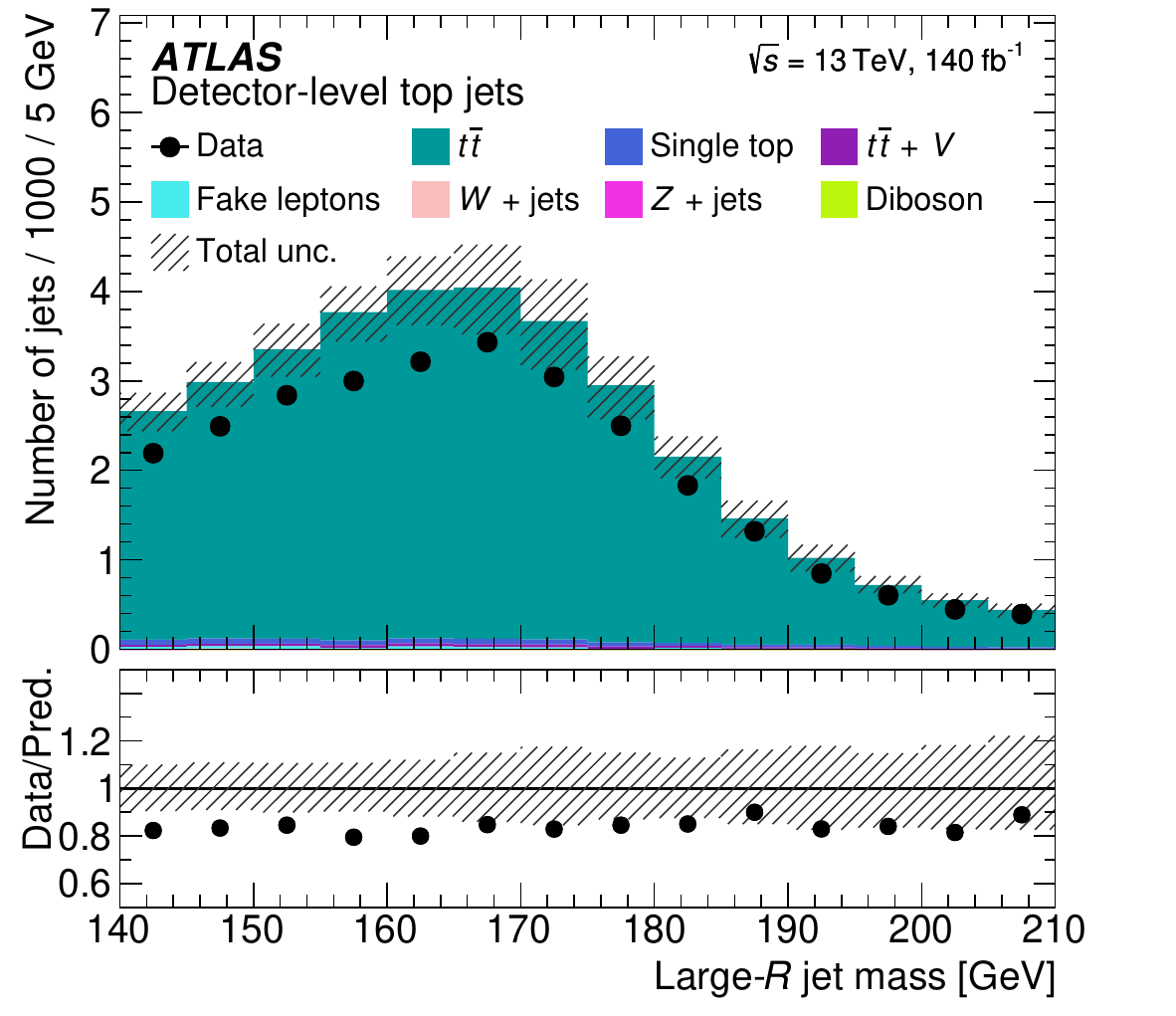}
\caption{}
\label{fig:recoplot:c}
\end{subfigure}
\begin{subfigure}[b]{\sfwidth}
\centering
\includegraphics[width=\textwidth,trim={0.3cm 0 1.5cm 0},clip]{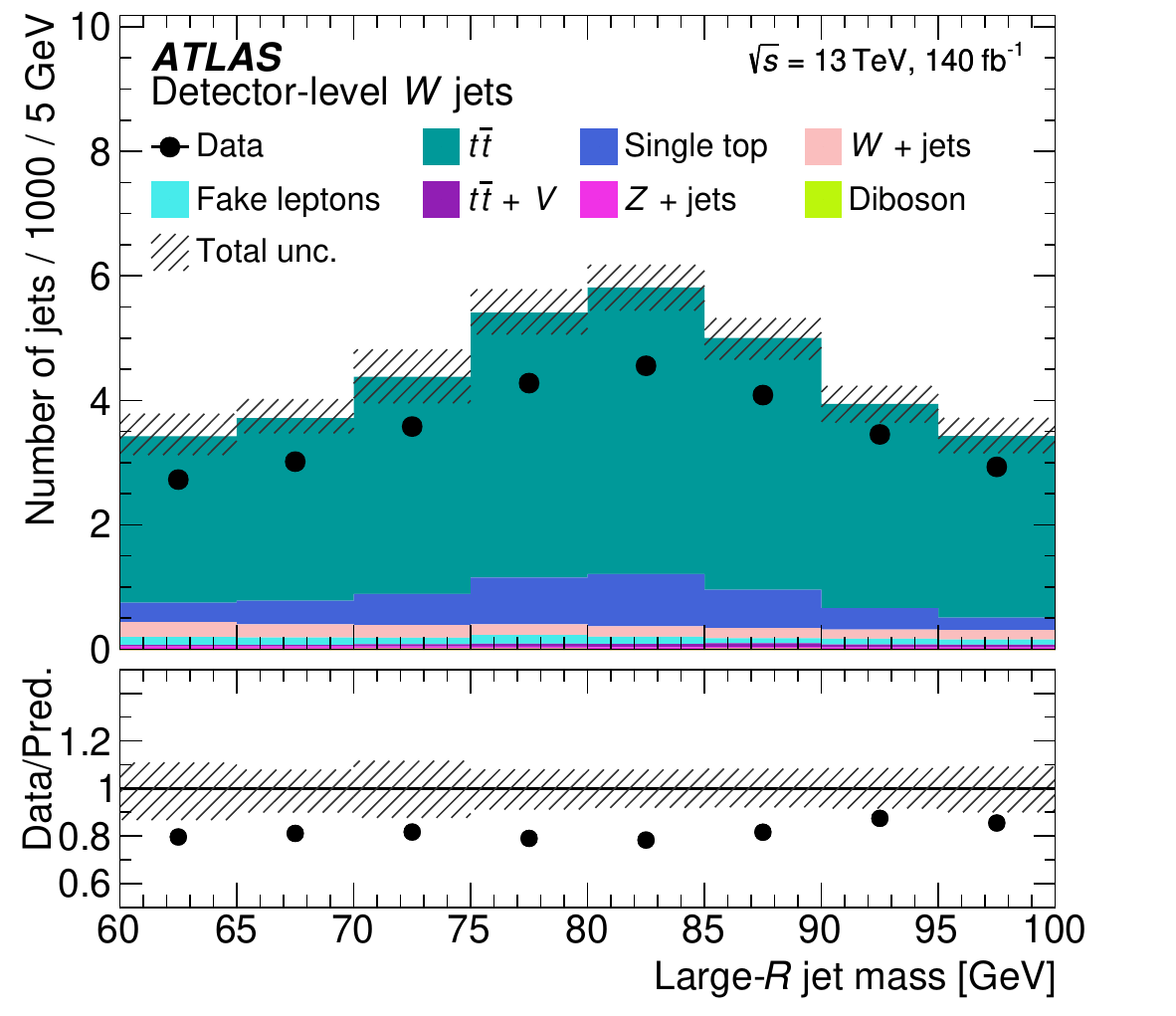}
\caption{}
\label{fig:recoplot:d}
\end{subfigure}
\caption{Detector-level number of events as a function of the \largeR\ jet \pT\ for (a) top jets and (b) $W$ jets and as a function of the \largeR\ jet mass for (c) top jets and (d) $W$ jets. The uncertainty band includes detector and background modelling uncertainty components. In the bottom pad, the ratio of the data to the prediction is shown. The discrepancy between MC predictions and the data is described in the text.}%
\label{fig:recoplot}
\end{figure}

Table~\ref{tab:yield} shows there is tension between the measured data and the predicted yield, with their ratio being about $84\%$ in each of the two selections. The same difference is visible in Figure~\ref{fig:recoplot}, where kinematic distributions of the \largeR\ jets in the data and MC samples are compared. This mismodelling increases with \largeR\ jet \pt, as is visible in Figures~\ref{fig:recoplot:a} and~\ref{fig:recoplot:b}. Here, the following trend can be observed: predictions overestimate the data by around $30\%$ in the tail of the distribution and by $15-20\%$ in the head of the distribution. This mismodelling has been observed in many ATLAS and CMS measurements of the \ttbar\ production cross-section performed in the high ($>350$~GeV) top quark \pt region and mostly derives from missing higher-order corrections to the normalised NLO matrix-element predictions with parton-shower matching (NLO+PS)~\cite{TOPQ-2018-11,TOPQ-2019-23,CMS-TOP-18-013}. In addition to the number of events passing each selection, Table~\ref{tab:yield} also displays the total number of reconstructed LJP emissions (as defined in Section~\ref{sec:LJPdescription}). The MC emission counts have not been corrected using unfolding or matching between the detector- and particle-level emissions.

Figures~\ref{fig:recoplot:c} and~\ref{fig:recoplot:d} show the measured distributions of the \largeR\ jet mass in the top jet and $W$ jet selections. The jet mass distributions peak close to the top quark and $W$ boson masses, as would be expected for jets containing the decay products of on-shell top quarks or $W$ bosons.

\sisetup{round-mode=places,round-precision=0}
\begin{table}[h]
\centering
\caption{Predicted and actual yields for the top jets and $W$ jets. The $t\bar{t}$ prediction is from the nominal \POWPY[8] sample. The second and fourth columns report the actual number of emissions that enter the LJP. The uncertainties are the total uncertainties estimated as described in \Cref{sec:uncertanities}.}
\label{tab:yield}
\begin{tabular}{l@{\hspace{2.5em}}r@{}c@{}r@{\hspace{2.5em}}r@{}c@{}r@{\hspace{2.5em}}r@{}c@{}r@{\hspace{2.5em}}r@{}c@{}r }\toprule %
& \multicolumn{6}{c}{\bfseries Top jets} & \multicolumn{6}{c}{\bfseries $W$ jets}   \\
\bfseries Sample & \multicolumn{3}{c}{Events} & \multicolumn{3}{c}{Emissions} & \multicolumn{3}{c}{Events} & \multicolumn{3}{c}{Emissions} \\
\midrule
$t\bar{t}$ & \num{33800} & $\;\pm\;$ & \num{3400} & \num{216000} &$\;\pm\;$& \num{22000} & \num{28000} &$\;\pm\;$& \num{2900} & \num{164000} &$\;\pm\;$& \num{17000} \\
Single top & \num{650} &$\;\pm\;$& \num{170} & \num{4200} &$\;\pm\;$& \num{1100} & \num{900} &$\;\pm\;$& \num{1000} & \num{22000} &$\;\pm\;$& \num{6000} \\
$t\bar{t}$ + $V$ & \num{330} &$\;\pm\;$& \num{50} & \num{2200} &$\;\pm\;$& \num{300} & \num{330} &$\;\pm\;$& \num{40} & \num{1850} &$\;\pm\;$& \num{250} \\
Fake leptons & \num{230} &$\;\pm\;$& \num{120} & \num{1400} &$\;\pm\;$& \num{700} & \num{900} &$\;\pm\;$& \num{400} & \num{5400} &$\;\pm\;$& \num{2800} \\
$W$ + jets & \num{110} &$\;\pm\;$& \num{40} & \num{760} &$\;\pm\;$& \num{290} & \num{1500} &$\;\pm\;$& \num{600} & \num{9100} &$\;\pm\;$& \num{3400} \\
$VV$ & \num{12} &$\;\pm\;$& \num{6} & \num{80} &$\;\pm\;$& \num{40} & \num{170} &$\;\pm\;$& \num{90} & \num{1000} &$\;\pm\;$& \num{500} \\
$Z$ + jets & \num{8} &$\;\pm\;$& \num{4} & \num{47} &$\;\pm\;$& \num{24} & \num{100} &$\;\pm\;$& \num{50} & \num{800} &$\;\pm\;$& \num{400} \\
\midrule
Total pred. & \num{35100} &$\;\pm\;$& \num{3400} & \num{224533} &$\;\pm\;$& \num{22000} & \num{35000} &$\;\pm\;$& \num{3100} & \num{204000} &$\;\pm\;$& \num{18000} \\
Data & \num{29328} & & & \num{189902} & & & \num{28686} & & & \num{166533} &
\\\bottomrule
\end{tabular}
\end{table}

\FloatBarrier


%
\section{Measurement of the Lund jet plane}
\label{ljp:measurement}
\subsection{Particle-level object definitions}
The LJP observable is reconstructed using stable particles.\footnote{Particles with $c\tau > 10$~mm, where $\tau$ represents the mean proper lifetime of the particle.} These are identified in the MC generator record without considering simulations of their interactions with the detector material.

Each particle-level charged lepton is \emph{dressed}, meaning its four-momentum is summed with that of any prompt photon within a cone of size $\Delta R=0.1$. The \met is calculated from the \emph{prompt} neutrinos originating from the decays of the on-shell $W$ bosons. Non-prompt charged leptons and neutrinos, which are typically produced in the decays of hadrons, or $\tau$-leptons produced in their decay chains, are not considered. Events where the leptons originate from the decay of a $\tau$-lepton originating from a $W$ boson decay are accepted provided they pass the kinematic selections outlined below.

Particle-level \smallR\ and \largeR\ jets are built from all stable particles in the event except the prompt neutrinos and the dressed charged leptons. Jets are constructed using the \antikt algorithm with a radius parameter value of $R=0.4$ or $1.0$. The overlap removal strategy for the particle-level jets and leptons is simpler than at the detector level: dressed muons and electrons are removed from the event if they are separated from a jet by $\Delta R<0.4$.

Particle-level $b$-tagging is performed by ghost-matching~\cite{Cacciari:2008} $b$-hadrons with $\pT > 5 $~GeV to the \smallR\ jets. In the ghost-matching method, the jet algorithm is used to cluster the $b$-hadrons together with the stable particles making up the jet's constituents while treating the momenta of the former as negligible.

With the particle-level objects passing all the selections above, particle-level signal events must contain exactly one charged lepton with $p_{\mathrm{T}} > 27$~GeV and $|\eta| < 2.5$ originating from the top quark decays, as well as \met, one \largeR\ jet and at least one \smallR\ jet with a ghost-associated $b$-hadron. The kinematic cuts on the \met and jets are the same as for the corresponding detector-level objects. The cuts that define the top jet and $W$ jet selections, as described for the detector-level objects in Section~\ref{sec:selections}, are also applied at the particle level.

\subsection{Reconstruction of the LJP observable}
At the detector level, this analysis uses charged-particle tracks as inputs for the CA algorithm to reconstruct the LJP. This approach takes advantage of the high angular resolution of the tracker to reduce the uncertainty of the $\Delta R$ coordinate in the LJP. This charged-particle LJP is what was studied in previous measurements~\cite{STDM-2018-57,CMS:2023lpp} and is very similar to the all-particle LJP because of the approximate isospin symmetry of QCD.
Only tracks ghost-matched~\cite{Cacciari:2008} to the leading trimmed large-$R$ jet are used to reconstruct the LJP. The topoclusters that make up the jet are clustered together with the tracks using the same jet definition. To leave the jet momentum unaffected, the magnitude of the momentum of each track is set to a value very close to zero. The tracks that are clustered with the constituent topoclusters are called \emph{ghost-tracks} and are subjected to the same trimming algorithms as the jet's constituents.

The particle-level LJP is built from the charged constituents of the particle-level \largeR\ jets. Like the detector-level tracks, these must fulfil the requirements $\pT\ > 500$ MeV and $|\eta| < 2.5$. Trimming is applied to the particle-level jet constituents with the same parameter values used at the detector level. Also, at particle level, only the leading trimmed large-$R$ is employed to reconstruct the LJP.

The LJP is built using the appropriate detector- or particle-level inputs by following the procedure described in Section~\ref{sec:LJPdescription}. It should be noted that the CA algorithm is used in the LJP reconstruction even though the \largeR\ jets were built from their constituents using the \antikt algorithm. The anti-$k_t$ algorithm is preferred in the reconstruction of the jet four-momentum because it is more resilient to soft radiation and produces regularly shaped jets that are easily calibrated. On the other hand, the CA algorithm is more suited to investigating the jet substructure because it imposes an angular ordering on the emissions in analogy with a parton shower. Reclustering the constituents using different jet definitions affects the observable. This was investigated in Ref.~\cite{ljpdensitytheory2020lifson} and the effect was found to be small except in the region of large $\Delta R$ at the edge of the jet.

\subsection{Structure of the Lund jet plane}
\label{sec:LJPstructure}
Various regions of the LJP are sensitive to the kinematics of the top quark and $W$ boson decay as well as the subsequent parton showering and hadronisation of the decay products. At small angles and small \pt fractions (upper right-hand corner), the structure of the plane resembles that of light jets while the hard and wide-angle emissions (lower left-hand corner) are sensitive to the top quark and $W$ boson decays. In the light-jet-like regime, the density of emissions is influenced by the running of $\alpha_\text{S}(k_T)$, rising with increasing $z \times \Delta R$ before falling sharply at the hadronisation scale. The quarks decay products of the top quark and $W$ boson have an opening angle proportional to $m/ \pt$, where $m$ is the mass of the top quark or $W$ boson. This gives rise to a second region of high emission densities at corresponding values of $\Delta R$.

\begin{figure}[h]
\centering
\def\sfwidth{ 0.495\textwidth }

\begin{subfigure}[b]{\sfwidth}
\centering
\includegraphics[width=\textwidth]{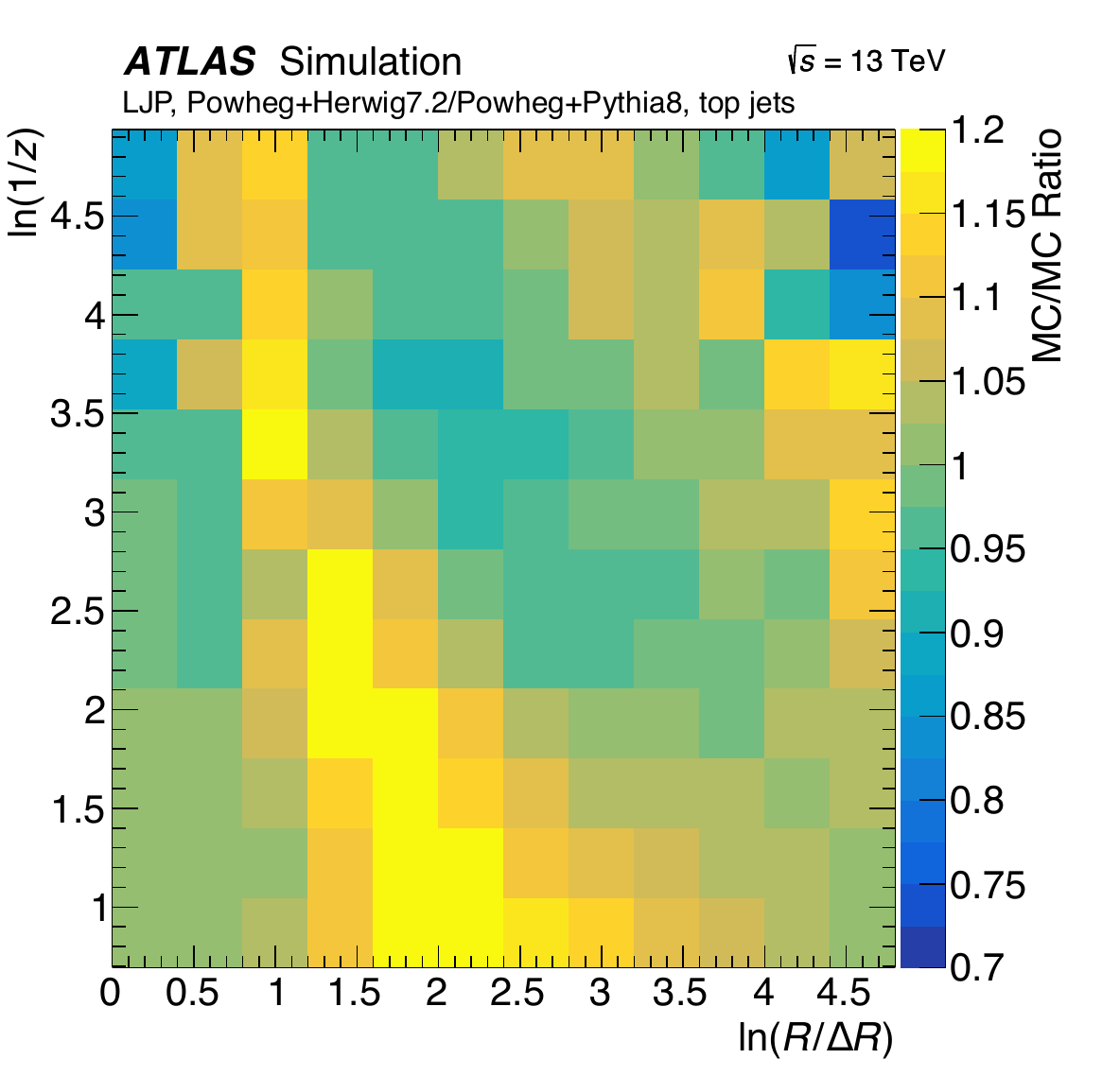}
\caption{}
\label{fig:mcaltmcratiostopsel:h72}
\end{subfigure}
\begin{subfigure}[b]{\sfwidth}
\centering
\includegraphics[width=\textwidth]{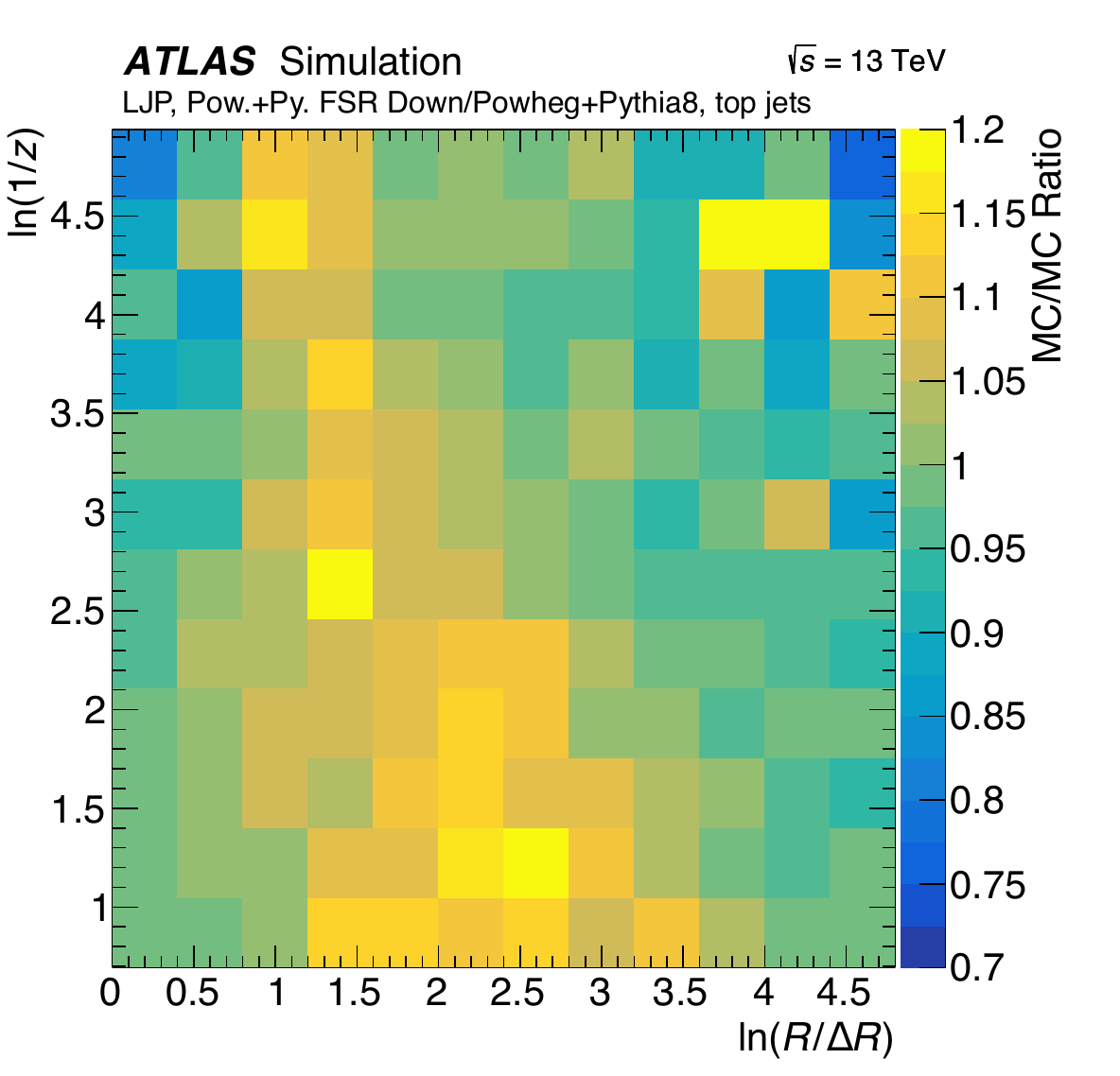}
\caption{}
\label{fig:mcaltmcratiostopsel:fsrdn}
\end{subfigure}
\begin{subfigure}[b]{\sfwidth}
\centering
\includegraphics[width=\textwidth]{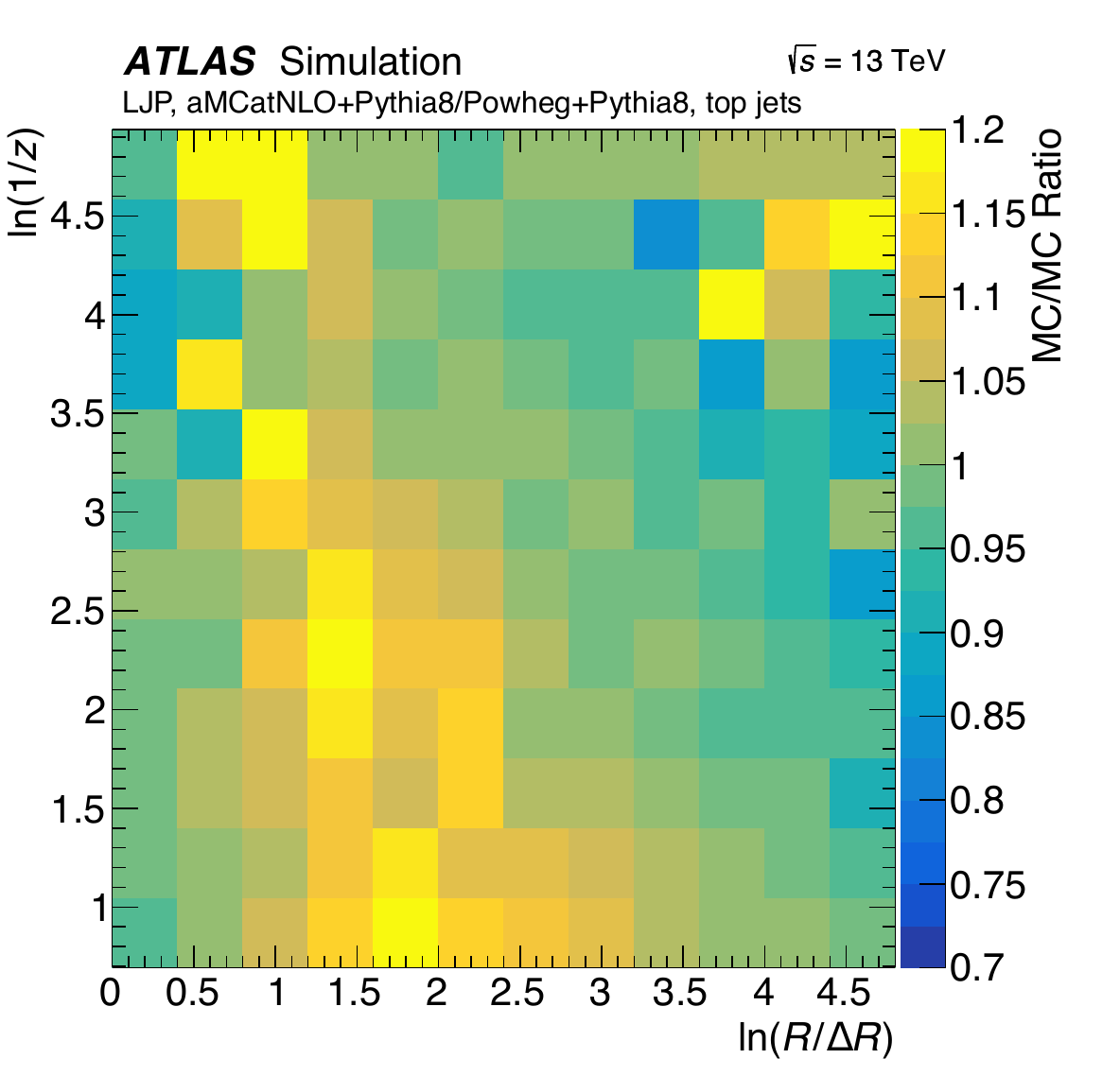}
\caption{}
\label{fig:mcaltmcratiostopsel:amc}
\end{subfigure}
\caption{Ratio of the Lund jet plane density predicted by alternative MC generators to that predicted by \POWPY[8], for jets including all the top quark decay products. The alternative predictions employed are: (a) \textsc{Powheg+Herwig} 7.2, (b) \textsc{Powheg+Pythia} 8 (FSR down) and (c) \textsc{aMC@NLO+Pythia} 8. Large deviations from 1 in the top left and top right corners of the plane are mostly due to statistical fluctuations. Definitions of the generator configurations are given in Section~\ref{sec:datamcsamples}}
\label{fig:mcaltmcratiostopsel}
\end{figure}

The LJPs sensitivity to parton shower effects can be demonstrated by evaluating the ratios of the LJP densities predicted by different MC generators, as shown in Figure~\ref{fig:mcaltmcratiostopsel}. MC generators using different parton-shower algorithms, such as \HERWIG[7.2]\ and \PYTHIA[8]\ in Figure~\ref{fig:mcaltmcratiostopsel:h72}, produce variations in the density of emissions along diagonals from the top left to the bottom right of the plane. Altering the amount of final-state radiation has a similar effect, as can be seen in Figure~\ref{fig:mcaltmcratiostopsel:fsrdn}.
A comparison of different treatments of the hard-scatter process and the matching to the parton shower, as with the \POWHEG\ and \AMCatNLO\ generators in Figure~\ref{fig:mcaltmcratiostopsel:amc}, shows smaller differences in the density of soft and collinear emissions. A detailed definition of the different generators is given in Section~\ref{sec:datamcsamples}.

\begin{figure}[ht]
\def\sfwidth{ 0.495\textwidth }
\centering
\begin{subfigure}[b]{\sfwidth}
\centering
\includegraphics[width=\textwidth,trim={0.1cm 0 0.3cm 0},clip]{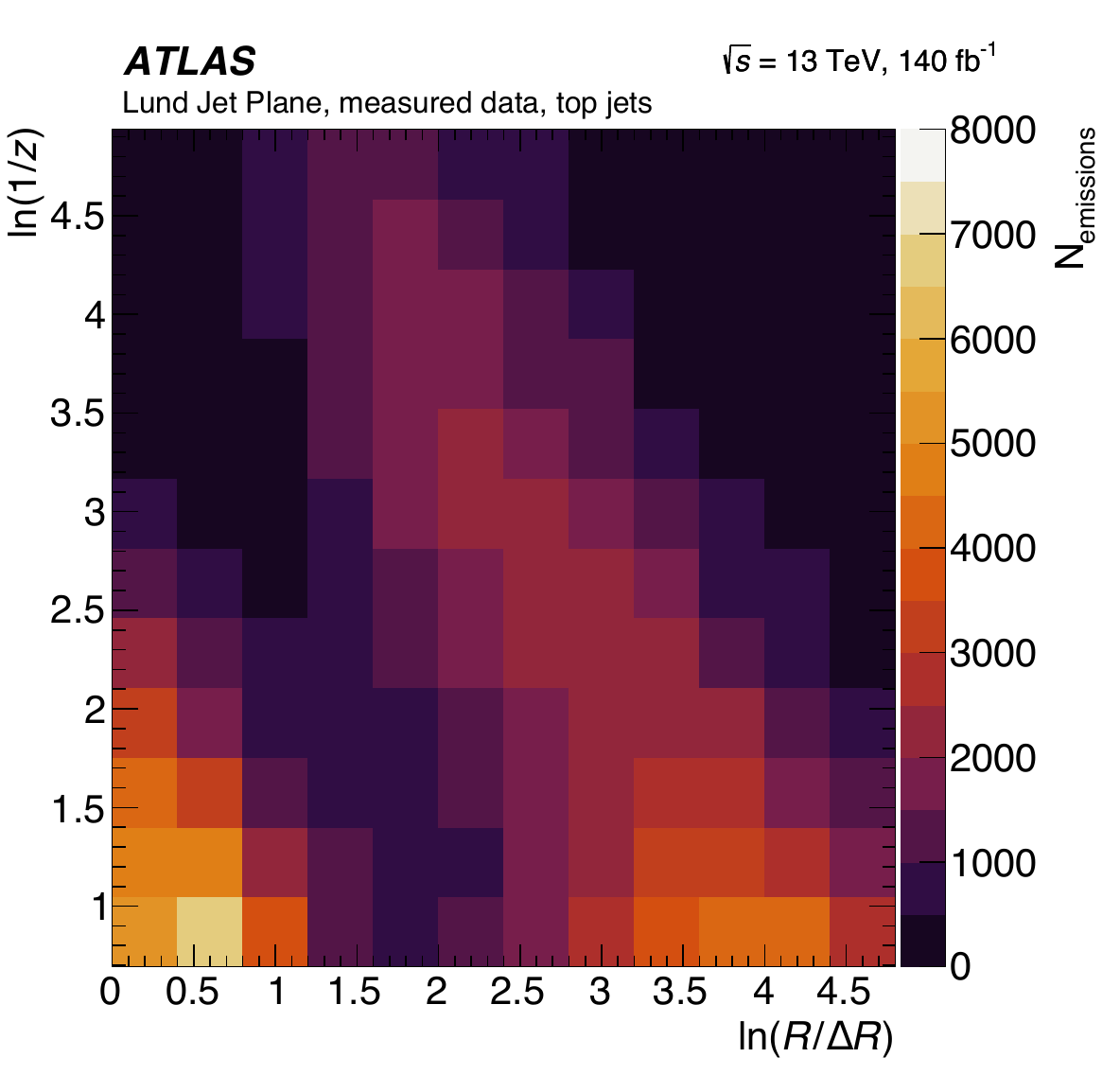} %
\caption{}
\end{subfigure}
\begin{subfigure}[b]{\sfwidth}
\centering
\includegraphics[width=\textwidth,trim={0.1cm 0 0.3cm 0},clip]{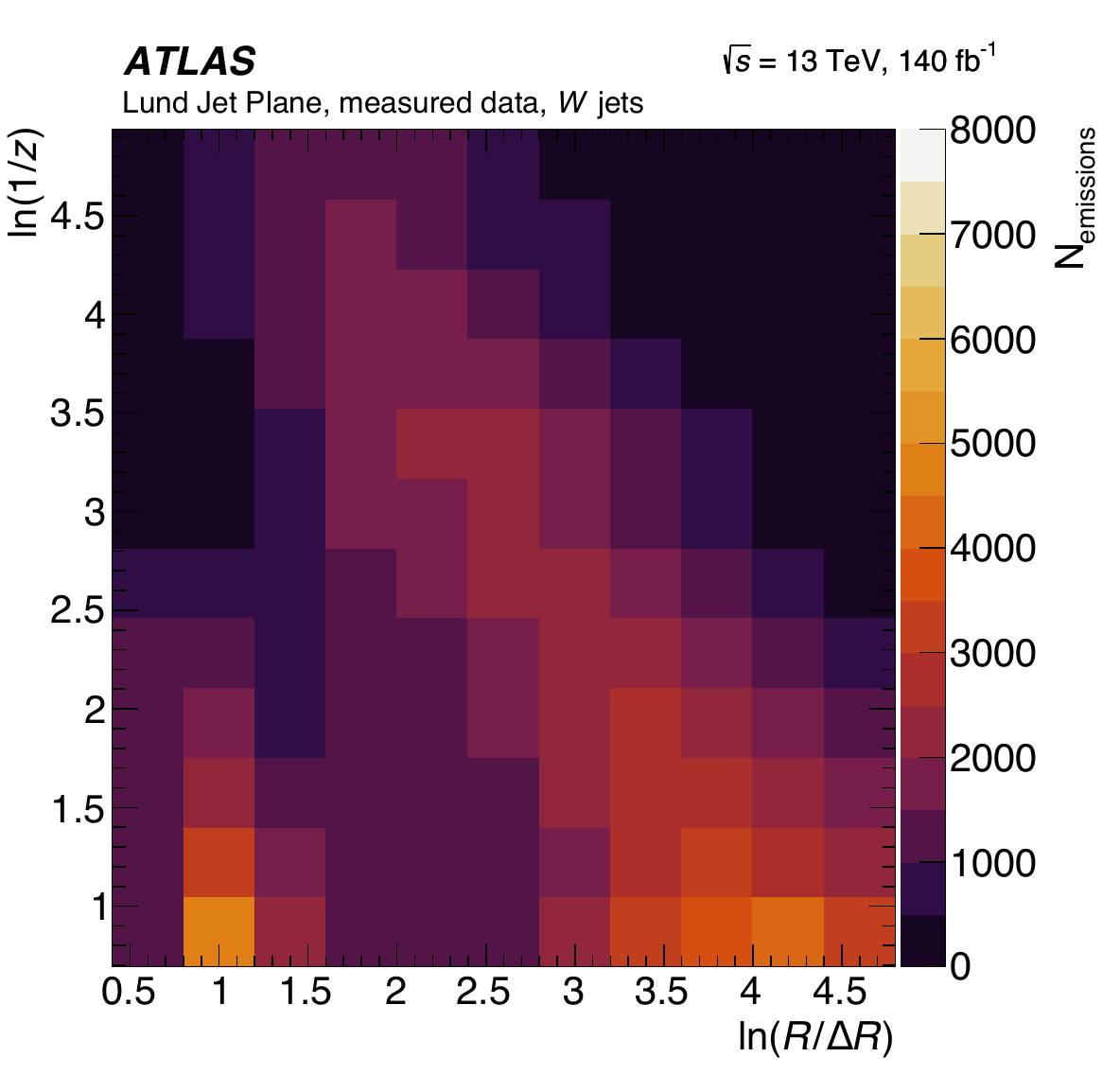}
\caption{}
\end{subfigure}
\caption{Reconstructed LJPs obtained by applying the (a) top jet and (b) $W$ jet selections to the data.}
\label{fig:LJPdetector}
\end{figure}

The detector-level LJP is shown in Figure~\ref{fig:LJPdetector} for the top jets and the $W$ jets. The quantity plotted is the number of emissions, including background processes. The 2D histogram that makes up the LJP for the top jets ($W$ jets) has equally spaced bins in 11 (12) rows along the $\ln(R/\Delta R)$ direction from $0$ ($0.4$) to $4.8$ and 11 columns along the $\ln(1/z)$ direction from $0.69$ to $4.9$. The fiducial LJP contains only bins with $>0$ entries. In both selections, one can distinguish two regions of the plane that are densely populated, one in the bottom left and one along the diagonal of the plane. On the other hand, the top left and top right regions of the plane are very sparsely populated. The separation between the densely populated regions is somewhat greater in the $W$ jet selection.

\FloatBarrier

\subsection{Unfolding}
\label{sec:unfolding}
The measured LJP is corrected for detector-level effects using the iterative Bayesian unfolding (IBU) procedure~\cite{mod:DAgostini:1995}. The whole 2D plane is unfolded at once, using corrections derived from the nominal \POWPY[8]\ simulated events. The unfolding can be represented by the following equation:
\begin{equation*}
\frac{\dif^{2} N_{\text{emissions}}}{ \dif \ln(1/z)~\dif \ln(R/ \Delta R)} =  \frac{1}{\Delta X \Delta Y \epsilon_{\text{eff}}^{I}} \sum_{J} {M^{-1}}^{I}_{J}  \cdot  \epsilon_{\text{acc}}^{J}  \cdot (N_{\text{obs}}^{J} - N_{\text{bkg}}^{J})
\end{equation*}
where $I$ is the index of the bin in the particle-level LJP,\footnote{Here, $I=i_{\Delta R} + j_{z} \times N_{\Delta R}$, where $N_{\Delta R}$ is the total number of bins along the $\ln(R/\Delta R )$ direction of the plane, $i_{\Delta R}$ is the bin number in the $\ln(R / \Delta R )$ direction of the plane and $j_{z}$ is the bin number in the $\ln(1/z)$ direction of the plane.} $J$ is the bin index at detector level, $\Delta X$ and $\Delta Y$ are the widths of the bins corresponding to $I$ across the two axes, $M$ is the response matrix (with its inversion representing the unfolding technique), and $N_{\text{obs}}$ and $N_{\text{bkg}}$ correspond to the number of observed emissions and the predicted background, respectively.
The factor $\epsilon_{\text{eff}}$ corrects for emissions passing the particle-level selection but not the detector-level requirements, while $\epsilon_{\text{acc}}$ removes the emissions passing the detector-level requirements and not matched to the particle-level ones. The latter factor corrects the LJPs for events that do not pass the particle-level selection, and also for individual detector-level emissions that do not have a corresponding particle-level emission. The matching between detector- and particle-level emissions is performed in the $\eta{-}\phi$ space after transforming the coordinates into the rest frame of the \largeR\ jet. Detector- and particle-level emissions with a closest match at a distance $\Delta R_{\text{match}} < 0.1$ are used to populate the response matrix. A dedicated uncertainty, described in Section~\ref{sec:uncertanities}, is estimated to capture the effect of the choice of $\Delta R_\text{match}$. The resulting uncertainty is $<1\%$ across all of the densely populated regions of the plane; hence the closeness of the matching has little effect on the unfolded distribution.

\begin{figure}[h]
\def\sfwidth{ 0.495\textwidth }
\centering
\begin{subfigure}[b]{\sfwidth}
\centering
\includegraphics[width=\textwidth,trim={0.1cm 0 0.5cm 0},clip]{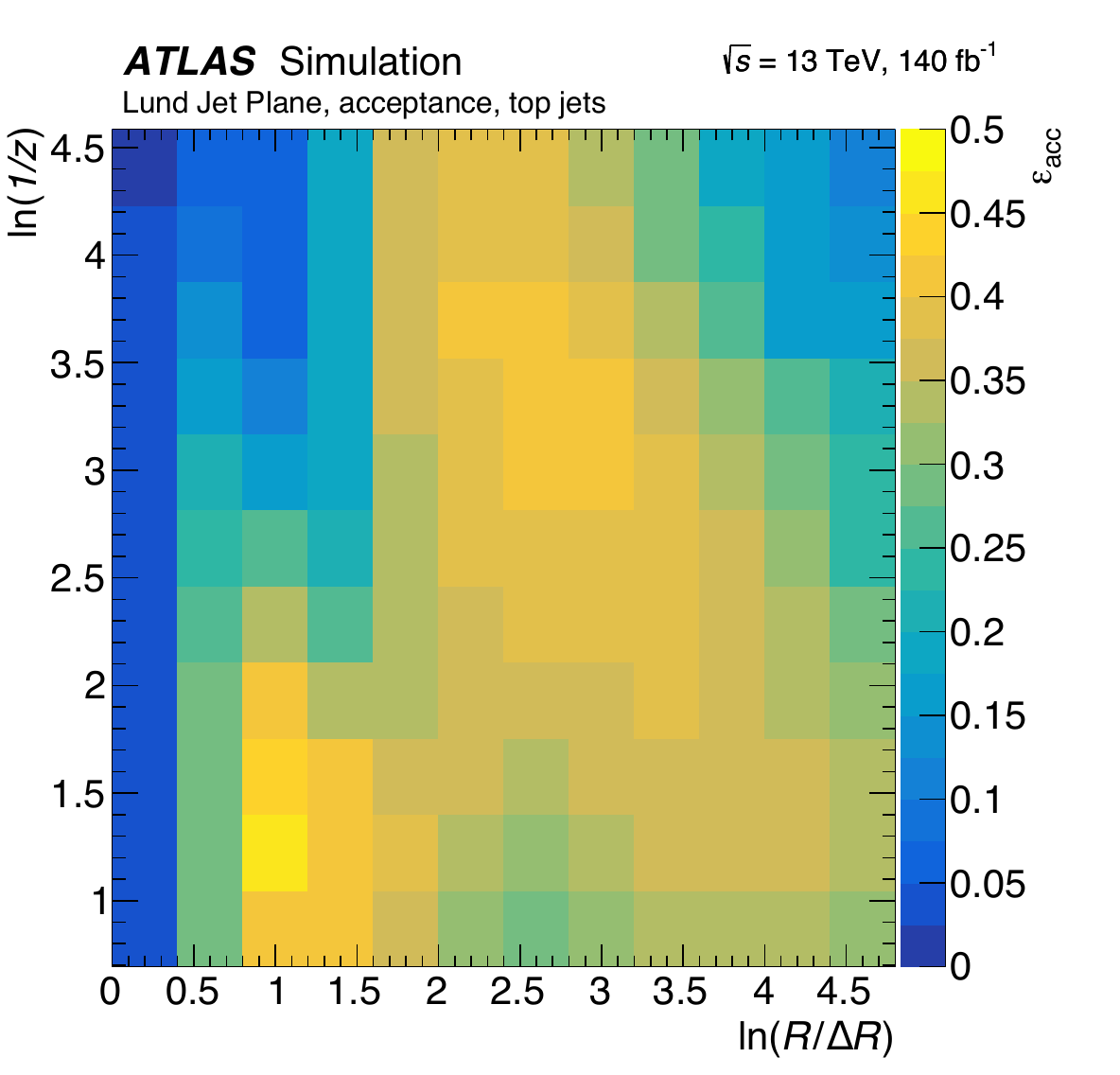}
\caption{}
\end{subfigure}
\begin{subfigure}[b]{\sfwidth}
\centering
\includegraphics[width=\textwidth,trim={0.1cm 0 0.5cm 0},clip]{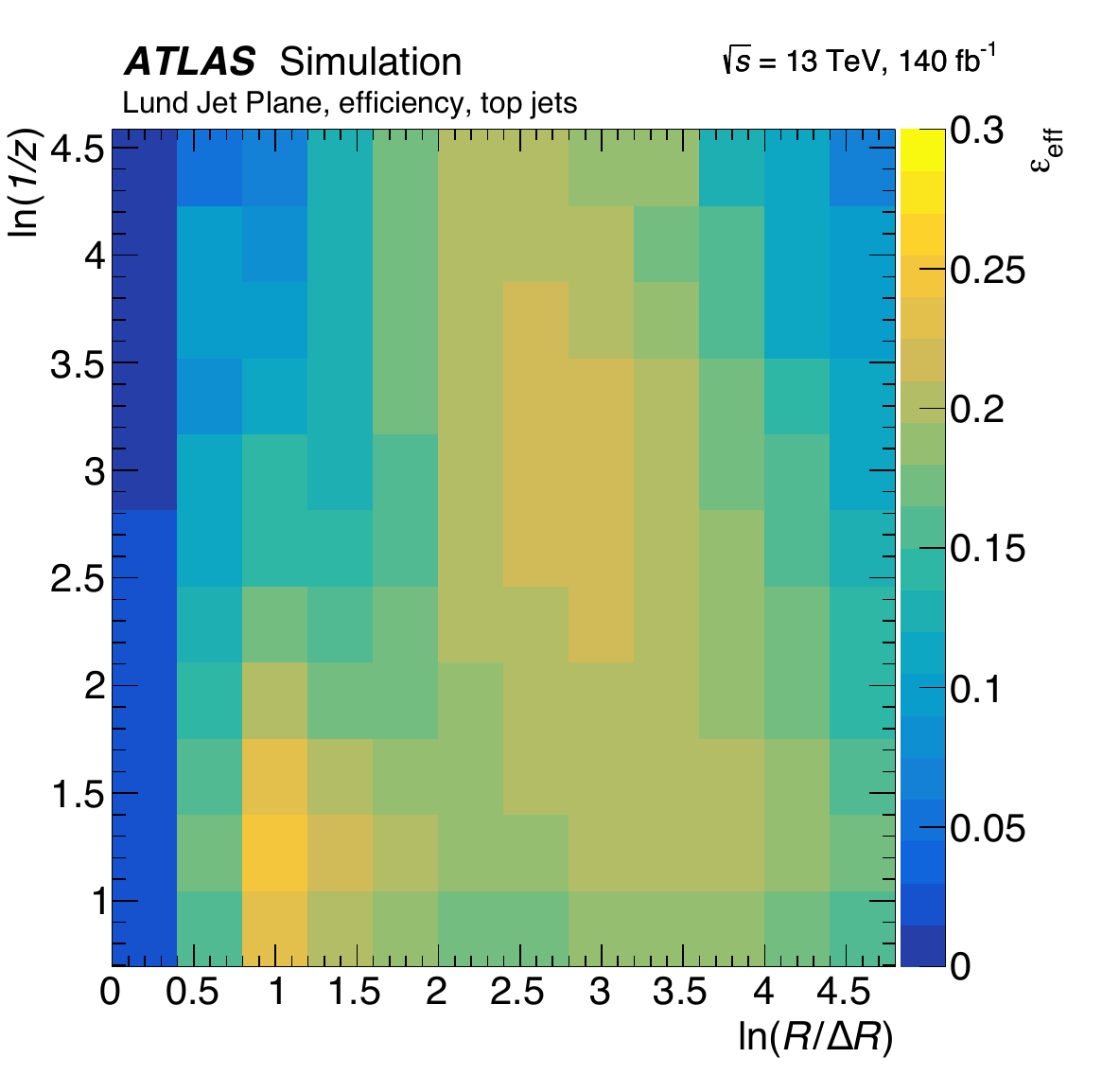}
\caption{}
\end{subfigure}
\begin{subfigure}[b]{\sfwidth}
\centering
\includegraphics[width=\textwidth,trim={0.1cm 0 0.5cm 0},clip]{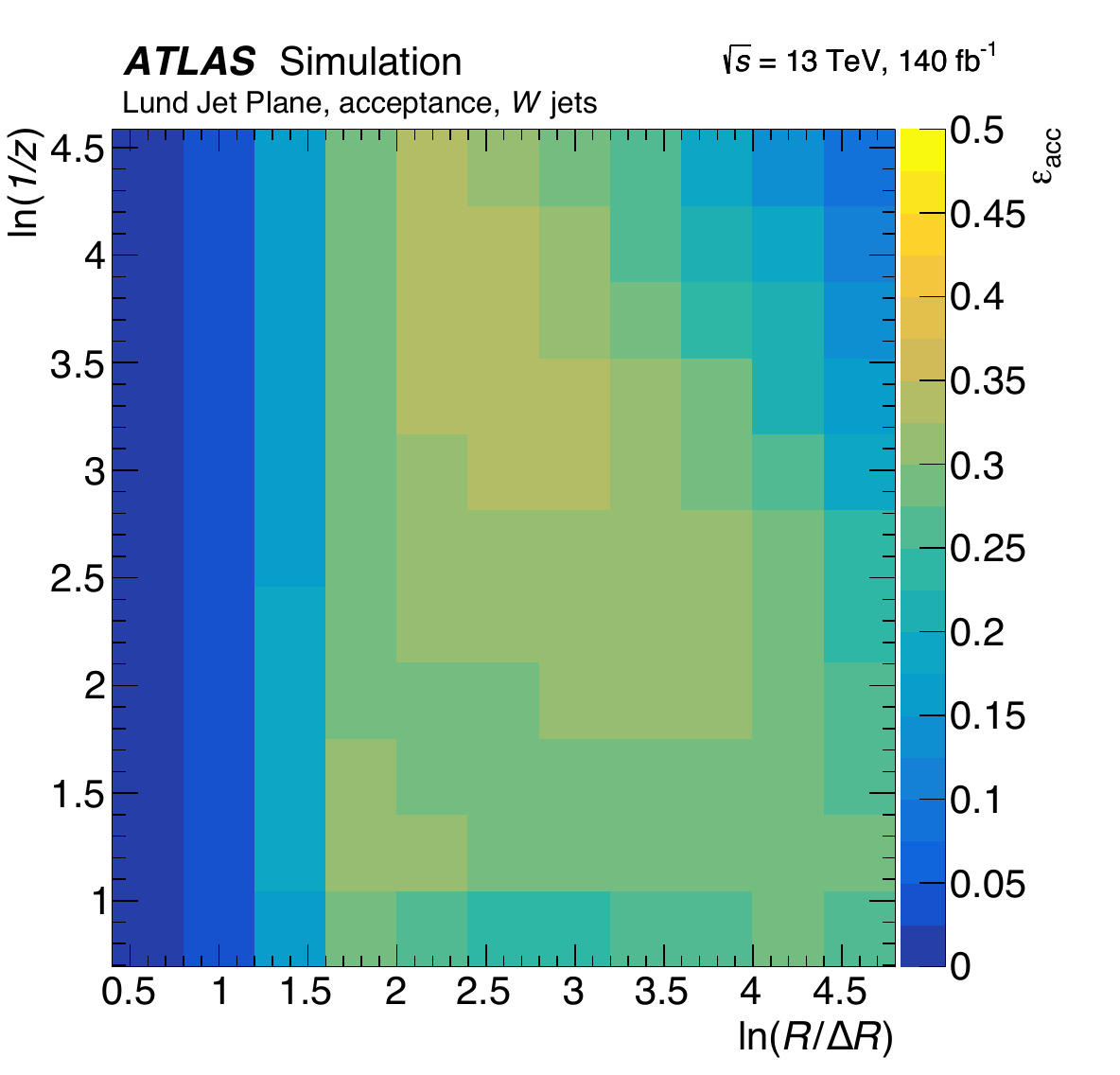}
\caption{}
\end{subfigure}
\begin{subfigure}[b]{\sfwidth}
\centering
\includegraphics[width=\textwidth,trim={0.1cm 0 0.5cm 0},clip]{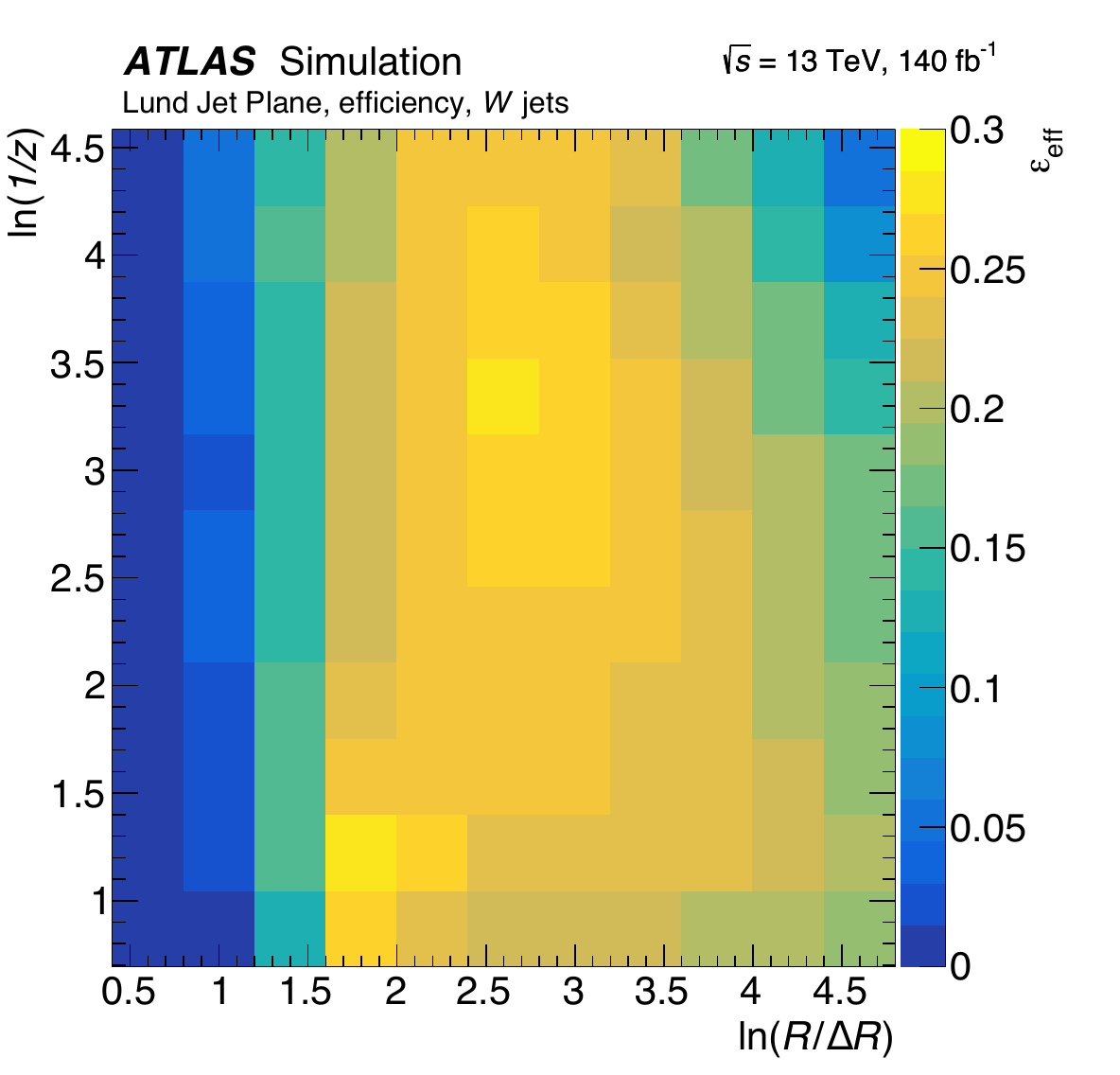}
\caption{}
\end{subfigure}
\caption{Unfolding corrections for the LJP in the (a,b) top jet and (c,d) $W$ jet selections. The first column reports the $\epsilon_{\text{acc}}$ factor, sensitive to the selection acceptance and matching between the detector- and particle-level emissions, while the second one shows the $\epsilon_{\text{eff}}$ corrections. The corrections are evaluated using events simulated with the nominal \textsc{Powheg+Pythia} 8 MC generator.}
\label{fig:unfoldingcorr}
\end{figure}

A response matrix is built for each selection. Bin-to-bin migrations between the detector level and the particle level occur mostly between adjacent bins. For $\ln(R/\Delta R )$ bins at identical $\ln(1/z)$ values, the response matrix is on average approximately $90\%$ diagonal. For the $\ln(1/z)$ bins, this value is around $65\%$. The unfolding corrections for the top jet and $W$ jet selections are shown in Figure~\ref{fig:unfoldingcorr}. Here, it can be observed that the $\epsilon_{\text{acc}}$ distribution, excluding some regions of low acceptance in the wide-angle emissions regions ($\ln(R / \Delta R ) < 0.4$), is relatively flat and around $45\%$ (35\%) for the top jet ($W$ jet) selection. The difference in acceptance between the two selections is driven by the difference in the mass cut applied. The $\epsilon_{\text{eff}}$ correction factor is affected by the inefficiencies induced by the $b$-tagging requirements, and is indeed lower than the acceptance correction. Moreover, one can observe a higher efficiency in the $W$ jet selection, where only one $b$-tagged jet is required.

\begin{figure}[h]
\def\sfwidth{ 0.495\textwidth }
\centering
\begin{subfigure}[b]{\sfwidth}
\centering
\includegraphics[width=\textwidth,trim={0.1cm 0 0.5cm 0},clip]{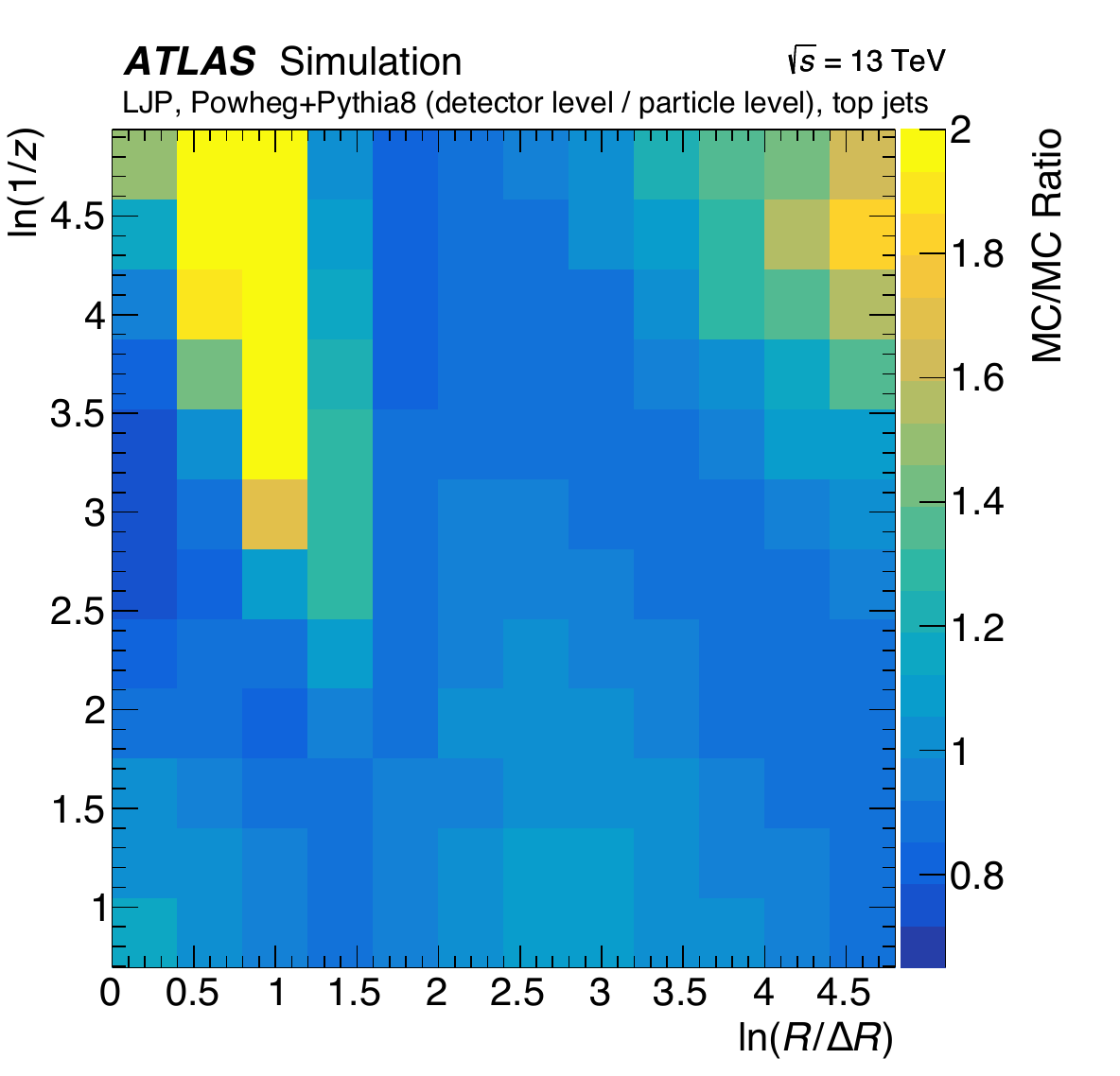}
\caption{}
\end{subfigure}
\begin{subfigure}[b]{\sfwidth}
\centering
\includegraphics[width=\textwidth,trim={0.1cm 0 0.5cm 0},clip]{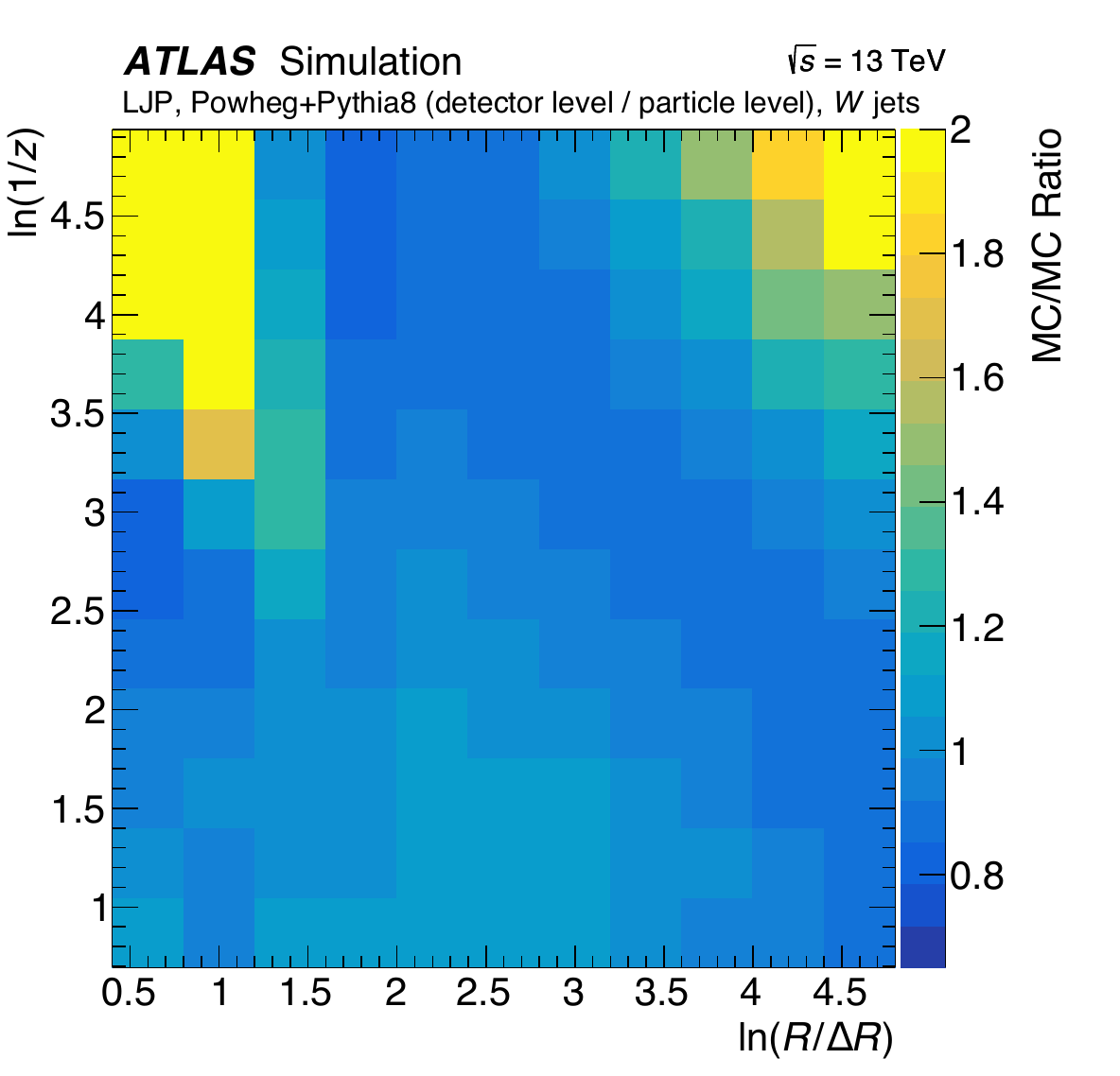}
\caption{}
\end{subfigure}
\caption{Ratios of the detector-level LJP to the particle-level LJP for (a) top jets and (b) $W$ jets. The LJPs are normalised to unity before evaluating the ratio. The LJPs are evaluated using events simulated with the nominal \textsc{Powheg+Pythia} 8 MC generator.}
\label{fig:response}
\end{figure}

Figure~\ref{fig:response} shows the ratio of the detector-level LJP to the particle-level LJP in the two selections, with the planes normalised to unity before taking their ratio. The effect of the unfolding procedure on the shape of the distributions can be observed in these ratios. These figures reveal that the main effect of the unfolding, in both selections, is to increase the separation between the hard and soft regimes in the LJP. The peak of the LJP, previously seen at detector level in the lower-left corner of the plane in Figure~\ref{fig:LJPdetector}, is shifted towards softer (i.e.\ smaller $k_\text{T} = z\;\!\Delta R (p_\text{T}^i + p_\text{T}^j)$) values at particle level.

The parameter that governs the regularisation in IBU is the number of iterations. Here, four iterations are chosen. This is a compromise that limits statistical fluctuations that grow due to repeated applications of the unfolding procedure, while controlling the statistical bias toward the particle-level MC predictions induced by the regularisation. The effect of this bias on the measurement is quantified by a dedicated uncertainty, introduced in Section~\ref{sec:uncertanities}. With the chosen number of iterations, this uncertainty is mostly negligible ($<1\%$).

After the histograms of emissions in the LJP are unfolded, they are scaled by the total number of \largeR\ jets $N_\text{jets}$ to obtain the per-jet density of emissions from Eq.~(\ref{eq:ljpdensity}). Just like the LJP emissions, the measured number of \largeR\ jets is corrected for acceptance and inefficiency effects to obtain a particle-level result.

\FloatBarrier


%
\section{Uncertainties}
\label{sec:uncertanities}
The measured LJP is affected by several systematic uncertainties related to the objects calibration, background and signal modelling, and analysis method.
All uncertainties are propagated through the unfolding procedure by creating pseudo-data that are then unfolded using the nominal corrections (background estimate, acceptance correction, migration matrix and efficiency correction). The signal modelling uncertainties are estimated as non-closure between the unfolded LJP and the particle-level prediction for the corresponding generator set-up. Detector systematic uncertainties are estimated by varying the relevant parameters in the nominal signal and background predictions, using these as pseudo-data. Background modelling uncertainties are estimated in the same way, using alternative background MC samples to produce the pseudo-data.

Correlations between the number of emissions in the LJP and the number of jets are taken into account in the evaluation of the uncertainties by generating pseudo-experiments for each systematic in which both quantities are varied simultaneously according to the size of the uncertainty. The standard deviation of the combined variations over the pseudo-experiments gives the total systematic uncertainty.

Uncertainties in the \smallR\ and \largeR\ jet energy scale and resolution~\cite{JETM-2018-05,JETM-2018-02} and \largeR\ jet mass scale and resolution~\cite{ATLAS-CONF-2020-022}, as well as in the $b$-tagging efficiency~\cite{FTAG-2018-01,FTAG-2020-08,FTAG-2019-02}, JVT cut efficiency~\cite{PERF-2014-03} and \met reconstruction~\cite{PERF-2016-07}, are included. Uncertainties associated with electrons and muons are estimated for the trigger, reconstruction, identification and isolation efficiencies~\cite{EGAM-2018-01, MUON-2018-03}, as well as the energy and momentum scales and resolutions~\cite{MUON-2022-01,EGAM-2018-01}. The measured LJPs are also affected by uncertainties associated with inner-detector tracks, related to mismodelling of the impact parameters, alignment, tracking efficiency, fake-track rate and reconstruction of tracks in a dense environment~\cite{ATL-PHYS-PUB-2015-051,PERF-2015-08}.
The detector-related uncertainties have a considerable impact on the total uncertainty, their size in the centre of the LJP being of the same order as the statistical uncertainty: $5\%$--$10\%$ for the top jets and $10\%$--$15\%$ for the $W$ jets.

\begin{figure}[h]
\def\sfwidth{ 0.495\textwidth }
\centering
\begin{subfigure}[b]{\sfwidth}
\centering
\includegraphics[width=\textwidth,trim={1cm 0 2cm 0},clip]{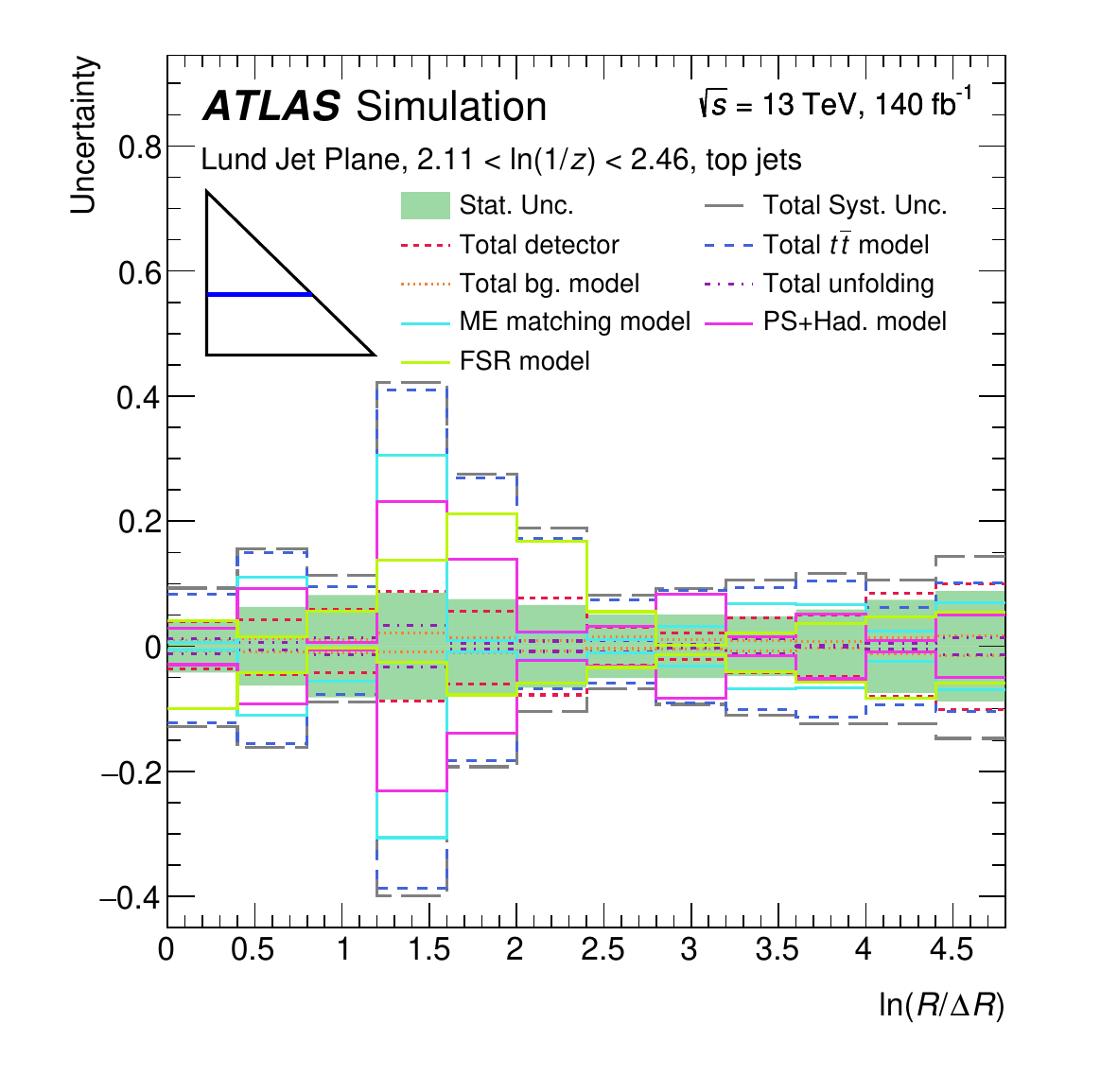}
\caption{}
\end{subfigure}
\hfill
\begin{subfigure}[b]{\sfwidth}
\centering
\includegraphics[width=\textwidth,trim={1cm 0 2cm 0},clip]{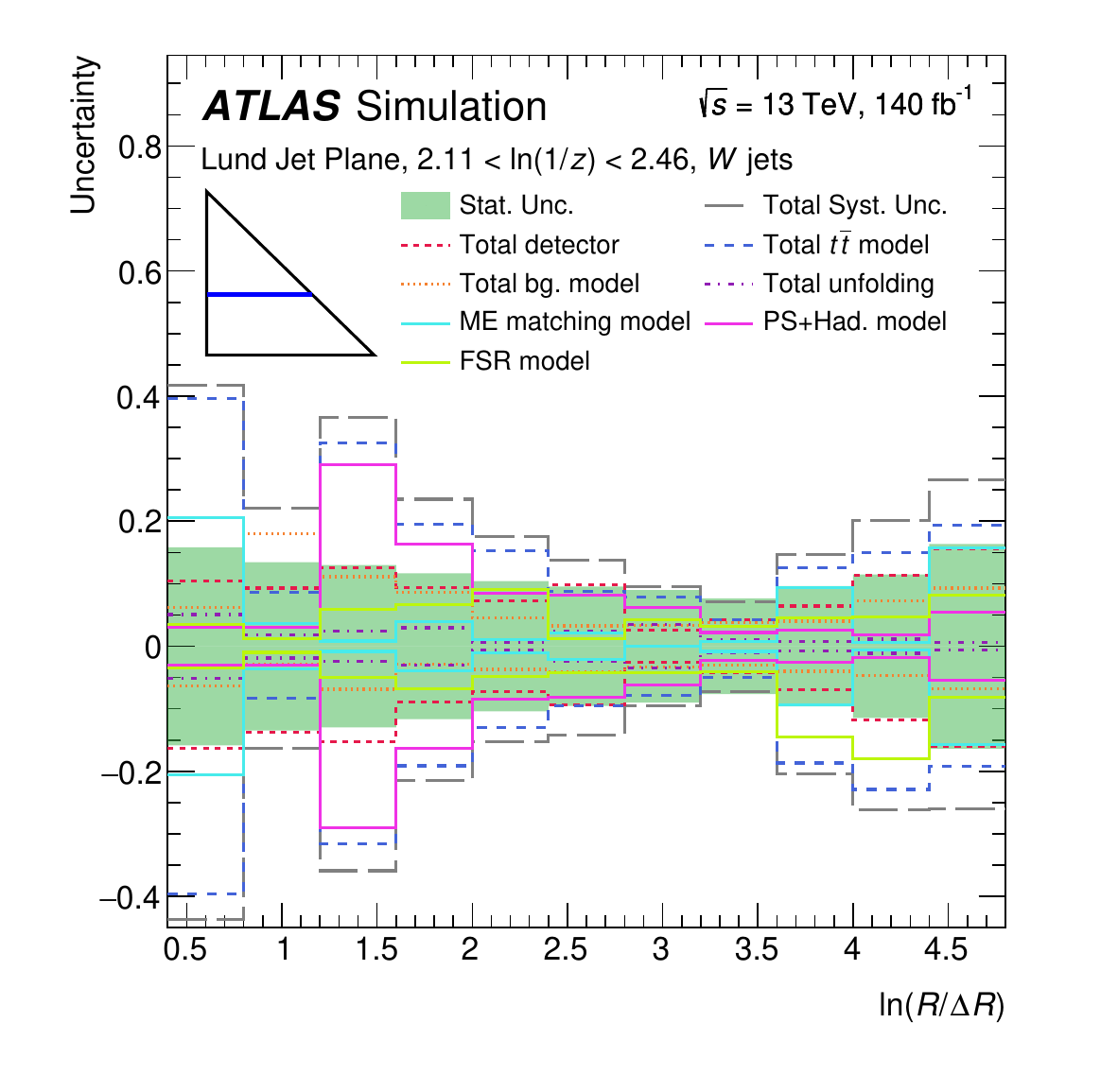}
\caption{}
\end{subfigure}
\caption{Breakdown of the systematic uncertainties on the unfolded density of emissions in a slice through the LJP along $\ln(R/ \Delta R)$ for the (a) top jet and (b) $W$ jet selections. The total systematic uncertainty is reported, as well as the total contributions from signal modelling, detector effects, background modelling and unfolding. The major sources of signal modelling uncertainties are also presented in the plot as separate contributions. Not shown are the RTT, PDF and ISR uncertainties, of which the RTT uncertainty contributes the most toward the total $t\bar{t}$ modelling uncertainty. The coloured band represents the total statistical uncertainty. }
\label{fig:systslice_x}
\end{figure}

\begin{figure}[h]
\def\sfwidth{ 0.495\textwidth }
\centering
\begin{subfigure}[b]{\sfwidth}
\centering
\includegraphics[width=\textwidth,trim={1cm 0 2cm 0},clip]{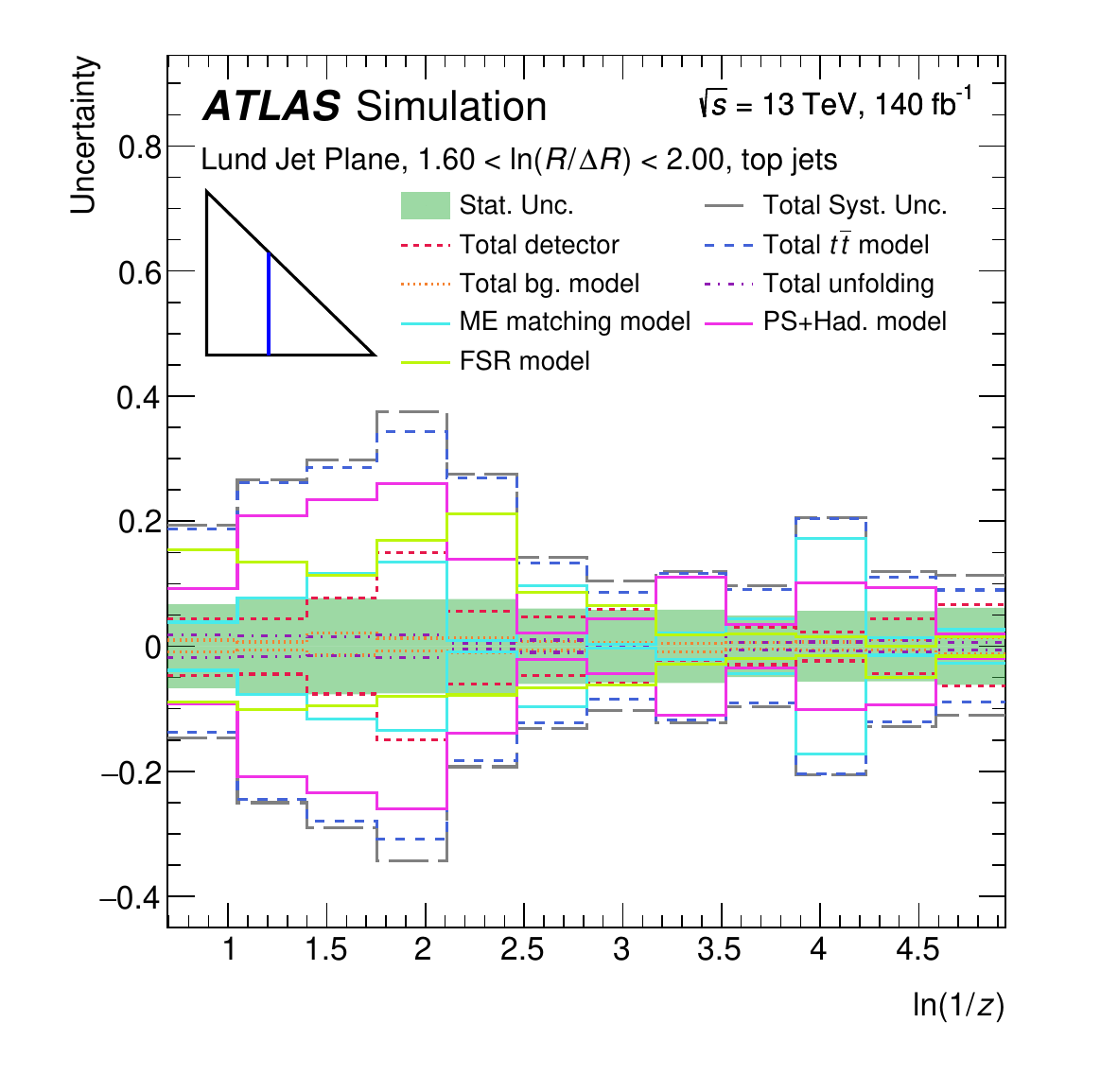}
\caption{}
\end{subfigure}
\hfill
\begin{subfigure}[b]{\sfwidth}
\centering
\includegraphics[width=\textwidth,trim={1cm 0 2cm 0},clip]{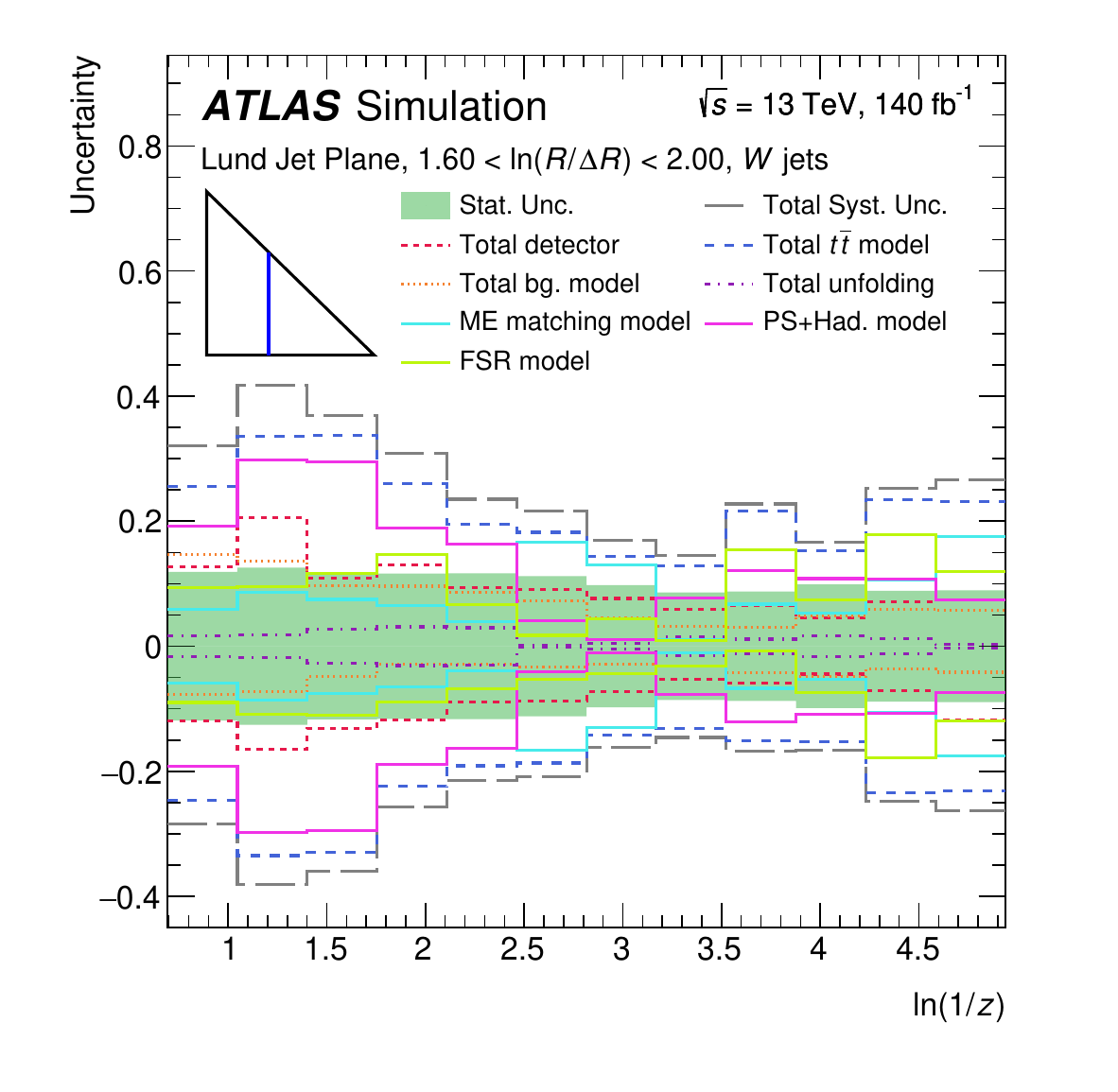}
\caption{}
\end{subfigure}
\caption{Breakdown of the systematic uncertainties in the unfolded density of emissions in a slice through the LJP along $\ln(1/z)$ for the (a) top jet and (b) $W$ jet selections. The total systematic uncertainty is reported, as well as the total contributions from signal modelling, detector effects, background modelling and unfolding. The major sources of signal modelling uncertainties are also presented in the plot as separate contributions. Not shown are the RTT, PDF and ISR uncertainties, of which the RTT uncertainty contributes the most toward the total $t\bar{t}$ modelling uncertainty. The coloured band represents the total statistical uncertainty.}
\label{fig:systslice_y}
\end{figure}

The systematic uncertainty related to the modelling of the \ttbar\ signal process is estimated using alternative Monte Carlo samples. Descriptions of these samples and the meaning of their configurations are presented in Section~\ref{sec:datamcsamples}. The components of the modelling systematic uncertainty relate to the matrix-element and matching (ME-matching), the parton shower and hadronisation modelling (PS+Had), the assumed recoil system for the decay products of the top quark (RTT), variation of the $h_\text{damp}$ parameter, FSR and ISR. Lastly, the uncertainty arising from the choice of PDF is estimated using the 30 eigenvectors of the Hessian matrix derived in the PDF4LHC15 uncertainty prescriptions~\cite{Butterworth:2015oua}.
The \ttbar\ signal modelling uncertainties are the dominant uncertainties of the measurement except in a few very sparsely populated bins at the edges of the plane. The dominant component depends on the region under consideration, as shown in Figures~\ref{fig:systslice_x} and~\ref{fig:systslice_y}.%

Several uncertainties are evaluated for the contributions from single top quark production, the dominant background process. These include the same ISR and FSR uncertainties considered for $t\bar{t}$, as well as a variation using an alternative treatment of the interference with \ttbar\ in the matrix element (DR/DS). The ISR and FSR uncertainties are treated as uncorrelated with the corresponding variations of the \ttbar\ signal. The inclusive cross-section used to normalise the samples also has an associated uncertainty: $4.2\%$ for the $t$-channel, $5.4\%$ for the $tW$-channel and $3.9\%$ for the $s$-channel process~\cite{Kidonakis:2010tc,Kidonakis:2005kz}. The uncertainties in modelling the single-top process are treated as uncorrelated with the modelling of the \ttbar\ signal. The uncertainty in the $W$+jets background is estimated by varying the \muR and \muF scales simultaneously by a factor of 2 or 0.5. A $13.3\%$ uncertainty is associated with the normalisation of the $t\bar{t}V$ background~\cite{deFlorian:2016spz}. A conservative $50\%$ uncertainty in the normalisation of the diboson, $Z$+jets and multijet background is considered, given the very small contribution of these backgrounds to the total number of events ($1\%$ and $3\%$ in the top jet and $W$ jet selections, respectively). The total uncertainty associated with the background modelling is below $5\%$ across the whole plane.

An unfolding bias uncertainty is estimated using a data-driven method~\cite{datadrivenclosure2009malaescu}. A set of pseudo-data is obtained by reweighting the predicted particle-level LJP by a smooth function following the ratio of the observed detector-level LJP to the predicted LJP and folding the result using the response matrix. The uncertainty is then evaluated as the non-closure between the unfolded data and the reweighted particle-level distribution. The impact of the uncertainty in the matching procedure between detector- and particle-level emissions on the unfolding procedure, as described in Section~\ref{sec:unfolding}, is evaluated by tightening the $\Delta R$ matching cut to $0.05$. Both uncertainties associated with the unfolding have a small effect ($1\%{-}3\%$) on the total uncertainty across the LJP fiducial region.

The uncertainty due to the limited numbers of observed and simulated events is estimated using the bootstrap procedure~\cite{bootstraps1979efron}, where weights distributed according to a Poisson distribution with mean 1 are associated with all events to create replicas of the measurement. Each replica is propagated through the unfolding procedure and then divided by the corresponding number of jets, which is also `bootstrapped' such that statistical correlations with the LJP emissions are preserved. The standard deviation of the unfolded distributions in each bin of the LJP represents the statistical uncertainty. As with the systematic uncertainties, the number of emissions in the LJP and the number of jets are varied simultaneously to account for the correlations induced in the calculation of the emission density. %

The uncertainty in the combined 2015--2018 integrated luminosity is $0.83\%$~\cite{DAPR-2021-01}, obtained using the LUCID-2 detector~\cite{LUCID2} for the primary luminosity measurements, complemented by measurements using the inner detector and calorimeters. The pile-up activity in each MC sample is reweighted to match the conditions in data, and a corresponding uncertainty is evaluated according to the uncertainty in the average number of interactions per bunch crossing.

\begin{figure}[h]
\def\sfwidth{ 0.495\textwidth }
\centering
\begin{subfigure}[b]{\sfwidth}
\centering
\includegraphics[width=\textwidth,trim={0.1cm 0 0.5cm 0},clip]{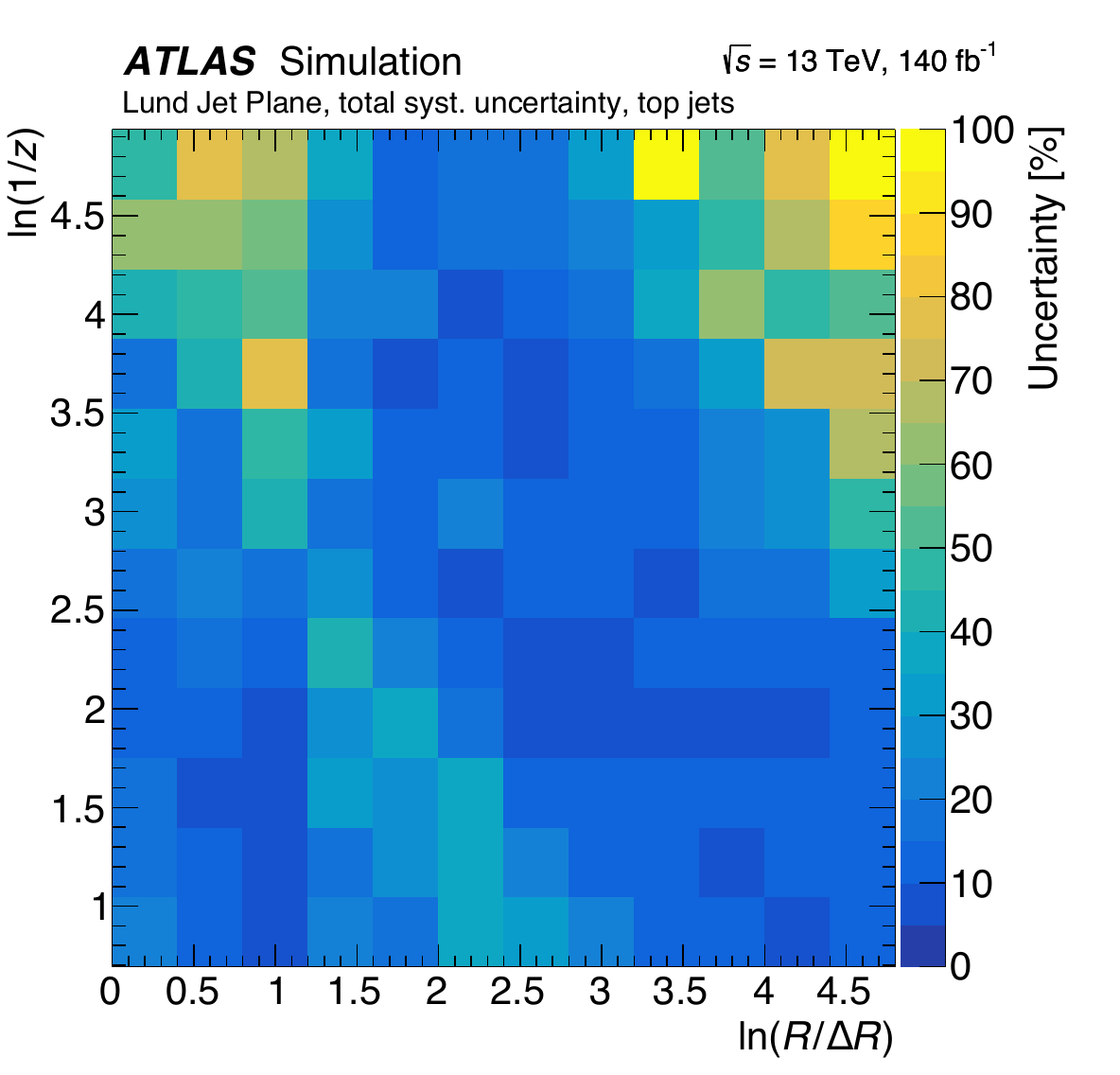}
\caption{}
\end{subfigure}
\begin{subfigure}[b]{\sfwidth}
\centering
\includegraphics[width=\textwidth,trim={0.1cm 0 0.5cm 0},clip]{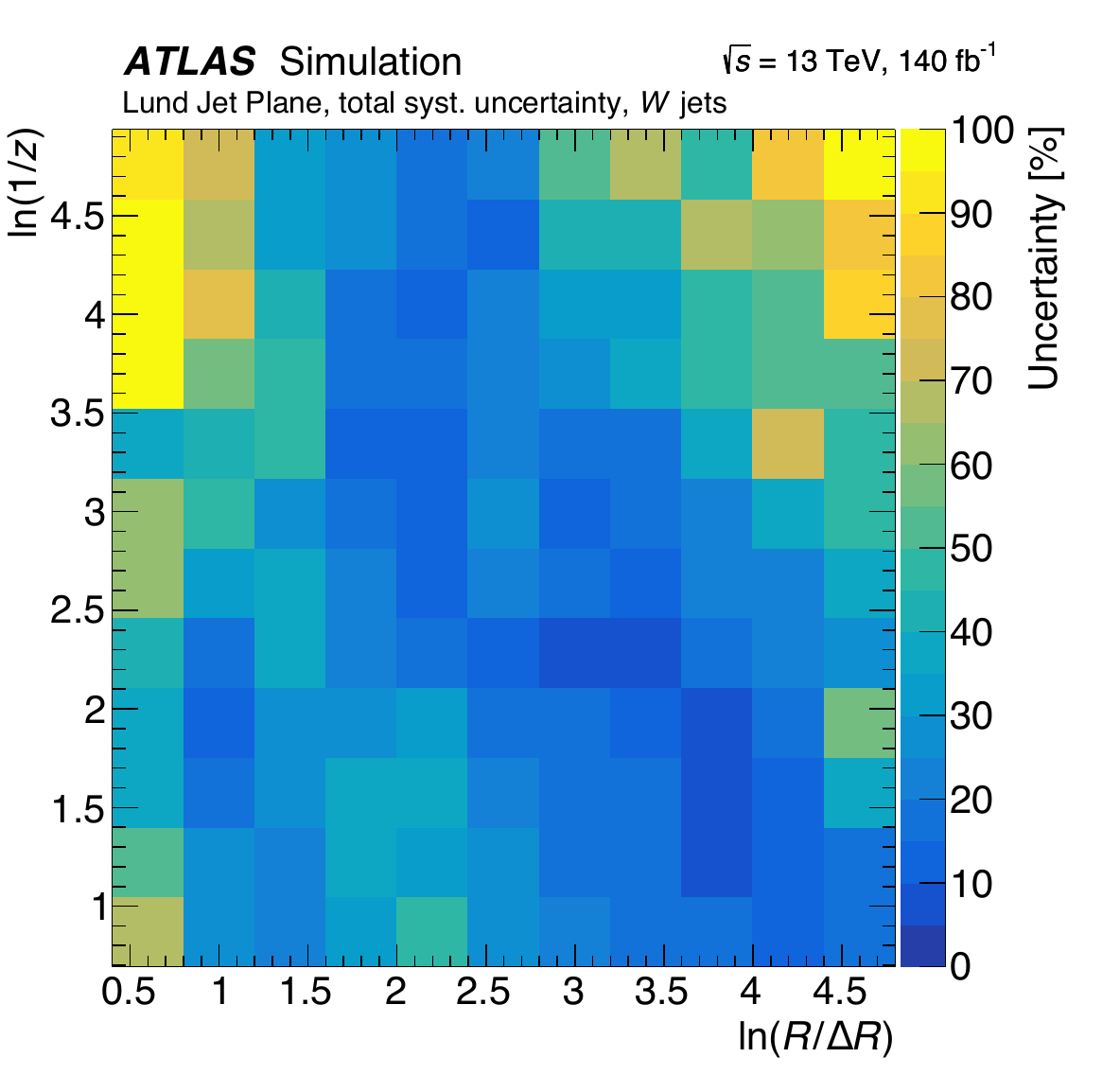}
\caption{}
\end{subfigure}
\begin{subfigure}[b]{\sfwidth}
\centering
\includegraphics[width=\textwidth,trim={0.1cm 0 0.5cm 0},clip]{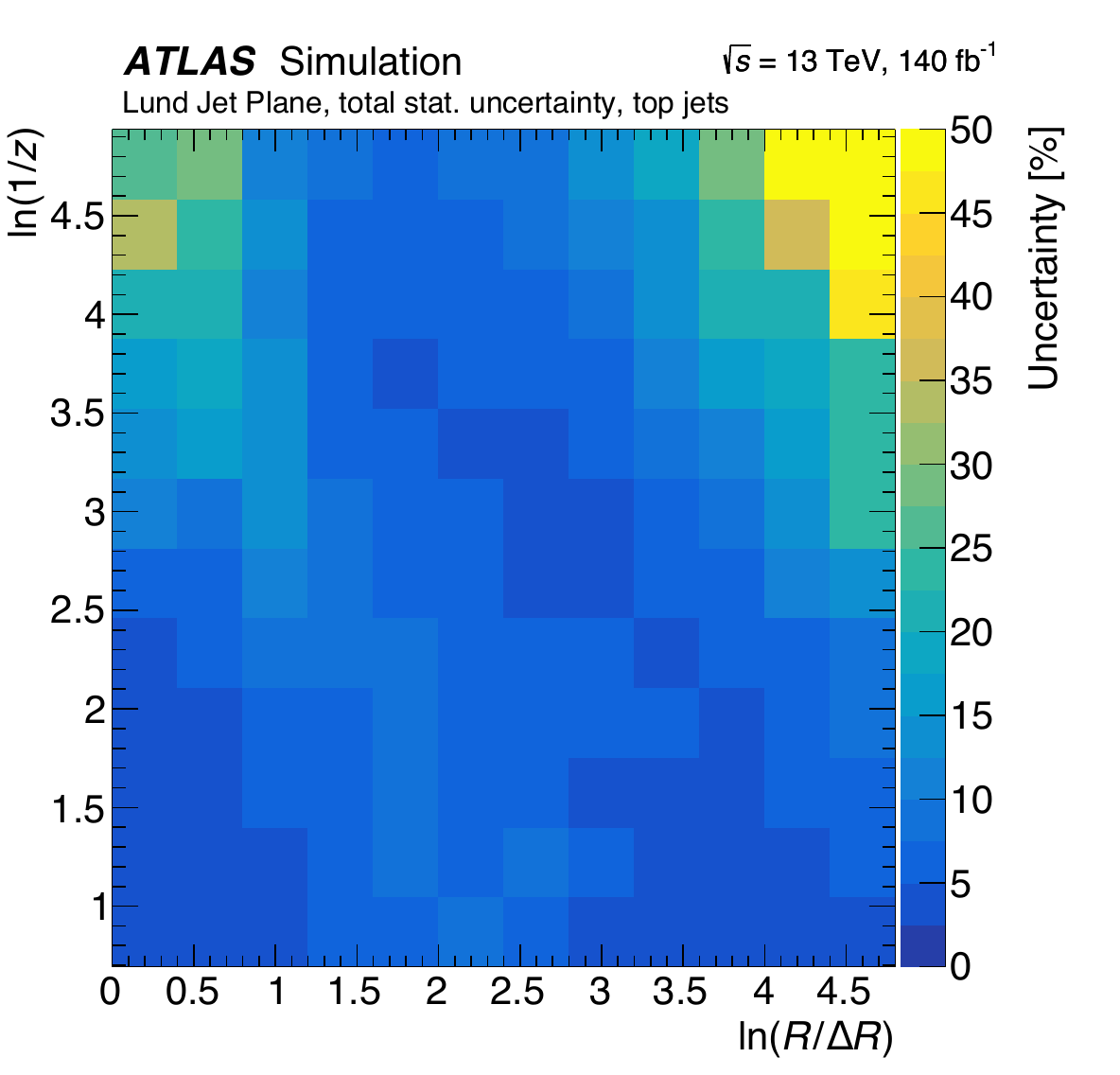}
\caption{}
\end{subfigure}
\begin{subfigure}[b]{\sfwidth}
\centering
\includegraphics[width=\textwidth,trim={0.1cm 0 0.5cm 0},clip]{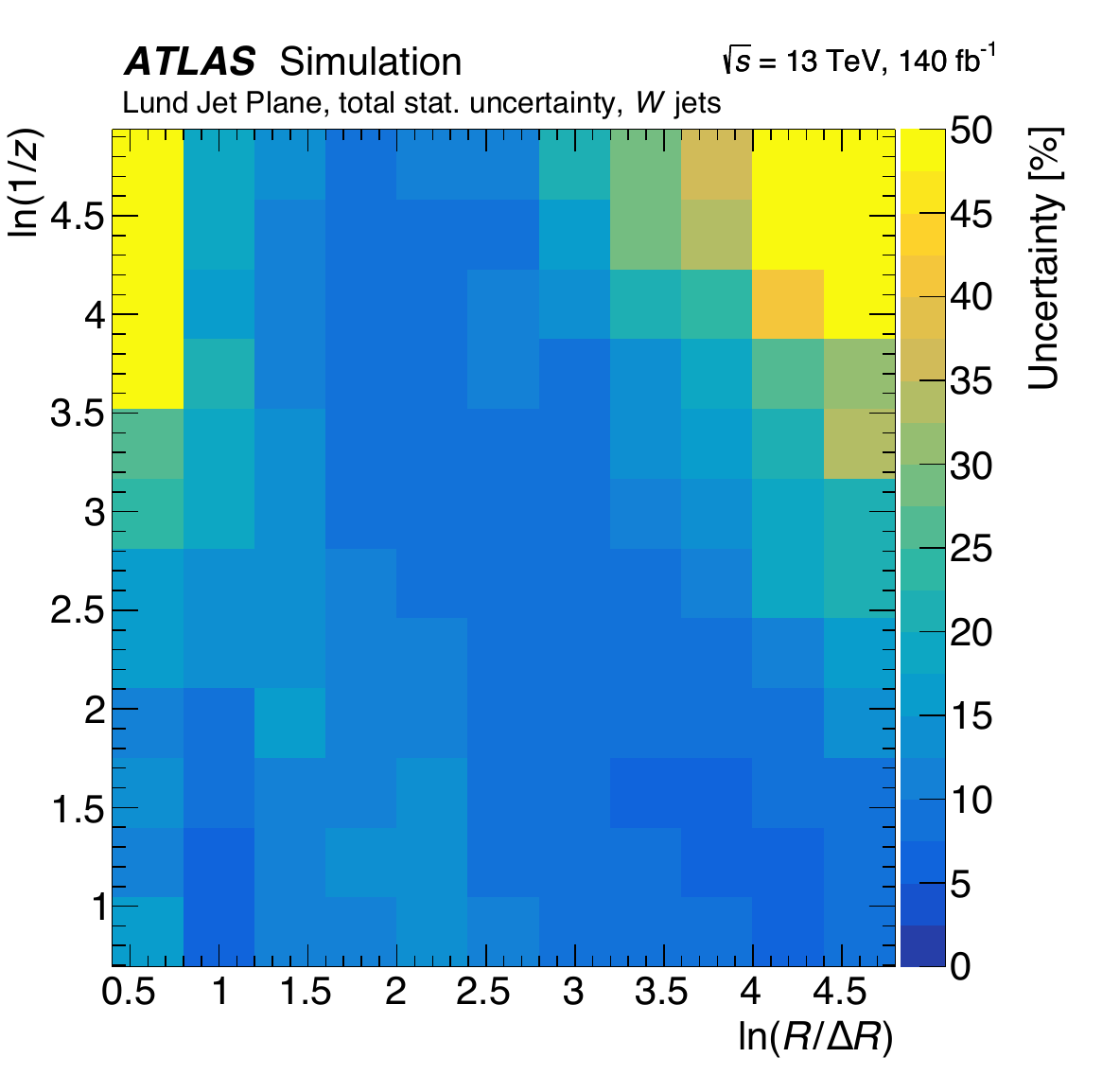}
\caption{}
\end{subfigure}
\caption{Total systematic (a, b) and statistical (c, d) uncertainty across the LJP. Top jets are shown in (a, c), $W$ jets in (b, d).}
\label{fig:2dsyststatunc}
\end{figure}

The total statistical uncertainty and total systematic uncertainty across the LJP are shown in Figure~\ref{fig:2dsyststatunc}. The statistical uncertainty for the top ($W$) jets is in the range $5\%{-}10\%$ ($10\%{-}20\%$) in the bulk of the plane, rising to $50\%$ in the very soft (upper-right) and trimmed (upper-left) edges of the plane. Systematic uncertainties dominate across the whole plane. The total systematic uncertainty outside of the very sparsely populated areas is mostly around $10\%{-}30\%$ ($10\%{-}40\%$). Peaks of up to $45\%$ ($50\%$) are observed where differences in modelling of the parton shower, hadronisation and ME matching are most pronounced, in the densely populated diagonal and lower-left corner of the plane (cf.\ the plots shown in Figure~\ref{fig:mcaltmcratiostopsel}). Large peaks of up to $70\%$ are also observed along the less densely populated left edge of the plane, at wide angles, where the $W$ decay products are less likely to be found due to their large boost along the jet's axis.

A summary of the effect of the various sources of uncertainty on a slice through the LJP in the top and $W$ jets is shown in Figures~\ref{fig:systslice_x} and~\ref{fig:systslice_y}. Here all the detector uncertainties are grouped together (Total detector), as are the background-related uncertainties (Total background model), signal modelling uncertainties (Total $t\bar{t}$ model) and unfolding bias and matching uncertainties (Total unfolding). The average size of the uncertainty in these slices is $15\%$ and $25\%$ for the top and $W$ jets, respectively, with peaks of up to $40\%$ in both selections. The largest components of the signal modelling uncertainty are shown separately. The modelling of the \ttbar\ signal is the largest source of uncertainty in both selections. The leading component depends on the region of the plane considered and is generally either the FSR, ME-matching or PS+Hadronisation uncertainty. The detector uncertainty components, and especially the tracking uncertainties, become relevant only at very small $\Delta R$ where LJP emissions are affected by the detector's spatial resolution. The background and unfolding uncertainties have a negligible impact across the whole plane.

\clearpage
\newpage

\clearpage
\newpage


%
\section{Results}
\label{sec:result}
\begin{figure}[ht]
\centering
\def\sfwidth{0.495\textwidth}
\begin{subfigure}[b]{\sfwidth}
\centering
\includegraphics[width=\textwidth,trim={0.1cm 0 0.5cm 0},clip]{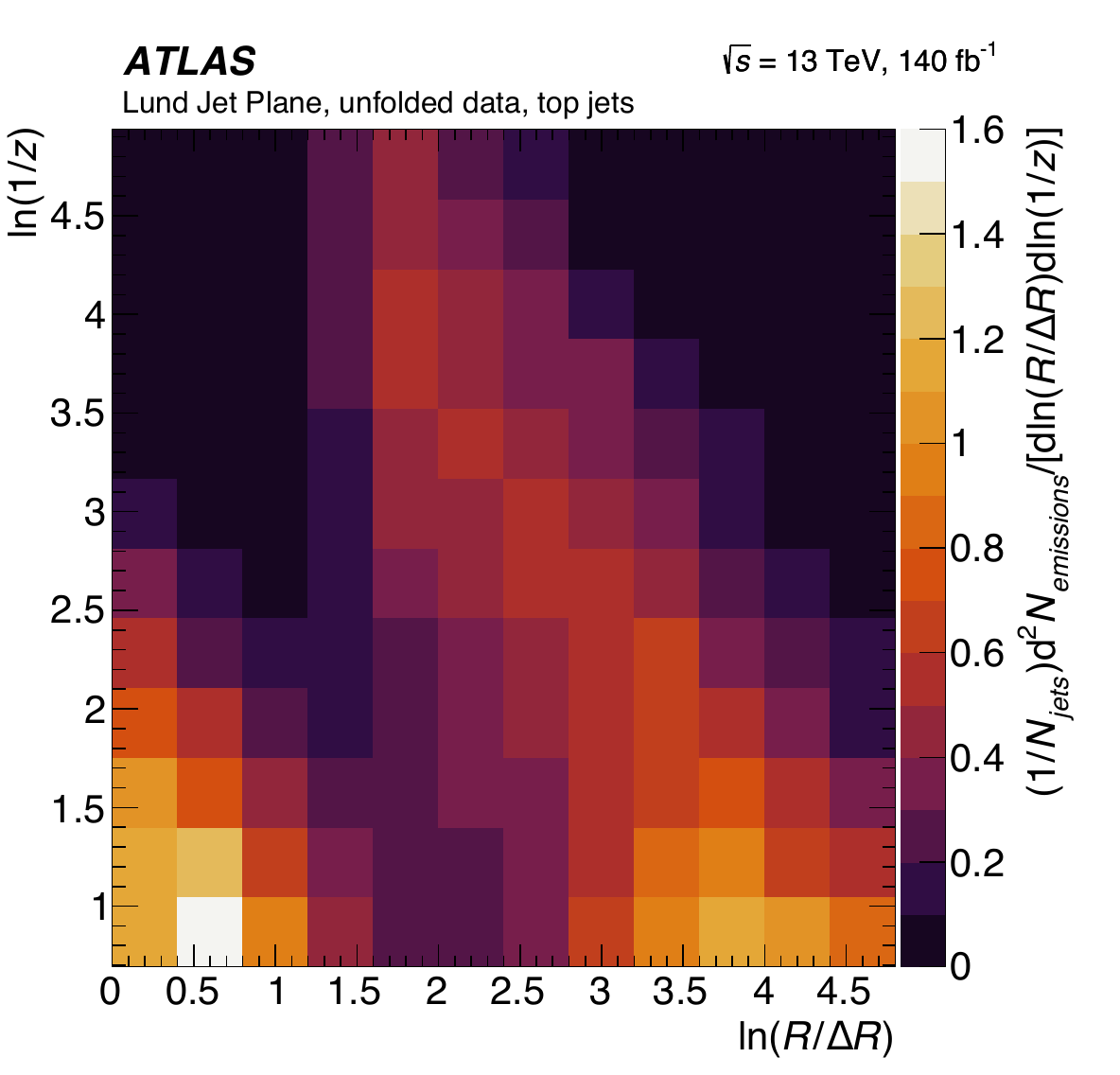}
\caption{}
\label{fig:ljpttbar2dtop}
\end{subfigure}
\begin{subfigure}[b]{\sfwidth}
\centering
\includegraphics[width=\textwidth,trim={0.1cm 0 0.5cm 0},clip]{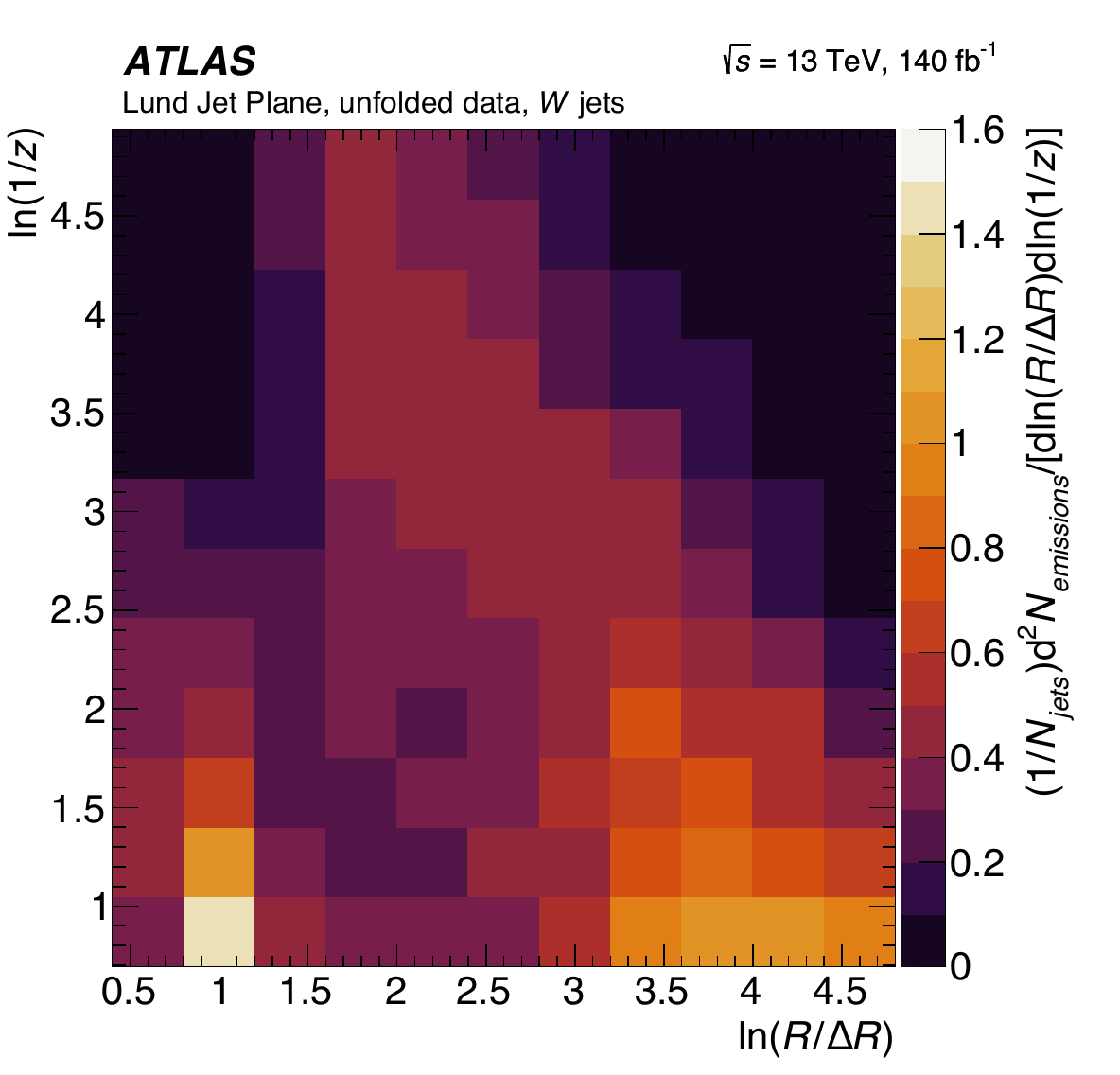}
\caption{}
\label{fig:ljpttbar2dw}
\end{subfigure}
\caption{Measured LJP for the (a) top jet and (b) $W$ jet selections.}
\label{fig:ljpttbar2d}
\end{figure}

The measured density of emissions in the Lund jet plane is shown in Figures~\ref{fig:ljpttbar2dtop} and~\ref{fig:ljpttbar2dw} for the top and $W$ jet selections respectively. As outlined in Section \ref{sec:LJPstructure}, the decays of the high-\pT top quarks and $W$ bosons can be distinguished in the LJP in the region of $\ln(1/z) < 1.5$, $\ln(R/\Delta R)<1.5$. In $\eta{-}\phi$ space, these same hard and wide-angle emissions are what give rise to the three- or two-pronged energy distribution inside the jets. Because the top quark has a larger mass than the $W$ boson, the peak in the lower-left corner of the LJP is shifted towards wider angles for the top jets. The average number of emissions per jet, equivalent to the total integrated density across the LJP shown in Figure~\ref{fig:ljpttbar2d}, is
$6.74 \pm 0.02\,\textrm{(stat.)} \pm 0.13\,\textrm{(syst.)}$ for the top jets and $6.02 \pm 0.04\,\textrm{(stat.)} \pm 0.22\,\textrm{(syst.)}$ for the $W$ jets. This implies that, on average, almost one more emission is reconstructed for the top jets, which contain all the decay products of the top quark, than for the $W$ jets. The estimated total uncertainty is greater for the $W$ jets, mainly due to two effects. The first effect is the background modelling uncertainty, which is about $4$--$5$ times greater in these jets. The second and more important effect, is the increased number of events migrating out of the fiducial region due to narrow mass-window cut, which increases the size of the non-closure uncertainties after unfolding corrections.

In the region where $\Delta R$ and $z$ are both small, the LJP for the top jets and the $W$ jets resembles that of the light-quark-initiated jets measured in Ref.~\cite{STDM-2018-57}. Here, a high density of emissions is observed in the transition region between the perturbative and non-perturbative regimes where $k_\text{T} (\approx z\,\Delta R p_\text{T}^\text{Jet}) = \Lambda_\text{QCD}$. In the upper-right corner, the number of emissions is suppressed. Here $ k_\text{T} < \Lambda_\text{QCD}$, so the density of emissions receives large corrections from non-perturbative terms proportional to powers of ($ k_\text{T} / \Lambda_\text{QCD}$)~\cite{lundplane2018dreyer}.

In the upper-left corner of the LJP, there is a low density of emissions due to the trimming procedure that was applied to the jets' constituents. Since the trimming algorithm proceeds by reclustering the jet constituents with a radius parameter $R_\text{trim} = 0.2$ and subsequently discarding subjets with a $p_\text{T}$-fraction smaller than $f_\text{trim} = 0.05$, one can expect this region to be bounded by $\ln(R/\Delta R) \lesssim \ln(R/R_\text{trim}) \approx 1.6$ and $\ln(1/z) \gtrsim \ln(1/f_\text{trim}) \approx 2.2$.

\subsection{Comparison with MC predictions}
The measured LJP is compared with several Monte Carlo simulations of top-pair production, generated at NLO in QCD and subsequently matched to the PS and hadronisation models, as described in Section~\ref{sec:datamcsamples}.
A quantitative comparison between the measurement and the MC set-ups is obtained by performing $\chi^2$ tests:
\begin{equation}
\label{eq:chi2formula}
\chi^2 = V_{j}^{\mathrm{T}} (C_\mathrm{syst}+C_\mathrm{stat}+C_{j})^{-1} V_{j}
\end{equation}
where $V$ is a vector containing the difference between the measurement and the prediction for each generator set-up $j$, and $C^{-1}$ is the inverse of the covariance matrix, which is calculated from the sum of the following: the statistical covariance ($C_\mathrm{stat}$) and systematic covariance ($C_\mathrm{syst}$) for the measurement, and the statistical covariance for the simulated sample, $C_{j}$. Starting from the systematic uncertainties of the measurement, the matrix $C_\mathrm{syst}$ is derived with the following assumptions: each component is considered uncorrelated with respect to the other systematic uncertainties and, for each component, the bins of the distribution are considered either fully correlated or uncorrelated, depending on the relative sign of the uncertainties in each bin. %
The statistical covariance $C_\mathrm{stat}$ is derived using the `bootstrap' technique, described for the statistical uncertainty in Section~\ref{sec:uncertanities}, where bin-to-bin correlations are calculated from an ensemble of simulated LJP histograms whose statistical variance matches that of the measurement. The statistical uncertainty of the predictions, contained in the matrix $C_{j}$, is only due to the number of simulated entries in each bin. Double-sided uncertainties are not symmetrised prior to creating the covariance matrix and enter just once in the calculation. Correlations between bins are evaluated from the sign of the up/down components of the uncertainties. Two bins are considered correlated if the signs of the components in each bin are the same and anti-correlated if they are opposite.

The $\chi^2$ formula in Eq.~(\ref{eq:chi2formula}) is used to evaluate how well the measurement agrees with the predictions across the whole regions shown in Figure~\ref{fig:ljpttbar2d}, as well as in smaller regions and slices thereof. In order to compare $\chi^2$ values for regions of different size, the number of degrees of freedom (NDF), which is equal to the number of bins in the region investigated, must also be considered. The $p$-value for a given $\chi^2$ and NDF is given by the formula $1 - \Gamma(\text{NDF}/2,\chi^2/2)$, where $\Gamma$ denotes the cumulative distribution function of the gamma distribution.

\begin{figure}[ht]
\centering
\def\sfwidth{ 0.495\textwidth }
\def\fnameprefix{}

\begin{subfigure}[b]{\sfwidth}
\centering
\includegraphics[width=\textwidth,trim={0.5cm 0 2cm 0},clip]{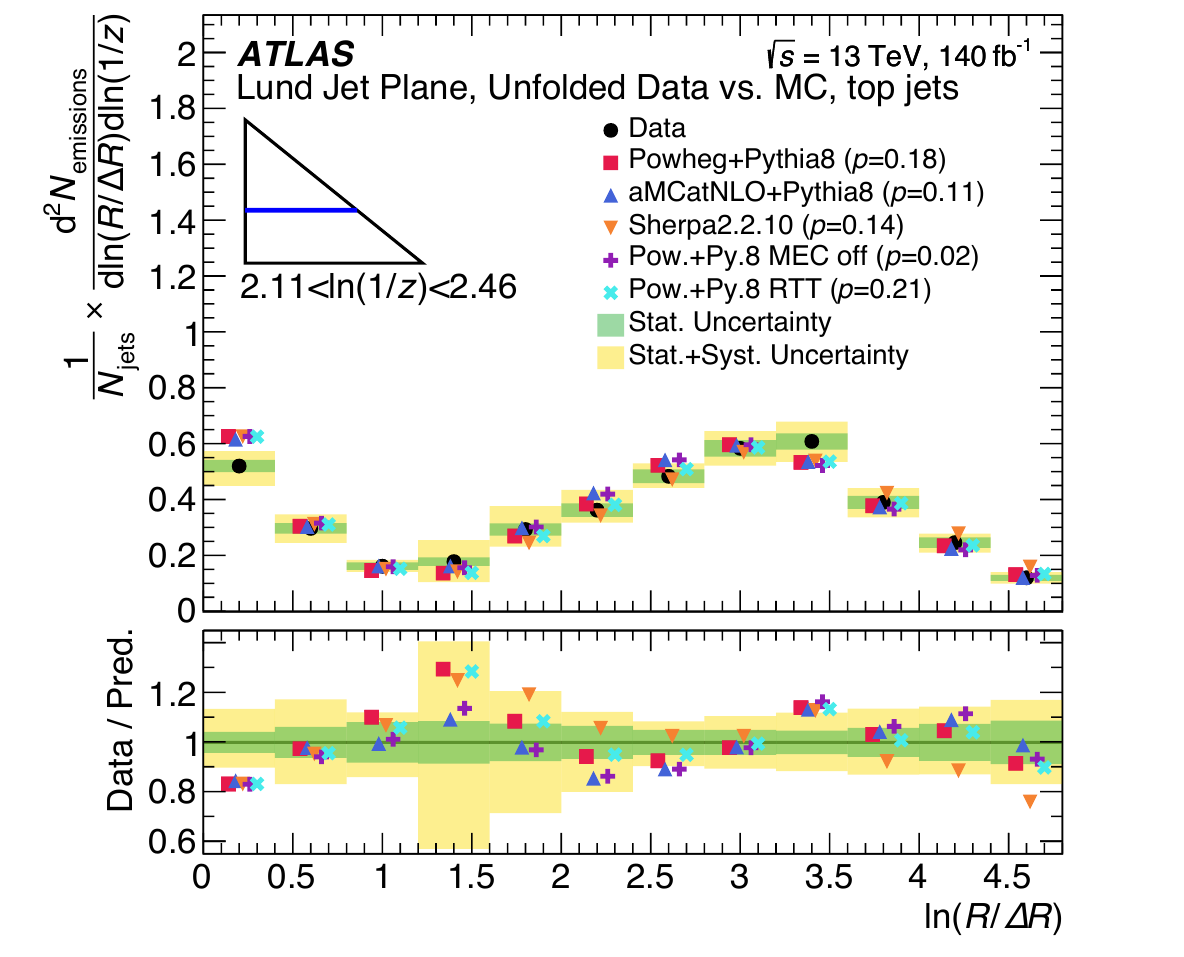}
\caption{}
\end{subfigure}
\begin{subfigure}[b]{\sfwidth}
\centering
\includegraphics[width=\textwidth,trim={0.5cm 0 2cm 0},clip]{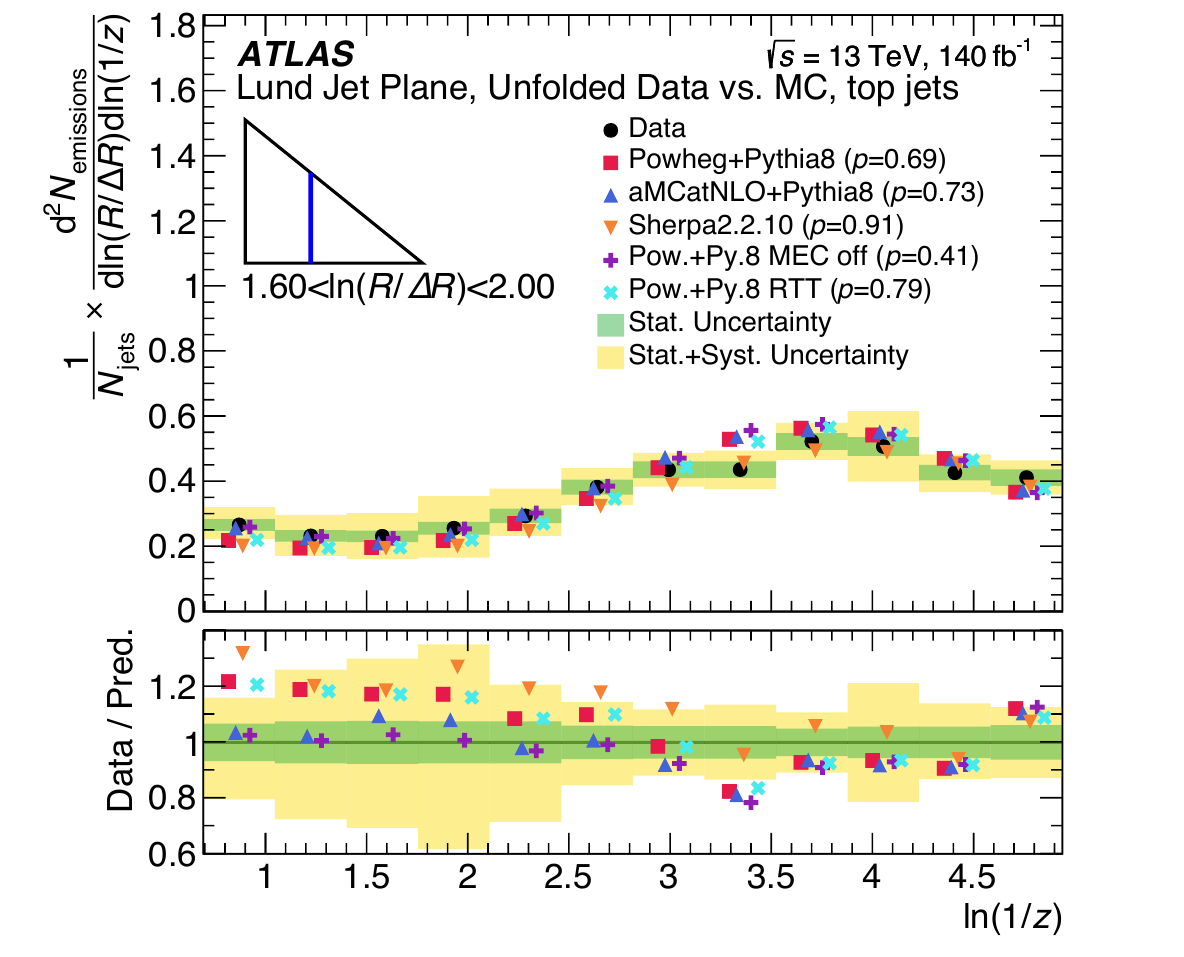}
\caption{}
\end{subfigure}

\caption{Slices through the measured top jet LJP, as a function of (a) $\ln(R/\Delta R)$ and (b) $\ln(1/z)$. The data are compared with several predictions, differing from the nominal \textsc{Powheg+Pythia} 8 predictions in both the hard scatter and showering generator setting. The bottom pad reports the ratio of the data to the different predictions, with both the total and statistical uncertainty of the measured LJP shown as a coloured band. Local $p$-values for the bins in the slice are shown for each prediction. The number of degrees of freedom in the $p$-value calculation is equal to the number of bins in the slice.}
\label{fig:topljpamcssl}
\end{figure}

\begin{figure}[ht]
\centering
\def\sfwidth{ 0.495\textwidth }
\begin{subfigure}[b]{\sfwidth}
\centering
\includegraphics[width=\textwidth,trim={0.5cm 0 2cm 0},clip]{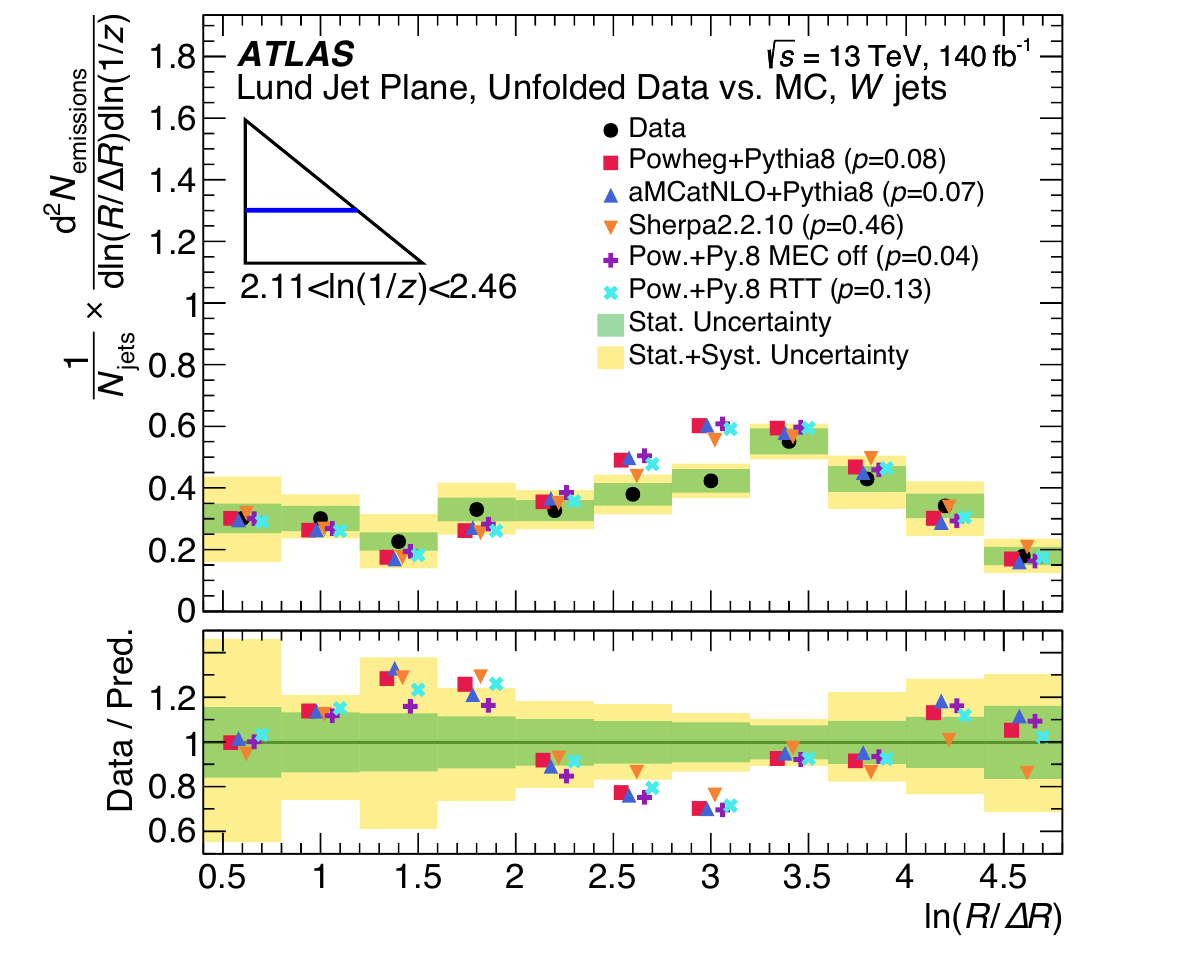}
\caption{}
\label{fig:wljpamcsslx5}
\end{subfigure}
\begin{subfigure}[b]{\sfwidth}
\centering
\includegraphics[width=\textwidth,trim={0.5cm 0 2cm 0},clip]{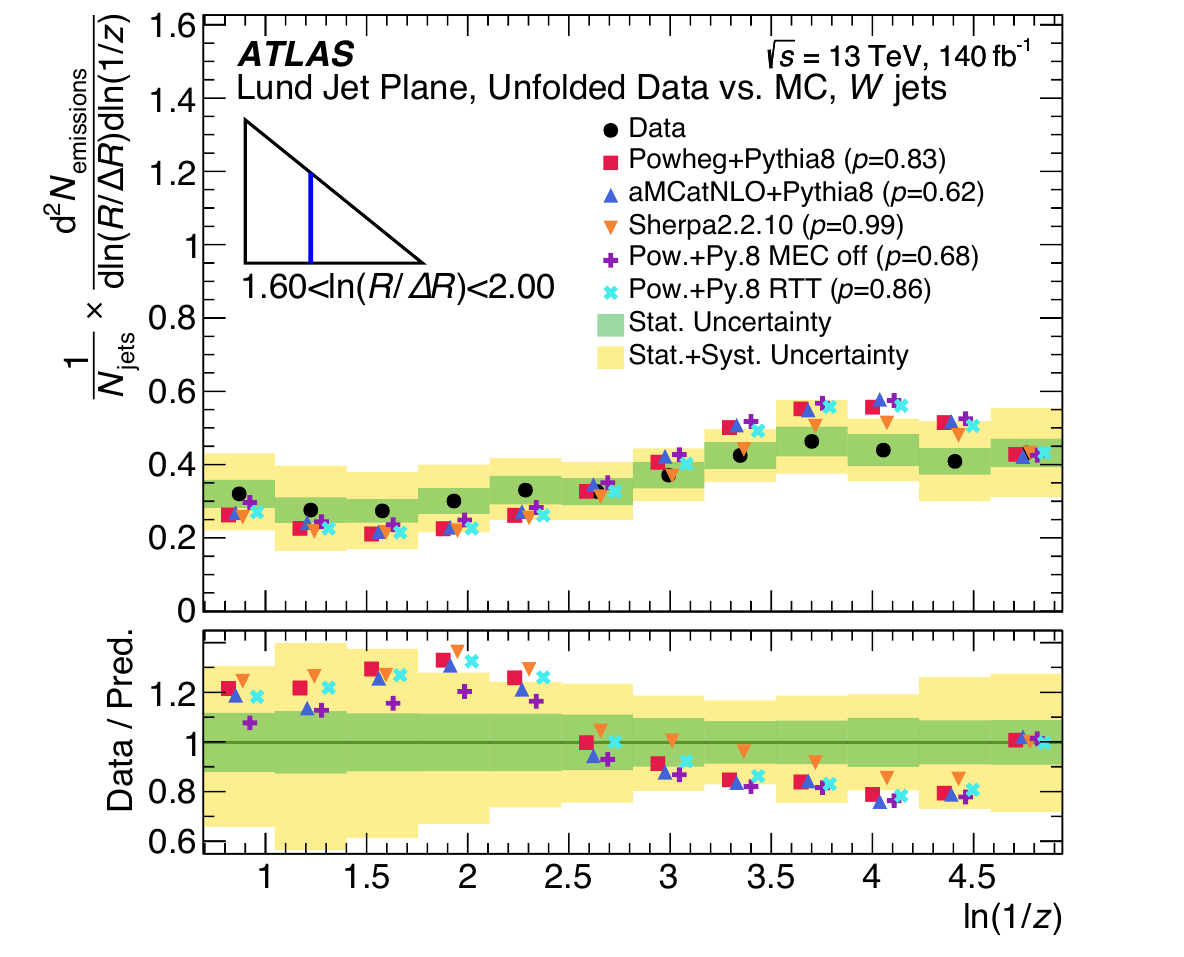}
\caption{}
\end{subfigure}

\caption{Slices through the measured $W$ jet LJP, as a function of (a) $\ln(R/\Delta R)$ and (b) $\ln(1/z)$. The data are compared with several predictions, differing from the nominal \textsc{Powheg+Pythia} 8 predictions in both the hard scatter and showering generator setting. The bottom pad reports the ratio of the data to the different predictions, with both the total and statistical uncertainty of the measured LJP shown as a coloured band. Local $p$-values for the bins in the slice are shown for each prediction. The number of degrees of freedom in the $p$-value calculation is equal to the number of bins in the slice.}
\label{fig:wljpamcssl}
\end{figure}

\begin{figure}[ht]
\centering
\def\sfwidth{ 0.495\textwidth }
\begin{subfigure}[b]{\sfwidth}
\centering
\includegraphics[width=\textwidth,trim={0.5cm 0 2cm 0},clip]{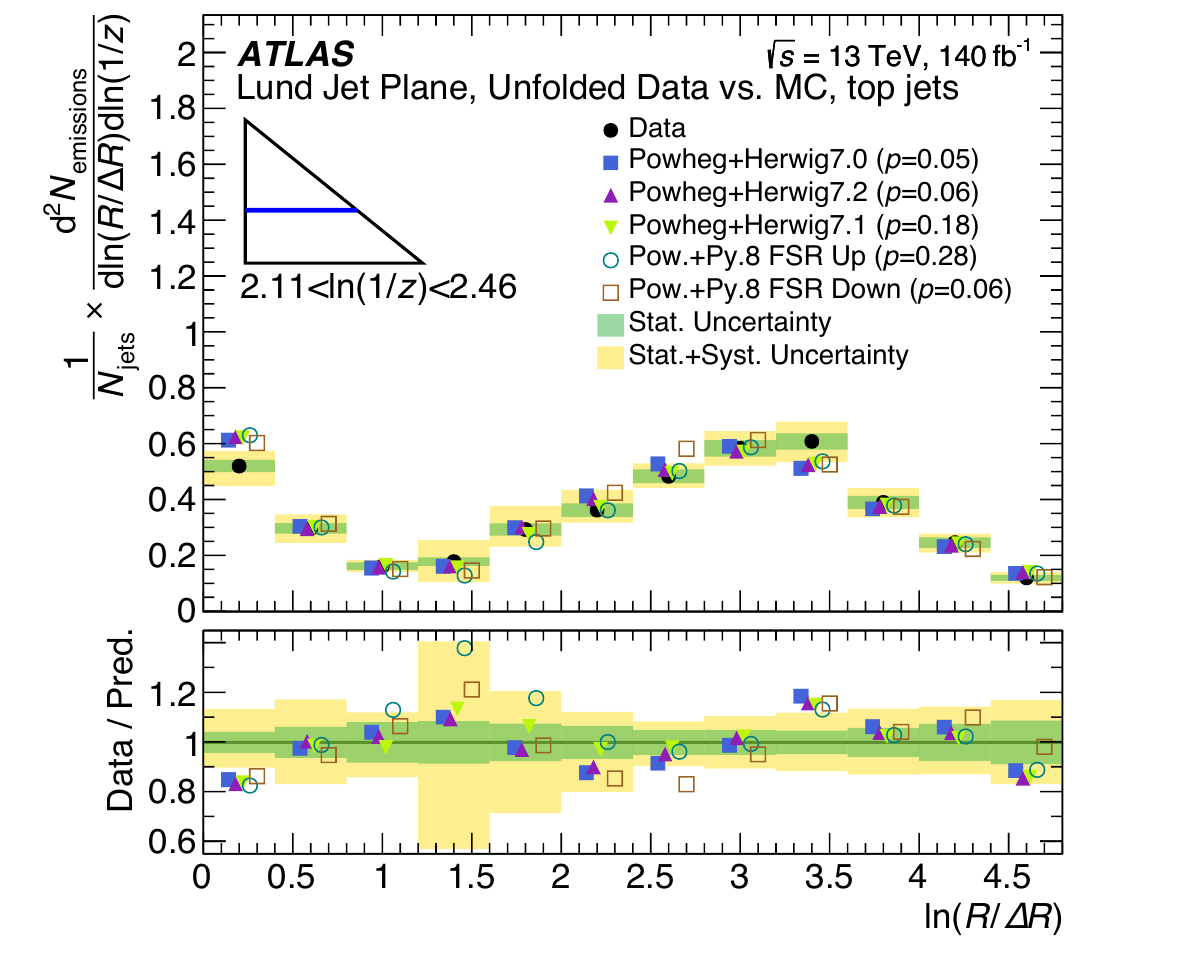}
\caption{}
\end{subfigure}
\begin{subfigure}[b]{\sfwidth}
\centering
\includegraphics[width=\textwidth,trim={0.5cm 0 2cm 0},clip]{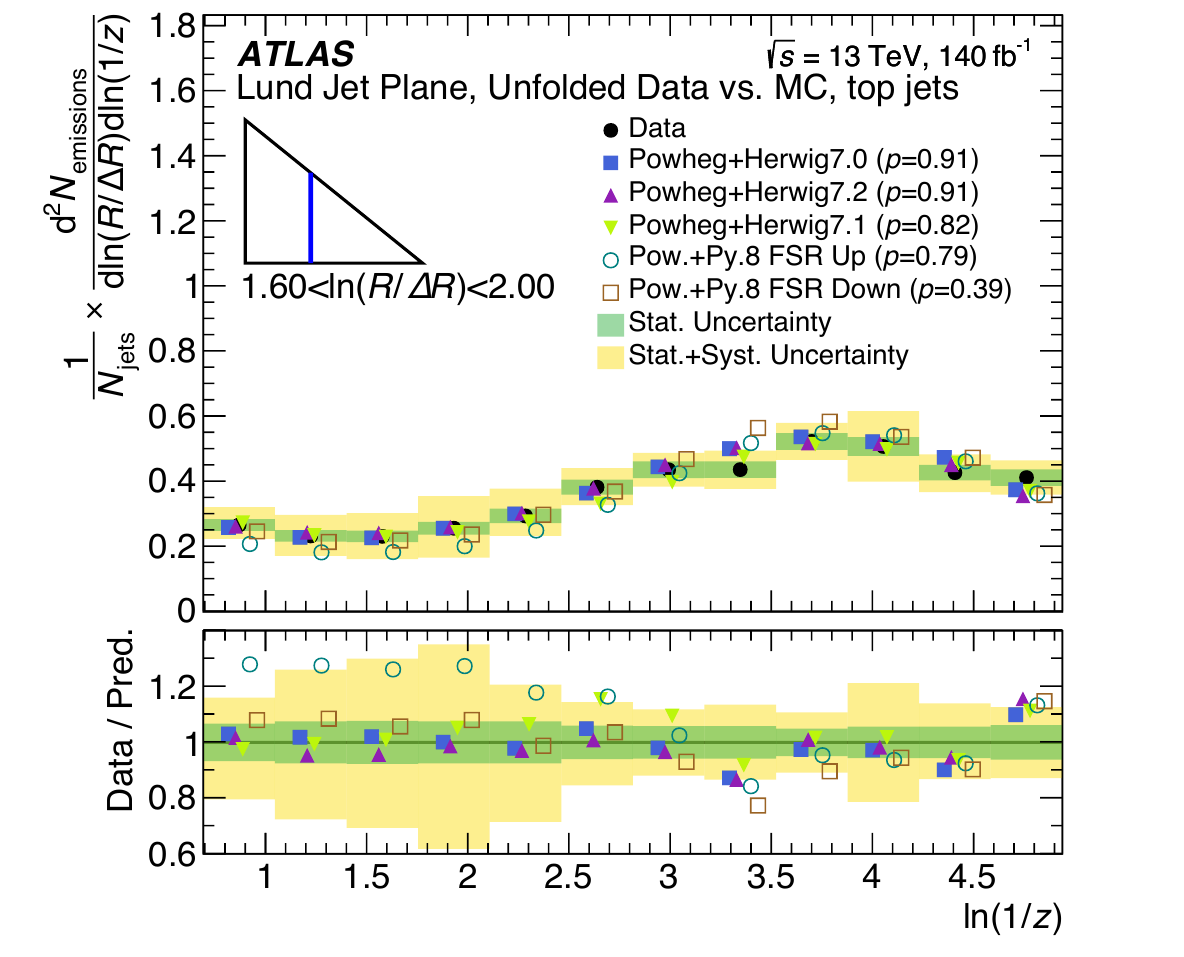}
\caption{}
\end{subfigure}

\caption{Slices through the measured top jet LJP, as a function of (a) $\ln(R/\Delta R)$ and (b) $\ln(1/z)$. The data are compared with several predictions, differing from the nominal \textsc{Powheg+Pythia} 8 predictions in both the hard scatter and showering generator setting. The bottom pad reports the ratio of the data to the different predictions, with both the total and statistical uncertainty of the measured LJP shown as a coloured band. Local $p$-values for the bins in the slice are shown for each prediction. The number of degrees of freedom in the $p$-value calculation is equal to the number of bins in the slice.}
\label{fig:topljpamcssl_fsr}
\end{figure}

\begin{figure}[ht]
\centering
\def\sfwidth{ 0.495\textwidth }
\begin{subfigure}[b]{\sfwidth}
\centering
\includegraphics[width=\textwidth,trim={0.5cm 0 2cm 0},clip]{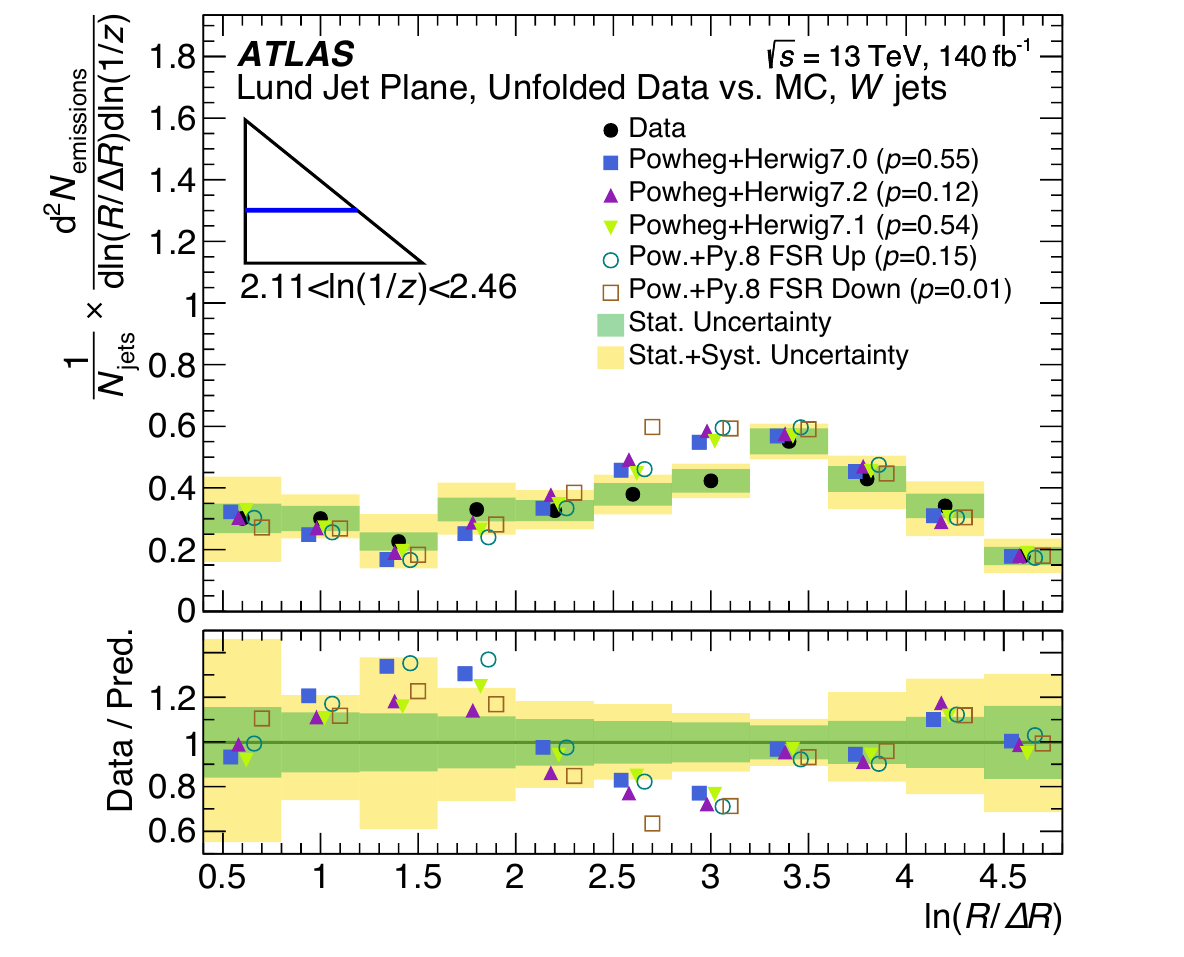}
\caption{}
\label{fig:wljpamcsslx5_fsr}
\end{subfigure}
\begin{subfigure}[b]{\sfwidth}
\centering
\includegraphics[width=\textwidth,trim={0.5cm 0 2cm 0},clip]{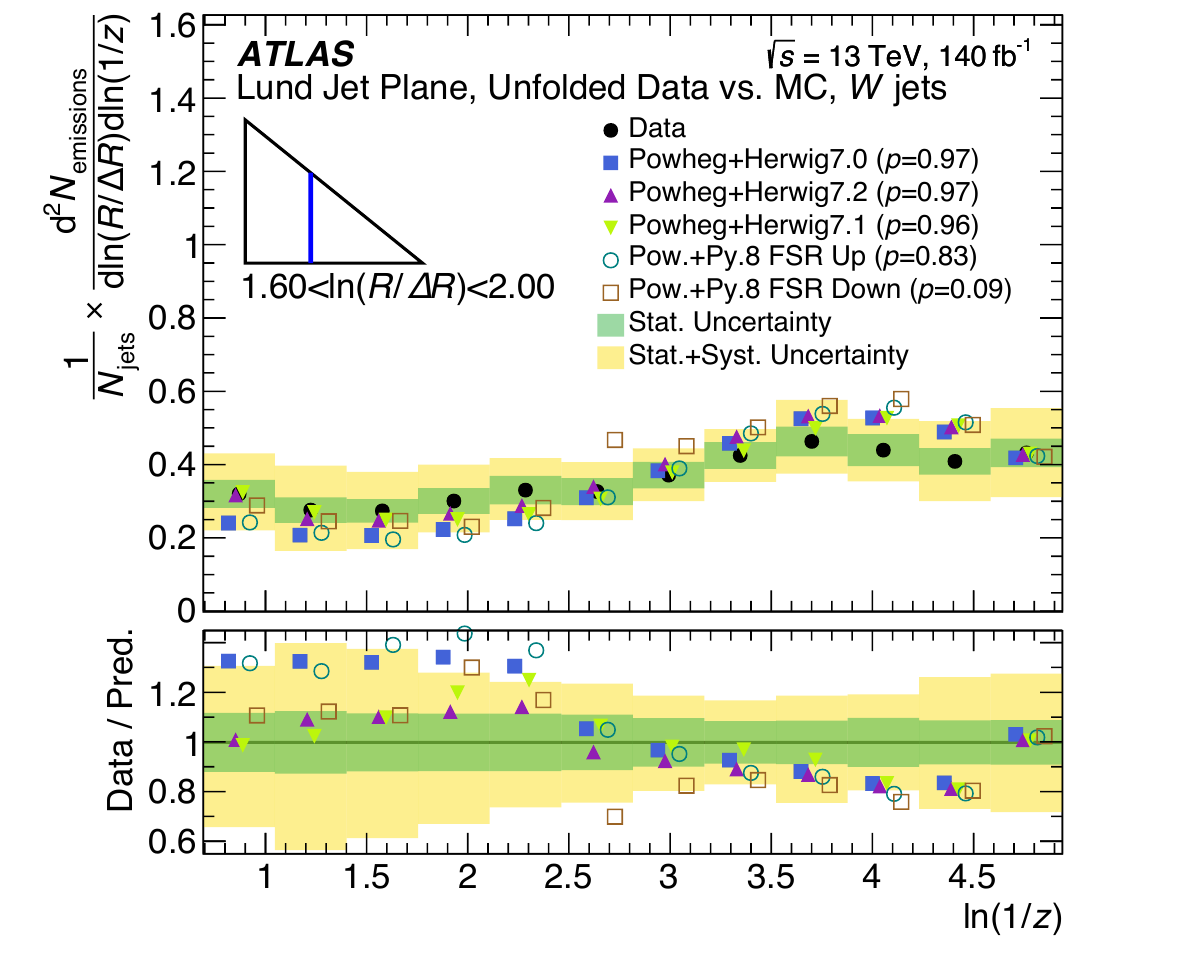}
\caption{}
\end{subfigure}

\caption{Slices through the measured $W$ jet LJP, as a function of (a) $\ln(R/\Delta R)$ and (b) $\ln(1/z)$. The data are compared with several predictions, differing from the nominal \textsc{Powheg+Pythia} 8 predictions in both the hard scatter and showering generator setting. The bottom pad reports the ratio of the data to the different predictions, with both the total and statistical uncertainty of the measured LJP shown as a coloured band. Local $p$-values for the bins in the slice are shown for each prediction. The number of degrees of freedom in the $p$-value calculation is equal to the number of bins in the slice.}
\label{fig:wljpamcssl_fsr}
\end{figure}

Comparisons with various generator set-ups are performed in central slices through the plane in Figures~\ref{fig:topljpamcssl} and~\ref{fig:wljpamcssl}. MC event generator predictions using different matrix-element, parton-shower and hadronisation models are investigated. In Figures~\ref{fig:topljpamcssl_fsr} and~\ref{fig:wljpamcssl_fsr} the measurement is compared with three versions of the \HERWIG[7] PS and hadronisation generator as well as with the \PYTHIA[8] generator with various amounts of FSR.
These slices cover the most densely populated region of the LJP, including the regions with the smallest total uncertainty, while also showing the transitions between regions dominated by ME-matching, parton-shower and hadronisation effects.
Differences between predictions and the data are found to lie within the estimated uncertainties in large sections of the spectra. Sizeable differences are found in the central region of the plane, particularly in the $W$ jet selection, as can be seen in the slices shown in Figures~\ref{fig:wljpamcsslx5} and~\ref{fig:wljpamcsslx5_fsr} where all predictions are found to lie outside the uncertainty band in the two bins covering $2.4 < \ln(R/\Delta R) < 3.2$. As statistical sources contribute more than $50\%$ of the total uncertainty in many bins ($45\%$ of bins in the slice shown in Figures~\ref{fig:wljpamcsslx5} and~\ref{fig:wljpamcsslx5_fsr}), the precision of the measurement could be expected to improve significantly if repeated with a larger dataset.

The level of agreement between the predictions and the data in the slices shown in Figures~\ref{fig:topljpamcssl} to~\ref{fig:wljpamcssl_fsr} is quantified by evaluating a local $p$-value using the method outlined above and considering only the bins within the slice. In both selections, agreement with the various predictions is better for slices along the $\ln(1/z)$ direction. It should be noted that the average uncertainty is larger for these slices than for the $\ln(R/\Delta R)$ slices. Along the $\ln(R/\Delta R)$ direction in the top jets, the $p$-values for the slice shown range from $2\%$ to $28\%$. For the $W$ jets in the same slice, they range from $1\%$ to $55\%$. Only the lowest of these $p$-values, $<5\%$ for \POWPY[8] (FSR Down), implies disagreement with the data. Along the $\ln(1/z)$ direction, $p$-values in the range $39\%{-}91\%$ are observed for the top jets. For the $W$ jets, $p$-values are in the range $62\%{-}99\%$ for all predictions except the \POWPY[8] (FSR Down) sample, which yields a somewhat lower $p$-value of $9\%$. This is likely due to the large differences between this prediction and the measurement around $\ln(1/z) \approx 2.7$ and $\ln(1/z) \approx 4.2$.

A global picture of how well the various generators agree with the measurement is given by the figures shown in Tables~\ref{tab:chi2ljptopjets} and~\ref{tab:chi2ljpwjets}, which report the $\chi^2$ values and corresponding $p$-values for the whole plane. In the top jet selection, disagreement with the measurement is observed for \POWPY[8] (MEC Off) and \POWPY[8] (FSR Down) as well as \POWHER[7.0] and \POWHER[7.2] with $p$-values $<5\%$. The lowest $p$-value, at $1\%$, is obtained for the \POWPY[8] (MEC Off) sample. \SHERPA[2.2.10] shows the highest degree of agreement with the data at a $p$-value of $33\%$. For the $W$ jets, disagreement is observed between all the tested predictions and the measurement, as the $p$-values are less than $1\%$. The greatest disagreement is again observed for \POWPY[8] (MEC Off).

\begin{figure}[ht]
\centering
\def\sfwidth{0.95\textwidth}
\begin{subfigure}[b]{\sfwidth}
\centering
\includegraphics[width=\textwidth,trim={0.2cm 0cm 1.5cm 0cm},clip]{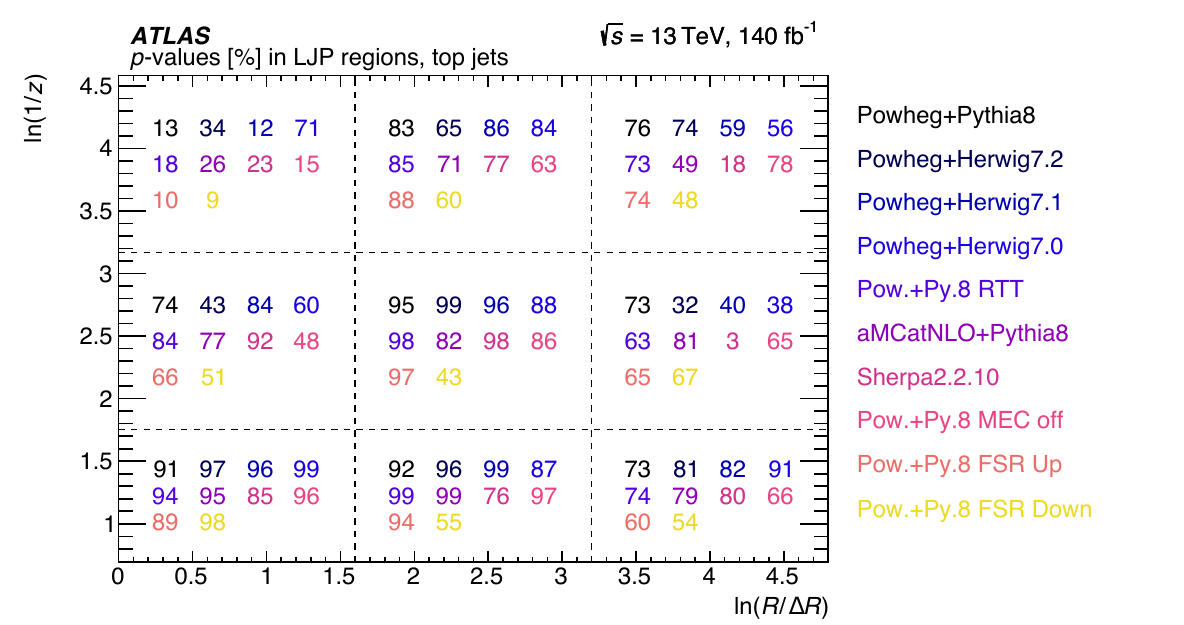}
\caption{}
\end{subfigure}
\begin{subfigure}[b]{\sfwidth}
\centering
\includegraphics[width=\textwidth,trim={0.2cm 0cm 1.5cm 0cm},clip]{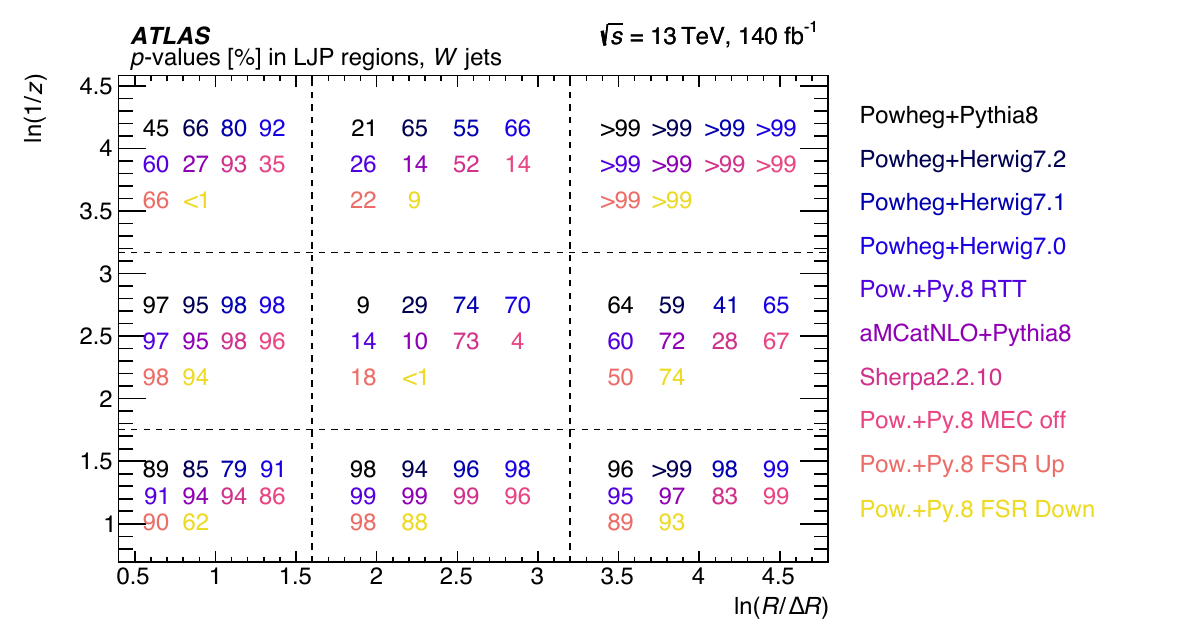}
\caption{}
\end{subfigure}

\caption{The $p$-values for the MC predictions in different regions of the LJP in the (a) top jet and (b) $W$ jet selections. The number of degrees of freedom used in calculating the $p$-values is equal to the number of bins in each region. }
\label{fig:allpvs}
\end{figure}

Local $p$-values are evaluated for nine subregions of the LJPs. These as well as the boundaries of the subregions are shown in Figure~\ref{fig:allpvs}. The values are calculated from the $\chi^2$ and NDF corresponding to the number of bins in each subregion. In the top jets, $p$-values $\leq 10\%$ are observed for \SHERPA[2.2.10], \POWPY[8]\ (FSR Down) and \POWPY[8]\ (FSR Up) in one subregion each. In the $W$ jets, \POWPY[8]\ (FSR Down) yields $p$-values $\leq 10\%$ in three subregions and \POWPY[8]\ (MEC Off) in one. Due to large bin-to-bin correlations across the whole plane, including bins with large data--MC differences, the global $p$-values in Table~\ref{tab:chi2ljptopjets} are generally much smaller than the smallest local $p$-values, which only consider correlations within the respective subregions.

In the lower-left corner of the plane, at wide angles and large momentum fractions, no $p$-value below $50\%$ is observed for either the top jets or the $W$ jets, showing no disagreement between the data and the predictions. The $p$-values for the top jets range from $85\%$ to $99\%$. In the $W$ jets, $p$-values of $85\%$ to $94\%$ are observed for all predictions except \POWPY[8]\ (FSR Down), which gives a $p$-value of $62\%$.

In the subregion at the centre of the plane, $p$-values $<50\%$ are observed for only one MC prediction in the top jets, namely \POWPY[8]\ (FSR Down) at $43\%$. For the $W$ jets, \POWPY[8] (FSR Down) and \POWPY[8]\ (MEC Off) disagree with the measurement with $p$-values of $<1\%$ and $4\%$ respectively, while \POWPY[8] and \AMCatNLO{}+\PYTHIA[8]\ score slightly better at $9\%$, respectively $10\%$.

Low $p$-values are also observed for the $W$ jets in the centre-top subregion at $\ln(1/z)>3.2$, the lowest being a $p$-value of $9\%$ from \POWPY[8] (FSR Down). All but four generators yield $p$-values of $26\%$ or less, the exceptions being \SHERPA[2.2.10] and the three versions of \HERWIG, all with $p$-values $>52\%$.

In the subregion at narrow angles and medium momentum fractions ($3.2<\ln(R/\Delta R) < 4.8$, $1.7<\ln(1/z)<3.2$) in the top jets, a $p$-value of $3\%$ indicates disagreement between \SHERPA[2.2.10] and the data.  Among the generators, \SHERPA[2.2.10] also yields the lowest $p$-value, at $18\%$, in the higher $\ln(1/z)$ subregion ($3.2<\ln(1/z)<4.6$). In the $W$ jets, the lowest degree of agreement is observed for \SHERPA[2.2.10] in the former of the two regions, but at a considerably higher $p$-value of $28\%$. This tendency towards local disagreement with the data may indicate mismodelling of narrow-angle emissions by the \SHERPA[2.2.10]\ parton-shower generator.

In the subregion at wide angles and small momentum fractions ($\ln(R/\Delta R) < 1.6$, $\ln(1/z) > 3.2$) in the top jets, all MC predictions except \POWHER[7.0]\ give $p$-values $\leq 50\%$, although only \POWPY[8]\ (FSR Down) and (FSR Up) are $\leq 10\%$ and none are $<5\%$. This is the subregion in which contributions from the underlying event are most likely to enter, provided that these were not already removed by the trimming algorithm. For the $W$ jets in the said subregion, \POWPY[8]\ (FSR Down) disagrees with the measurement at a $p$-value $<1\%$, while \POWPY[8], \AMCatNLO{}+\PYTHIA[8] and \POWPY[8]\ (MEC Off) yield $p$-values $\leq 50\%$.

\FloatBarrier

\FloatBarrier

\sisetup{round-mode=figures,round-precision=3}
\begin{table}[ht]
\centering
\caption{Values of $\chi^2$ and corresponding $p$-values for the top jets, calculated for the full LJP and the various \ttbar\ Monte Carlo samples. }
\label{tab:chi2ljptopjets}
\begin{tabular}{l@{\hspace{3em}}S S S[round-mode=places,round-precision=0] @{}c }\toprule
\bfseries Sample name & \multicolumn{1}{c}{$\chi^2$} & \multicolumn{1}{c}{$\chi^2/\text{NDF}$ (NDF\,=\,132)} & \multicolumn{1}{c}{$p$-value [$\%$]} & \\
\midrule
Powheg + Pythia 8 & 149 & 1.13 & 15 & \\
aMCatNLO + Pythia 8 & 149 & 1.13 & 14 & \\
Sherpa 2.2.10 & 139 & 1.05 & 33 & \\
Powheg + Herwig 7.0 & 169 & 1.28 & 2 & \\
Powheg + Herwig 7.2 & 165 & 1.25 & 3 & \\
Powheg + Herwig 7.1 & 150 & 1.14 & 14 & \\
Powheg + Pythia 8 MEC Off & 176 & 1.34 & 1 & \\
Powheg + Pythia 8 RTT & 145 & 1.10 & 20 & \\
Powheg + Pythia 8 FSR Up & 148 & 1.12 & 17 & \\
Powheg + Pythia 8 FSR Down & 162 & 1.23 & 4 &
\\\bottomrule
\end{tabular}
\end{table}
\vfill

\sisetup{round-mode=figures,round-precision=3}
\begin{table}[ht]
\centering
\caption{Values of $\chi^2$ for the $W$ jets, calculated for the full LJP and the various \ttbar\ Monte Carlo samples. The corresponding $p$-values are all smaller than $1 \%$.}
\label{tab:chi2ljpwjets}
\begin{tabular}{l@{\hspace{3em}}S S @{}c }\toprule
\bfseries Sample name & \multicolumn{1}{c}{$\chi^2$} & \multicolumn{1}{c}{$\chi^2/\text{NDF}$ (NDF\,=\,121)} & \\
\midrule
Powheg + Pythia 8 & 233 & 1.92 & \\
aMCatNLO + Pythia 8 & 204 & 1.69 & \\
Sherpa 2.2.10 & 168 & 1.39 & \\
Powheg + Herwig 7.0 & 170 & 1.41 & \\
Powheg + Herwig 7.2 & 168 & 1.39 & \\
Powheg + Herwig 7.1 & 177 & 1.46 & \\
Powheg + Pythia 8 MEC Off & 240 & 1.99 & \\
Powheg + Pythia 8 RTT & 227 & 1.88 & \\
Powheg + Pythia 8 FSR Up & 230 & 1.90 & \\
Powheg + Pythia 8 FSR Down & 213 & 1.76 &
\\\bottomrule
\end{tabular}
\end{table}
\vfill

\FloatBarrier

\FloatBarrier


%
\section{Conclusion}
\label{sec:conclusion}
The primary Lund jet plane density has been measured in large hadronic jets initiated by the decays of top quarks or $W$ bosons in 140~fb$^{-1}$ of 13~TeV $pp$ collision data collected with the ATLAS detector at the LHC. The jets were selected from events matching the topology of $t\bar{t}$ pair production in which one of the (anti-)top candidates decays fully hadronically while the other produces a lepton, a neutrino and a $b$-tagged jet. Jets containing the full decay products of either a top quark or its daughter $W$ boson were studied separately by refining the selections on the large jets. This is the first measurement of the LJP for jets produced in the decays of boosted heavy particles, and it is the first LJP measurement by ATLAS to use $R=1.0$ jets. The measurement builds on and complements previous analyses by ATLAS and other LHC experiments, including measurements of the LJP for small-radius jets initiated by light quarks or gluons, as well as many measurements of related jet-substructure observables in jets initiated by top quarks or $W$ bosons.

The LJP is measured at the level of stable particles after removing the effects of limited detector resolution and reconstruction efficiency. The statistical procedure for the removal of detector effects makes use of Monte Carlo event generator predictions. The total uncertainty in the measured observable is dominated by systematic uncertainties due to modelling of the $t\bar{t}$ signal across most of the measured phase space. In the centre of the Lund jet plane, where it is most densely populated (around $0.5 < \ln(R/\Delta R) < 3.5$, $\ln(1/z) < 3.5$), the size of the uncertainty is in the range $15\%{-}40\%$ for both the top and $W$ jets. The modelling uncertainty is dominated by components related to the choice of parton-shower and hadronisation model, and value of $\alphas^{\text{FSR}}$.
The measured LJP has a structure distinctly different from the LJP for jets initiated by light quarks or gluons. Two densely populated regions are visible: one mainly related to the decays of the top quarks and $W$ bosons into highly boosted quarks, and the other corresponding to subsequent emissions of softer QCD radiation.

The measurement is compared with a wide range of predictions using different Monte Carlo generator settings. Alternative MC generator samples were considered for the $t\bar{t}$ production at NLO in QCD and for the parton shower, while the remaining samples were obtained using alternative configurations of the nominal \POWPY[8]\ set-up. A quantitative comparison was performed both in the whole LJP and in multiple subregions of it. For the $W$ jets, no generator agrees with the measurement across the whole LJP, although there is better agreement in the subregions. In the top jets, the highest level of agreement is observed for the \SHERPA[2.2.10] prediction, with a $p$-value of $33\%$. Disagreement with the data is observed for \POWHER[7.0], \POWHER[7.2], \POWPY[8] (MEC Off) and \POWPY[8] (FSR Down), with $p$-values $<5\%$. The extent to which the various generators agree with the measurement was found to be highly dependent on the LJP coordinate ranges considered, as local $p$-values close to 1, and more rarely $< 10\%$, are found when restricting the tests to one of nine subregions of the plane. The lack of overall compatibility between the measurement and the MC predictions, particularly in the $W$ jets, can be compared with results of previous measurements of the LJP in dijet events \cite{STDM-2018-57,CMS:2023lpp}, where differences of more than four times the total uncertainty were observed across many bins making small statistical contributions to the uncertainty.

The results show that many aspects of hadronic top quark or $W$ boson decays, including parton-shower and hadronisation effects, can be investigated by measuring a single two-dimensional jet-substructure observable. The results could be useful for improving the tuning of MC $t\bar{t}$ event generators by targeting the parameters controlling sources of radiation that are poorly modelled in the LJP. They will also be a valuable resource for the ongoing development of new, highly performant top and $W$ jet taggers that make use of the LJP or related jet-substructure observables.


%
\section*{Acknowledgements}
%

%
%

%
%

We thank CERN for the very successful operation of the LHC and its injectors, as well as the support staff at
CERN and at our institutions worldwide without whom ATLAS could not be operated efficiently.

The crucial computing support from all WLCG partners is acknowledged gratefully, in particular from CERN, the ATLAS Tier-1 facilities at TRIUMF/SFU (Canada), NDGF (Denmark, Norway, Sweden), CC-IN2P3 (France), KIT/GridKA (Germany), INFN-CNAF (Italy), NL-T1 (Netherlands), PIC (Spain), RAL (UK) and BNL (USA), the Tier-2 facilities worldwide and large non-WLCG resource providers. Major contributors of computing resources are listed in Ref.~\cite{ATL-SOFT-PUB-2023-001}.

We gratefully acknowledge the support of ANPCyT, Argentina; YerPhI, Armenia; ARC, Australia; BMWFW and FWF, Austria; ANAS, Azerbaijan; CNPq and FAPESP, Brazil; NSERC, NRC and CFI, Canada; CERN; ANID, Chile; CAS, MOST and NSFC, China; Minciencias, Colombia; MEYS CR, Czech Republic; DNRF and DNSRC, Denmark; IN2P3-CNRS and CEA-DRF/IRFU, France; SRNSFG, Georgia; BMBF, HGF and MPG, Germany; GSRI, Greece; RGC and Hong Kong SAR, China; ISF and Benoziyo Center, Israel; INFN, Italy; MEXT and JSPS, Japan; CNRST, Morocco; NWO, Netherlands; RCN, Norway; MNiSW, Poland; FCT, Portugal; MNE/IFA, Romania; MESTD, Serbia; MSSR, Slovakia; ARIS and MVZI, Slovenia; DSI/NRF, South Africa; MICIU/AEI, Spain; SRC and Wallenberg Foundation, Sweden; SERI, SNSF and Cantons of Bern and Geneva, Switzerland; NSTC, Taipei; TENMAK, T\"urkiye; STFC/UKRI, United Kingdom; DOE and NSF, United States of America.

Individual groups and members have received support from BCKDF, CANARIE, CRC and DRAC, Canada; PRIMUS 21/SCI/017, CERN-CZ and FORTE, Czech Republic; COST, ERC, ERDF, Horizon 2020, ICSC-NextGenerationEU and Marie Sk{\l}odowska-Curie Actions, European Union; Investissements d'Avenir Labex, Investissements d'Avenir Idex and ANR, France; DFG and AvH Foundation, Germany; Herakleitos, Thales and Aristeia programmes co-financed by EU-ESF and the Greek NSRF, Greece; BSF-NSF and MINERVA, Israel; Norwegian Financial Mechanism 2014-2021, Norway; NCN and NAWA, Poland; La Caixa Banking Foundation, CERCA Programme Generalitat de Catalunya and PROMETEO and GenT Programmes Generalitat Valenciana, Spain; G\"{o}ran Gustafssons Stiftelse, Sweden; The Royal Society and Leverhulme Trust, United Kingdom.

In addition, individual members wish to acknowledge support from CERN: European Organization for Nuclear Research (CERN PJAS); Chile: Agencia Nacional de Investigaci\'on y Desarrollo (FONDECYT 1190886, FONDECYT 1210400, FONDECYT 1230812, FONDECYT 1230987); China: Chinese Ministry of Science and Technology (MOST-2023YFA1605700, MOST-2023YFA1609300), National Natural Science Foundation of China (NSFC - 12175119, NSFC 12275265, NSFC-12075060); Czech Republic: Czech Science Foundation (GACR - 24-11373S), Ministry of Education Youth and Sports (FORTE CZ.02.01.01/00/22\_008/0004632), PRIMUS Research Programme (PRIMUS/21/SCI/017); EU: H2020 European Research Council (ERC - 101002463); European Union: European Research Council (ERC - 948254, ERC 101089007), Horizon 2020 Framework Programme (MUCCA - CHIST-ERA-19-XAI-00), European Union, Future Artificial Intelligence Research (FAIR-NextGenerationEU PE00000013), Italian Center for High Performance Computing, Big Data and Quantum Computing (ICSC, NextGenerationEU); France: Agence Nationale de la Recherche (ANR-20-CE31-0013, ANR-21-CE31-0013, ANR-21-CE31-0022, ANR-22-EDIR-0002), Investissements d'Avenir Labex (ANR-11-LABX-0012); Germany: Baden-Württemberg Stiftung (BW Stiftung-Postdoc Eliteprogramme), Deutsche Forschungsgemeinschaft (DFG - 469666862, DFG - CR 312/5-2); Italy: Istituto Nazionale di Fisica Nucleare (ICSC, NextGenerationEU), Ministero dell'Università e della Ricerca (PRIN - 20223N7F8K - PNRR M4.C2.1.1); Japan: Japan Society for the Promotion of Science (JSPS KAKENHI JP21H05085, JSPS KAKENHI JP22H01227, JSPS KAKENHI JP22H04944, JSPS KAKENHI JP22KK0227); Netherlands: Netherlands Organisation for Scientific Research (NWO Veni 2020 - VI.Veni.202.179); Norway: Research Council of Norway (RCN-314472); Poland: Ministry of Science and Higher Education (IDUB AGH, POB8, D4 no 9722), Polish National Agency for Academic Exchange (PPN/PPO/2020/1/00002/U/00001), Polish National Science Centre (NCN 2021/42/E/ST2/00350, NCN OPUS nr 2022/47/B/ST2/03059, NCN UMO-2019/34/E/ST2/00393, NCN \& H2020 MSCA 945339, UMO-2020/37/B/ST2/01043, UMO-2021/40/C/ST2/00187, UMO-2022/47/O/ST2/00148, UMO-2023/49/B/ST2/04085); Slovenia: Slovenian Research Agency (ARIS grant J1-3010); Spain: Generalitat Valenciana (Artemisa, FEDER, IDIFEDER/2018/048), Ministry of Science and Innovation (MCIN \& NextGenEU PCI2022-135018-2, MICIN \& FEDER PID2021-125273NB, RYC2019-028510-I, RYC2020-030254-I, RYC2021-031273-I, RYC2022-038164-I), PROMETEO and GenT Programmes Generalitat Valenciana (CIDEGENT/2019/023, CIDEGENT/2019/027); Sweden: Carl Trygger Foundation (Carl Trygger Foundation CTS 22:2312), Swedish Research Council (Swedish Research Council 2023-04654, VR 2018-00482, VR 2022-03845, VR 2022-04683, VR 2023-03403, VR grant 2021-03651), Knut and Alice Wallenberg Foundation (KAW 2018.0157, KAW 2018.0458, KAW 2019.0447, KAW 2022.0358); Switzerland: Swiss National Science Foundation (SNSF - PCEFP2\_194658); United Kingdom: Leverhulme Trust (Leverhulme Trust RPG-2020-004), Royal Society (NIF-R1-231091); United States of America: U.S. Department of Energy (ECA DE-AC02-76SF00515), Neubauer Family Foundation.

%
%


%
%
%
%
%
%
%

%

%
%
%
%
%
%
%

%
%
%
\printbibliography
%
%
%

%

%
%
%
 
\begin{flushleft}
\hypersetup{urlcolor=black}
{\Large The ATLAS Collaboration}

\bigskip

\AtlasOrcid[0000-0002-6665-4934]{G.~Aad}$^\textrm{\scriptsize 103}$,
\AtlasOrcid[0000-0001-7616-1554]{E.~Aakvaag}$^\textrm{\scriptsize 16}$,
\AtlasOrcid[0000-0002-5888-2734]{B.~Abbott}$^\textrm{\scriptsize 121}$,
\AtlasOrcid[0000-0002-0287-5869]{S.~Abdelhameed}$^\textrm{\scriptsize 117a}$,
\AtlasOrcid[0000-0002-1002-1652]{K.~Abeling}$^\textrm{\scriptsize 55}$,
\AtlasOrcid[0000-0001-5763-2760]{N.J.~Abicht}$^\textrm{\scriptsize 49}$,
\AtlasOrcid[0000-0002-8496-9294]{S.H.~Abidi}$^\textrm{\scriptsize 29}$,
\AtlasOrcid[0009-0003-6578-220X]{M.~Aboelela}$^\textrm{\scriptsize 44}$,
\AtlasOrcid[0000-0002-9987-2292]{A.~Aboulhorma}$^\textrm{\scriptsize 35e}$,
\AtlasOrcid[0000-0001-5329-6640]{H.~Abramowicz}$^\textrm{\scriptsize 153}$
% ...
\bigskip
\\

% ...
$^{16}$Department for Physics and Technology, University of Bergen, Bergen; Norway.\\
% ...
$^{29}$Physics Department, Brookhaven National Laboratory, Upton NY; United States of America.\\
% ...
$^{35}$$^{(a)}$Facult\'e des Sciences Ain Chock, Universit\'e Hassan II de Casablanca;$^{(b)}$Facult\'{e} des Sciences, Universit\'{e} Ibn-Tofail, K\'{e}nitra;$^{(c)}$Facult\'e des Sciences Semlalia, Universit\'e Cadi Ayyad, LPHEA-Marrakech;$^{(d)}$LPMR, Facult\'e des Sciences, Universit\'e Mohamed Premier, Oujda;$^{(e)}$Facult\'e des sciences, Universit\'e Mohammed V, Rabat;$^{(f)}$Institute of Applied Physics, Mohammed VI Polytechnic University, Ben Guerir; Morocco.\\
% ...
$^{44}$Physics Department, Southern Methodist University, Dallas TX; United States of America.\\
% ...
$^{49}$Fakult\"{a}t Physik , Technische Universit{\"a}t Dortmund, Dortmund; Germany.\\
% ...
$^{55}$II. Physikalisches Institut, Georg-August-Universit\"{a}t G\"ottingen, G\"ottingen; Germany.\\
% ...
$^{103}$CPPM, Aix-Marseille Universit\'e, CNRS/IN2P3, Marseille; France.\\
% ...
$^{117}$$^{(a)}$New York University Abu Dhabi, Abu Dhabi;$^{(b)}$United Arab Emirates University, Al Ain; United Arab Emirates.\\
% ...
$^{121}$Homer L. Dodge Department of Physics and Astronomy, University of Oklahoma, Norman OK; United States of America.\\
% ...
$^{153}$Raymond and Beverly Sackler School of Physics and Astronomy, Tel Aviv University, Tel Aviv; Israel.\\
% ...
$^{a}$ Also Affiliated with an institute covered by a cooperation agreement with CERN.\\
$^{b}$ Also at An-Najah National University, Nablus; Palestine.\\
$^{c}$ Also at Borough of Manhattan Community College, City University of New York, New York NY; United States of America.\\
$^{d}$ Also at Center for High Energy Physics, Peking University; China.\\
$^{e}$ Also at Center for Interdisciplinary Research and Innovation (CIRI-AUTH), Thessaloniki; Greece.\\
$^{f}$ Also at Centro Studi e Ricerche Enrico Fermi; Italy.\\
% ...

\end{flushleft}

 %
%

\appendix

\end{document}